\documentclass[12pt]{article}
\usepackage{amsfonts}
\usepackage{graphicx}
\usepackage{epstopdf}
\usepackage{epsfig}
\usepackage{amssymb}
\usepackage{setspace}
\usepackage{caption}
\usepackage{amsfonts}
\usepackage{color}
\usepackage{amsmath}
\usepackage{float}
\usepackage{comment}
\newcommand {\be}{\begin{equation}}
\newcommand {\nn}{\nonumber}
\newcommand {\ee}{\end{equation}}
 \newcommand {\bea}{\begin{array}}
 
 \newcommand {\eea}{\end{array}}

 \newcommand {\RN}{Reissner–Nordstrom~}

\textwidth=6.5in \hoffset=-.75in \voffset=-1in \numberwithin{equation}{section}
\numberwithin{figure}{section}

\begin{document}

\begin{titlepage}
	\vspace{1cm}
	\begin{center}
		{\Large \bf {Magnetized Kerr-Newman-Taub-NUT spacetimes}}\\
	\end{center}
	\vspace{2cm}
	\begin{center}
		\renewcommand{\thefootnote}{\fnsymbol{footnote}}
		Masoud Ghezelbash{\footnote{amg142@campus.usask.ca}} , Haryanto M. Siahaan{\footnote{haryanto.siahaan@unpar.ac.id}}\\$^*$Department of Physics and Engineering Physics,\\
		University of Saskatchewan, Saskatoon, Saskatchewan S7N 5E2, Canada\\
		and\\
		$^\dagger$Center for Theoretical Physics,\\
		Department of Physics, Parahyangan Catholic University,\\
		Jalan Ciumbuleuit 94, Bandung 40141, Indonesia
		\renewcommand{\thefootnote}{\arabic{footnote}}
	\end{center}
	
	\begin{abstract}
		We find a new class of exact solutions in the Einstein-Maxwell theory by employing the Ernst magnetization process to the Kerr-Newman-Taub-NUT spacetimes. We study the solutions and find that they are regular everywhere. We also find the quasilocal conserved quantities for the spacetimes, the corresponding Smarr formula and the first law of thermodynamics.
		
	\end{abstract}
\end{titlepage}\onecolumn
\bigskip

\section{Introduction}
\label{sec:intro}

Finding the exact solutions to  the Einstein-Maxwell  theory is always  fascinating, as it opens a door to explore the new aspects of the gravitational physics. The exact solutions to the aforementioned theory contain the black hole solutions, such as the Kerr-Newman family, to a more general spacetime solutions of Plebanski-Demianski \cite{Griffiths:2009dfa}. Different aspects of those solutions have been studied and reported, in which, some can be related to the real astrophysical phenomena, and others are still in vague. Among the latter, is the spacetime solutions with the NUT parameter, which is considered as the extension of the mass parameter. We note that the conserved quantities, such as the mass and angular momentum in a spacetime  with a particular boundary, can be computed, which are related to the  symmetry of the spacetime. However, the NUT parameter is not associated to any symmetry of the spacetime, and yet it also leads to some peculiar properties in the spacetime, such as conical singularity and the regular invariants such as  squared Riemann tensor, at the origin of the coordinate system. Nevertheless, spacetimes with the NUT parameter has helped to shape our understanding of some gravitational and thermodynamical aspects of gravity theories \cite{Jefremov:2016dpi,Cebeci:2015fie,Pradhan:2013hqa,Mukherjee:2018dmm,Bini:2003bm,Paganini:2017qfo,AlonsoAlberca:2000cs,Liu:2010ja,Chen:2006ea,Aliev:2008wv,Sakti:2020jpo,Sakti:2019zix,Ghezelbash:2007kw}

The Kerr-Newman spacetime is a well known black hole solution in the Einstein-Maxwell theory. The solutions can be extended to contain the  NUT parameter, and usually referred to, as the Kerr-Newman-Taub-NUT spacetimes. Despite the conical singularities in the spacetimes,  there are many research works to explore the different aspects of the Kerr-Newman-Taub-NUT black holes. We note the presence of the NUT parameter in the spacetime, leads to the loss of asymptotic flatness, if the corresponding null NUT counterpart has this asymptotic \cite{Griffiths:2009dfa}. It can be shown that the Kerr-Newman-Taub-NUT spacetime is a special case of the Plebanski-Demianski spacetime which is considered as one of the most general solution in Einstein-Maxwell theory that can contain black holes \cite{Plebanski:1976gy}.

In Einstein-Maxwell theory, it also exists an exact solution describing a universe filled by a homogeneous magnetic field known as the Melvin universe \cite{Melvin:1963qx}. A black hole solution in this Melvin universe can be obtained by using the Ernst magnetization \cite{Ernst} applied to a known black hole spacetime in Einstein-Maxwell theory as a seed. In fact, performing Ernst magnetization to the Minkowski spacetime can give us the Melvin universe. In general, the magnetization can be done in two ways, namely at the level of perturbation as Wald introduced in \cite{Wald:1974np}, and as a strong field as Ernst proposed in \cite{Ernst}. In the Wald prescription, Maxwell field is introduced perturbatively by using the Killing vectors associated to the spacetime, while the presence of homogeneous magnetic field does not change the spacetime solution. We can infer that the magnetization by Wald does not change the asymptotic structure of the magnetized spacetime. The superradiant instability in this weakly magnetized black hole had been investigated in \cite{Konoplya:2008hj}, and this type of magnetization for Kerr-NUT-AdS spacetime had been performed in \cite{Frolov:2017bdq}.

This Ernst magnetization itself can be viewed as a type of Harrison transformation \cite{Harrison} which maps an old solution to a new one in the theory. A number of aspects of the known magnetized black hole solutions had been reported in literature \cite{Bicak:2015lxa,Siahaan:2015xia,Astorino:2015naa,Brito:2014nja,Kolos:2015iva,Tursunov:2014loa,Astorino:2016hls,Orekhov:2016bpc,Astorino:2015lca,Booth:2015nwa,Gibbons:2013yq,Gibbons:2013dna,Aliev:1989sw,Galtsov:1978ag,Siahaan:2016zjw,Aliev:1989wx,Aliev:1989wz}, and this shows the importance of such solution in shaping our knowledge on gravity. The most recent ones are the magnetization to \RN-Taub-NUT \cite{Siahaan:2021ags} and Kerr-Taub-NUT \cite{Siahaan:2021uqo}. The work presented in this paper extends the previous works to magnetizing the Kerr-Newman-Taub-NUT (KNTN) spacetimes, which we refer to as the Melvin-Kerr-Newman-Taub-Nut (MKNTN) spacetimes. Though the idea is straightforward and the mechanism is well understood, but incorporating the functions in the solution are quite challenging. 

In this paper, we perform the magnetization procedure to the Kerr-Newman-Taub-NUT spacetimes. We expect to get the magnetized KNTN spacetime solution, whose massless, null NUT, static, and neutral limit,  is the Melvin magnetic universe \cite{Melvin:1963qx}. Some aspects of the spacetime are discussed, such as the deformation of the horizon and the quasilocal conserved quantities associated to the solution. 

The organization of this paper is as follows. In section \ref{sec:MRNTNconstruction}, after reviewing the Ernst magnetization process, we construct the MKNTN solutions by employing the Ernst magnetization to the KNTN metric as the seed solution. In section \ref{sec:prop}, we study some properties of the MKNTN spacetimes. In section \ref{sec:thermo}, we obtain the quasilocal thermodynamical quantities for the MKNTN black holes, as well as the Smarr equation for the MKNTNblack holes and verify the first law of thermodynamics.  We consider the natural units $c={\hbar} = k_B = G_4 = 1$.

\section{Construction of the magnetized spacetimes}
\label{sec:MRNTNconstruction}

\subsection{Ernst magnetization}\label{sec:ErnstMag}

Ernst magnetization is a transformation acting on a set of Ernst potentials which can be defined by using some functions appearing in the seed spacetime solution and the accompanying vector field in Einstein-Maxwell theory. The seed solution is typically expressed in the Lewis-Papapetrou-Weyl (LPW) form
\be\label{metricLPW} 
ds^2  = - f^{ - 1} \left( {\rho ^2 dt^2  - e^{2\gamma } d{\chi} d{\chi} ^* } \right) + f\left( {\omega dt-d\phi } \right)^2,
\ee
where $f$, $\gamma$, and $\omega$ are function of ${\chi}$. Here we have used the $-+++$ signs convention for the spacetime, and $^*$ notation representing the complex conjugation. Using the $f$ function in the LPW line element above, accompanied by the vector ${\bf A} = A_\mu {\rm{d}}x^\mu$, the gravitational Ernst potential,
\be \label{Ernst.potential.Grav}
{\cal E} = f + {{\Phi }{\Phi }^*}   - i\Psi,
\ee 
and the electromagnetic one
\be \label{Ernst.potential.EM}
\Phi  = A_\phi   + i\tilde A_\phi ,
\ee 
can be constructed. The $A_t$ component can be obtained after solving
\be \label{eqA}
\nabla A_t  +\omega \nabla A_\phi + i\frac{\rho }{f}\nabla \tilde A_\phi  =0.
\ee 
Note that the imaginary part of $\Phi$ is the vector field which constructs the dual field strength tensor
\be \label{dualF}
{{\tilde F}_{\mu \nu }} =\frac{1}{2} {\varepsilon _{\mu \nu \alpha \beta }}{F^{\alpha \beta }},
\ee 
where ${{\tilde F}_{\mu \nu }} = {\partial _\mu }{{\tilde A}_\nu } - {\partial _\nu }{{\tilde A}_\mu }$. 

In equation above, the twist potential $\Psi$ is given by the relation
\be \label{eq.Psi}
\nabla \Psi  =i \frac{{ f^2 }}{\rho }\nabla \omega + 2i\Phi ^* \nabla \Phi.
\ee 
Using the Ernst potentials, the following equations can be extracted from the equations of motion in Einstein-Maxwell theory,
\be \label{eq.Ernst.grav}
\left( {{\cal E} +{\cal E}^* + {\Phi \Phi^*}} \right)\nabla^2 {\cal E} = 2\left( {\nabla {\cal E} + 2{\Phi ^*}\nabla \Phi } \right) \cdot \nabla {\cal E},
\ee 
\be \label{eq.Ernst.EM}
\left( {{\cal E} +{\cal E}^* + {\Phi \Phi^*}} \right)\nabla^2 {\Phi} = 2\left( {\nabla {\cal E} + 2{\Phi ^*}\nabla \Phi } \right) \cdot \nabla {\Phi}.
\ee 
The last equation is known as the Ernst equations, and is invariant under some transformation \cite{Kinnersley:1977pg}. 
{\textcolor{black} {We note that all the incorporating functions in the metric (\ref{metricLPW}) depend on $\rho$ and $z$ only, then the operator $\nabla$, in equations (\ref{eqA}), (\ref{eq.Psi}), (\ref{eq.Ernst.grav}) and (\ref{eq.Ernst.EM}) can be defined in the flat Euclidean space
\be\label{metric2rho.z}
d{\chi} d{\chi} ^*  = d\rho ^2  + dz^2,
\ee 
as $\nabla  = \partial _\rho   + i\partial _z$ , where we have set the complex coordinate $d{\chi}  = d\rho  + idz$. 
Moreover, as we explain explicitly in appendix A, we find the following differential equations for the function $\gamma$, 
\be \label{dzgamma1}
{\partial _z}\gamma  = \frac{1}{{2{f^2}\rho }}\left\{ {{\rho ^2}{\partial _\rho }f{\partial _z}f - {f^4}{\partial _\rho }\omega {\partial _z}\omega  + 2f{\rho ^2}\left( {{\partial _\rho }\Phi {\partial _z}{\Phi ^*} + {\partial _\rho }{\Phi ^*}{\partial _z}\Phi } \right)} \right\},
\ee 
and
\be \label{drhogamma1}
{\partial _\rho }\gamma  = \frac{1}{{4{f^2}\rho }}\left\{ {{\rho ^2}\left( {{{\left( {{\partial _\rho }f} \right)}^2} - {{\left( {{\partial _z}f} \right)}^2}} \right) + {f^4}\left( {{{\left( {{\partial _z}\omega } \right)}^2} - {{\left( {{\partial _\rho }\omega } \right)}^2}} \right) + 4f{\rho ^2}\left( {{\partial _\rho }\Phi {\partial _\rho}{\Phi ^*} - {\partial _z }{\Phi ^*}{\partial _z}\Phi } \right)} \right\}.
\ee 
}

According to Ernst, one can magnetized the seed solution described by the line element (\ref{metricLPW} ) and vector solution ${\bf A}$ above by transforming the corresponding Ernst potentials 
\be \label{magnetization}
{\cal E} \to {\cal E}' = \Lambda ^{ - 1} {\cal E}~~~{\rm and}~~~\Phi  \to \Phi ' = \Lambda ^{ - 1} \left( {\Phi  - b {\cal E}} \right),
\ee
where 
\be \label{LambdaDEF}
\Lambda  = 1 - 2b\Phi  + b^2 {\cal E}.
\ee 
Here, the constant $b$ is interpreted as the external magnetic field strength in the spacetime\footnote{For economical reason, we prefer to express the magnetic parameter as $b$ instead of $B/2$ as appeared in \cite{Ernst}. The relation is $B=2b$.}. The transformation (\ref{magnetization}) leaves equations (\ref{eq.Ernst.grav}) and (\ref{eq.Ernst.EM}) unchanged for the new potentials ${\cal E}'$ and $\Phi'$. In other words, the new metric consisting the functions $f'$ and $\omega'$, together with the new vector potentials $A'_t$ and $A'_\phi$ are also solutions to the Einstein-Maxwell field equations.

In particular, the transformed line element (\ref{metricLPW}) resulting from the magnetization (\ref{magnetization}) has the components
\be \label{fp}
f' = {\rm Re}\left\{{\cal E'}\right\} - \left| {\Phi '} \right|^2  =\left| \Lambda \right|^{-2}  f,
\ee 
and
\be \label{wp}
\nabla \omega ' = \left| \Lambda \right|^{2} \nabla \omega  - \frac{\rho }{f}\left( {\Lambda ^* \nabla \Lambda  - \Lambda \nabla \Lambda ^* } \right),
\ee 
while the function $\gamma$ remains unchanged. {\textcolor{black}{In appendix \ref{app.gamma}, we present an example, which shows the differential equations for the function $\gamma$, and the invariance of the function $\gamma$ under the Ernst magnetization process.
}

{\textcolor{black}{We note that typical black hole solutions in the Einstein-Maxwell theory, are more compact where they are expressed in the Boyer-Lindquist type coordinates $\left\{ {t,r,x = \cos \theta ,\phi } \right\}$.} Consequently, the LPW type metric (\ref{metricLPW}) with stationary and axial Killing symmetries will have the metric function that depend on $r$ and $x$, and the corresponding flat metric line element reads
\be \label{metric2rx}
{\rm{d}}{\chi} {\rm{d}}{\chi} ^*  = \frac{{{\rm{d}}r^2 }}{{\Delta _r }} + \frac{{{\rm{d}}x^2 }}{{\Delta _x }},
\ee 
where $\Delta _r = \Delta _r \left(r\right)$ and $\Delta _x = \Delta _x \left(x\right)$. Therefore, the corresponding operator $\nabla$ will read $\nabla  = \sqrt {\Delta _r } \partial _r  + i\sqrt {\Delta _x } \partial _x $. Furthermore we can have $\rho^2 = \Delta_r\Delta_x$, then eq. (\ref{eqA}) gives us
\be \label{drAt}
\partial _r A_t  =  - \omega \partial _r A_\phi   + \frac{{\Delta _x }}{f}\partial _x \tilde A_\phi,
\ee 
and
\be \label{dxAt}
\partial _x A_t  =  - \omega \partial _x A_\phi   - \frac{{\Delta _r }}{f}\partial _r \tilde A_\phi.
\ee 
The last two equations are useful later in obtaining the $A_t$ component associated to the magnetized spacetime according to (\ref{magnetization}). To end some details on magnetization procedure, another equations which will be required to complete the metric are
\be \label{drwp}
\partial _r \omega ' = \left| \Lambda  \right|^2 \partial _r \omega  + \frac{{\Delta _x }}{f}{\mathop{\rm Im}\nolimits} \left\{ {\Lambda ^* \partial _x \Lambda  - \Lambda \partial _x \Lambda ^* } \right\},
\ee 
and
\be \label{dxwp}
\partial _x \omega ' = \left| \Lambda  \right|^2 \partial _x \omega  - \frac{{\Delta _r }}{f}{\mathop{\rm Im}\nolimits} \left\{ {\Lambda ^* \partial _r \Lambda  - \Lambda \partial _r \Lambda ^* } \right\}.
\ee 
In the following section, we employ this magnetization scheme to the Taub-NUT spacetime.

\subsection{The Melvin-Kerr-Newman-Taub-NUT spacetimes}\label{sec:MKNTNsol}

To obtain the desired magnetized solution, we use the Ernst potentials that belong to Kerr-Newman-Taub-NUT system,
\be 
d{s^2} =  - \frac{{{\Delta _r}}}{{{\Sigma}}}{\left( {dt - \left( {a{\Delta _x} - 2lx} \right)d\phi } \right)^2} + {\Sigma}\left( {\frac{{d{r^2}}}{{{\Delta _r}}} + \frac{{d{x^2}}}{{{\Delta _x}}}} \right) + \frac{{{\Delta _x}}}{{{\Sigma}}}{\left( {adt - \left( {{a^2} + {l^2} + {r^2}} \right)d\phi } \right)^2},
\ee 
and 
\be 
{A_\mu }d{x^\mu } = \frac{{qr}}{{{\rho ^2}}}\left( {dt + \left( {2lx - a{\Delta _x}} \right)d\phi } \right),
\ee 
where $\Delta_r = r^2 -2mr +a^2+q^2-l^2$, $\Delta_x = 1-x^2$, and $\Sigma = r^2 + \left(ax+l\right)^2$. In the form of LPW line element (\ref{metricLPW}), the above spacetime metric, associates to the functions
\[ 
f = \Sigma^{-1} \left\{\left[3{x}^{2}+1\right] l^4 -4ax\Delta_x l^3 +\left[{a}^{2}{x}^{4}+ \left(8mr -8{a}^{2}-4{q}^{2}-6{r}^{2} \right) {x}^{2}+3{a}^{2}+2{r}^{2}\right] l^2 \right.
\]
\be 
\left. +4ax\Delta_x\left[\Delta_r+l^2\right] l + \Delta_x\left[r^4 + r^2 a^2 \left(1+x^2\right) +2ra^2 m \Delta_x+{a}^{2} \left( {a}^{2}{x}^{2}+{q}^{2}{x}^{2}-{q}^{2} \right)  \right]\right\},
\ee 
\[
\omega = \left[\Delta_x a \left(2l^2+2mr-q^2\right)+2l\Delta_r x\right] \left\{\left[3{x}^{2}+1\right] l^4 -4ax\Delta_x l^3  \right.
\]
\[
+\left[{a}^{2}{x}^{4}+ \left(8mr -8{a}^{2}-4{q}^{2}-6{r}^{2} \right) {x}^{2}+3{a}^{2}+2{r}^{2}\right] l^2 +4ax\Delta_x\left[\Delta_r+l^2\right] l 
\]
\be 
\left. + \Delta_x\left[r^4 + r^2 a^2 \left(1+x^2\right) +2ra^2 m \Delta_x+{a}^{2} \left( {a}^{2}{x}^{2}+{q}^{2}{x}^{2}-{q}^{2} \right)  \right]\right\}^{-1},
\ee 
\[
{e^{2\gamma }} = \Delta_x r^4 +\left[2{l}^{2}-{a}^{2}{x}^{4}-4al{x}^{3}-6{l}^{2}{x}^{2}+4alx+{a}^{2}
\right] r^2 + 2m \left[ a{x}^{2}+2lx-a \right] ^{2} r
\]
\[ 
-{a}^{2} \left[ {a}^{2}-{l}^{2}+{q}^{2} \right] x^4  +\left[{a}^{4}-8{a}^{2}{l}^{2}+2{a}^{2}{q}^{2}+3{l}^{4}-4{l}^{2}{q}^{2}\right] x^2 
\] 
\be
-4al \left[ {a}^{2}-{l}^{2}+{q}^{2} \right] x^3 +4 al \left[ {a}^{2}-{l}^{2}+{q}^{2} \right] x +3{a}^{2}{l}^{2}-{a}^{2}{q}^{2}+{l}^{4},\label{e2gamma}
\ee
and $\rho^2 = \Delta_x \Delta_r$. 

From this seed solution, one can construct the corresponding Ernst potentials as follows
\be\Phi  = \frac{{qrx + iq\left( {lx - a} \right)}}{{l + ax + ir}},\ee
and
\be
{\cal E} = \frac{{\cal E}_R + i {\cal E}_I}{l+ax+ir},
\ee
where
\be
{\cal E}_R = -a\Delta_r x^3 -3l\Delta_r x^2 +a \left({a}^{2}-5{l}^{2}-6mr+2{q}^{2}+{r}^{2} \right) x + l \left( 3{a}^{2}+{l}^{2}-{r}^{2} \right),
\ee
\be
{\cal E}_I =\left( 2{a}^{2}m-{a}^{2}r+2{l}^{2}m-3{l}^{2}r+{q}^{2}r-{r}^{3} \right) {x}^{2}-2al \left( m-2r \right) x+2{a}^{2}m+{a}^{2}r
+3{l}^{2}r+{r}^{3} .
\ee

The magnetized Ernst potentials can be obtained from the seed ones above, which yields to the magnetized metric with the new functions $f'$ and $\omega'$, while $\gamma$ is unchanged.

{\textcolor{black}{ In fact, the differential equations for the function $\gamma(r,x)$, are given by
\begin{eqnarray}
&&{f^2}\left( {2{\Delta _r}x{\partial _r}\gamma  - {\Delta _x}{\partial _r}{\Delta _r}{\partial _x}\gamma } \right) + 2f\rho^2\left( {{\partial _r}\Phi {\partial _x}{\Phi ^*} + {\partial _r}{\Phi ^*}{\partial _x}\Phi } \right)\nonumber\\
 \label{eq.dgamma11} 
 &-& {f^2}\left( {{f^2}{\partial _r}\omega {\partial _x}\omega  + x{\partial _r}{\Delta _r}} \right) + \rho^2{\partial _r}f{\partial _x}f = 0,
 \end{eqnarray}
and
\begin{eqnarray} 
&& 2{\rho ^2}{f^2}\left( {{\partial _r}{\Delta _r}{\partial _r}\gamma  + 2x{\partial _x}\gamma } \right) - {\rho ^2}\left( {{\Delta _r}{{\left( {{\partial _r}f} \right)}^2} - {\Delta _x}{{\left( {{\partial _x}f} \right)}^2}} \right)
\label{eq.dgamma22}\nonumber\\
&+& {f^4}\left( {{\Delta _r}{{\left( {{\partial _r}\omega } \right)}^2} - {\Delta _x}{{\left( {{\partial _x}\omega } \right)}^2}} \right) -4{\rho ^2}f\left( {{\Delta _r}{\partial _r}\Phi {\partial _r}{\Phi ^*} - {\Delta _x}{\partial _x}\Phi {\partial _x}{\Phi ^*}} \right) = 0.
\end{eqnarray}
Similar equations hold for $\gamma'(r,x)$ with $f \rightarrow f'$, $\omega \rightarrow \omega'$ and $\Phi \rightarrow \Phi'$. We explicitly check that equations (\ref{eq.dgamma11}) and (\ref{eq.dgamma22}) and their counterparts for $\gamma'$ imply the metric function $\gamma'(r,x)$ is the same as $\gamma(r,x)$, which is given by (\ref{e2gamma}).
}

To summarize the results, the Melvin-Kerr-Newman-NUT black hole is given by
\be\label{MAG} 
{\rm{d}}s^2  = - f'^{ - 1}(r,x) \left\{{\rho ^2(r,x) {\rm{d}}t^2  - e^{2\gamma(r,x) } \left(\frac{{{\rm{d}}r^2 }}{{\Delta _r }} + \frac{{{\rm{d}}x^2 }}{{\Delta _x }}\right)}\right\} + f'(r,x)\left( {\omega'(r,x) {\rm{d}}t-{\rm{d}}\phi } \right)^2,
\ee
together with the Maxwell's field
\be
A'=A'_\phi(r,x) d\phi+A'_t (r,x)dt,\label{AMAG}
\ee
where  the metric functions $f'(r,x)$ and $\omega'(r,x) $ are given by
\be \label{FPMAG}
f'(r,x) = \frac{{\sum\limits_{k = 0}^4 {{c_k(r)}{x^k}} }}{{\sum\limits_{j = 0}^6 {{d_j(r)}{x^j}} }},
\ee 
and
\be \label{OPMAG}
\omega '(r,x) = -\frac{{\sum\limits_{k = 0}^6 {{{\tilde c}_k(r)}{x^k}} }}{{\sum\limits_{j = 0}^4 {{c_j(r)}{x^j}} }}.
\ee 

Moreover, the components of the Maxwell's field (\ref{AMAG}) are given by
\be
A'_\phi(r,x)=\frac{\sum \limits_{i=0}^6 a_i(r) x^i}{\sum \limits_{i=0}^6 b_i(r) x^i},\label{APMAG}
\ee
and
\be
A'_t(r,x)=\frac{\sum \limits_{i=0}^{11} e_i(r) x^i}{\sum \limits_{i=0}^{10} f_i(r) x^i}.\label{ATMAG}
\ee
\
The expressions for $a_i,b_i,\cdots , f_i$, as functions of $r$, are given in appendix B.


\section{Some properties of the Melvin-Kerr-Newman-Taub-NUT spacetimes}
\label{sec:prop}

The largest root of the metric function $\Delta_r$  describes the outer event horizon of the black hole (\ref{MAG}), which is given by
\be
r_H=m+\sqrt{m^2+l^2-a^2-q^2},\label{HORIZON}
\ee
{\textcolor{black}{
which implies $m^2+l^2  \geq  a^2+q^2$, to have a real value for the outer event horizon. The inner event horizon is located at $r_-=m-\sqrt{m^2+l^2-a^2-q^2}$. We note that the inner event horizon exists at a real positive (or zero) value $r_- \geq 0$, if
$a^2+q^2  \geq   l^2$. Combining the former and the latter inequalities, we find the following range for the summation of squares of the rotational parameter and the electric charge of the spacetime (\ref{MAG}).
\be
 l^2 \leq  a^2+q^2 \leq m^2+l^2.
\ee
}}
The trace of the energy-momentum tensor for the Maxwell's field (\ref{AMAG}) is identically zero. We have verified exactly that the metric (\ref{MAG}) with the Maxwell's field  (\ref{AMAG}) satisfy exactly all the Einstein-Maxwell  field equations. The Ricci scalar of the spacetime is identically zero and the Ricci square invariant  ${\cal R}=R^{\mu\nu}R_{\mu\nu}$  is regular everywhere including  $r=0$.  The expression for the ${\cal R}$ is very long and so we don't present it here. We also find the Kretschmann invariant ${\cal K}=R^{\mu\nu\rho\sigma}R_{\mu\nu\rho\sigma}=\frac{{\cal K}_1(r,x)}{{\cal K}_2(r,x)}$ where ${\cal K}_1(r,x)$ and ${\cal K}_2(r,x)$ are two functions with coefficients of the black hole parameters.  Though the expression for ${\cal K}_1(r,x)$ is very complicated, however 
\be
{\cal K}_2(r,x)=4f'^2(r,x)\rho^8(r,x)e^{12\gamma(r,x)}.\label{K2}
\ee

The location of curvature/coordinate singularities for the black hole (\ref{MAG}) can be determined by the equation ${\cal K}_2(r,x)=0$. In fact, beside $\rho(r,x)$=0, we find the following equation for the location of singularities, which is expectedly independent of the magnetic field
\begin{eqnarray}
&&3\,{a}^{2}{l}^{2}-{a}^{2}{q}^{2}+{l}^{4}+ \left( 4\,{a}^{3}l-4\,a{l}^
{3}+4\,al{q}^{2} \right) x+ \left( {a}^{4}-8\,{a}^{2}{l}^{2}+2\,{a}^{2
}{q}^{2}+3\,{l}^{4}-4\,{l}^{2}{q}^{2} \right) {x}^{2}\nonumber\\
&+& \left( -4\,{a}^
{3}l+4\,a{l}^{3}-4\,al{q}^{2} \right) {x}^{3}+ \left( -{a}^{4}+{a}^{2}
{l}^{2}-{a}^{2}{q}^{2} \right) {x}^{4}+2\,m \left( a{x}^{2}+2\,lx-a \right) ^{2}r\nonumber\\
&+& \left( -{a}^{2}{x}^{4}-4\,al{x}^{3}-6\,{l}^{2}{x}^{2}+4\,alx+{a}^{2}
+2\,{l}^{2} \right) {r}^{2}+\left( 1-x^2\right) {r
}^{4}=0.\label{singu}
\end{eqnarray}
The event horizon and $x=\pm 1$ are the roots of $\rho(r,x)=0$, where ${\cal K}$ is regular and finite. Moreover, it seems $f'(r,x)=0$ or equation (\ref{singu}) leads to other singular points, however, we verify that at those points, the Kretschmann invariant ${\cal K}$ remains completely finite. 

We notice from equation (\ref{singu}) and table \ref{tab:table-name}, that the only magnetized spacetimes with the point singularity, at $r=0$, are Melvin-Schwarzschild and Melvin-Reissner-Nordstrom space-times. All other magnetized spacetimes, i.e.  Melvin, Melvin-Kerr, Melvin-NUT, Melvin-Kerr-Newmann, Melvin-Kerr-NUT, Melvin-Reissner-Norstrom-NUT and MKNTN are completely regular at $r=x=0$.   

In the special case, where all the black hole parameters $q,\,m,\,a,\,l$ approach zero, we find the metric (\ref{MAG}) reduces to
\be
ds_0^2=\Theta (r,x) \left(-dt^2+dr^2+\frac{r^2}{\Delta_x}dx^2\right)+\frac{r^2\Delta_x}{\Theta(r,x)}d\phi ^2,\label{MEL}
\ee
{\textcolor{black}{where 
		\be
		\Theta (r,x)=(1+b^2r^2\Delta_x)^2.
		\ee}}
{The metric (\ref{MEL}) describes the spacetime of axisymmetric universe filled by parallel magnetic forcelines known as the Melvin magnetic universe \cite{Melvin:1963qx}. Interestingly, these forcelines do not contract and collapse under their own gravity, and the corresponding stability against radial perturbations had been investigated in \cite{thorne}.} In fact, the Maxwell's field (\ref{AMAG}) becomes
\be
A=-\frac{br^2\Delta_x}{\sqrt{\Theta(r,x)}}d\phi,
\ee
 which generates the magnetic field
 \be
 \vec B=\frac{2br^2x}{\Theta(r,x)}\hat r+\frac{2br\Delta_x}{\Theta(r,x)}\hat \theta\label{BMelvin}.
 \ee
\begin{center}
\begin{table}
\begin{tabular}{ |c|c||c|c| } 
 \hline
 Spacetime & Is r=0=x singularity?& M-Spacetime & Is r=0=x singularity?\\ 
 \hline\hline
  Minkowski     & No  & Melvin & No \\ 
 \hline
 Schwarzschild & Yes & M-Schwarzschild &Yes \\ 
 \hline
Kerr & Yes & M-Kerr & No \\ 
 \hline
NUT & No & M-NUT & No \\ 
 \hline
Reissner-Nordstrom & Yes & M-Reissner-Nordstrom & Yes \\ 
 \hline \hline
Kerr-Newman & Yes & M-Kerr-Newman & No \\
\hline
 Kerr-NUT & No & M-Kerr-NUT & No \\
 \hline
 Reissner-Nordstorm-NUT & No & M-Reissner-Nordstrom-NUT & No \\ 
 \hline\hline
 Kerr-Newman-NUT & No & M-Kerr-Newman-NUT & No\\
 \hline
\end{tabular}
\caption{\label{tab:table-name}}
\end{table}
\end{center}
The Ricci scalar of the Melvin space-time (\ref{MEL}) is identically zero, while the Kretschmann invariant is given by
 \be
 {\cal K}=64b^4\frac{3\,{b}^{4}{r}^{4}\Delta_x^2-6\,{b}^{2}{r}^{2}\Delta_x+5 }{\Theta^4(r,x)}.\label{KMel}
 \ee
 Moreover, we find the asymptotic behaviour of the black hole (\ref{MAG}), where $r \rightarrow \infty$, by analyzing all the metric functions at large values of the radial coordinate. We find the asymptotic expressions for the metric functions, as
 \begin{eqnarray}
 f' &\sim& \frac{\epsilon^2}{b^4\Delta_x}+{\cal O}(\epsilon^3),\label{a1}\\
 e^{2\gamma} &\sim & \frac{\Delta_x}{\epsilon^4}+{\cal O}({\epsilon^{-2}}),\\
 \rho^2 &\sim & \frac{\Delta_x}{\epsilon^2}+{\cal O}({\epsilon^{-1}}),\\
 \omega' &\sim & \frac{2b^4lx(x^2+3)}{\epsilon^2}+{\cal O}({\epsilon^{-1}}),\label{a2}
 \end{eqnarray}
 where $\epsilon \rightarrow 0$. 
 
 Using expressions (\ref{a1})-(\ref{a2}), we find the asymptotic metric of the black hole (\ref{MAG}), as
 \be
 d{s}_{r \rightarrow \infty}^2= d{s}_{0,r \rightarrow \infty}^2+\frac{4lx(x^4+2x^2+3)}{\Delta_x^2}dtd\phi, \label{MAGINF}
 \ee
 where $d{s}_{0,r \rightarrow \infty}^2$ is the asymptotic of the Melvin universe (\ref{MEL}). We note that the presence of the NUT charge makes an off-diagonal term to the asymptotic metric of the black hole (\ref{MAG}).
 
We consider now the set of 3-dimensional surfaces at a fixed value for the radial coordinate $r=\varrho$. The induced metric on the surface $r=\varrho$, is given by
 \be
 {\rm{d}}s^2_{r=\varrho}  = - f'^{-1}(\varrho,x) \left( {\rho ^2(\varrho,x) {\rm{d}}t^2  - e^{2\gamma(\varrho,x) } \frac{dx^2}{\Delta _x} } \right) + f'(\varrho,x)\left( {\omega' (\varrho,x){\rm{d}}t-{\rm{d}}\phi } \right)^2.\label{slice3d}
 \ee
The determinant of the metric (\ref{slice3d}) is 
\be
\det (g_{ij})=-\frac{f'(\varrho,x)}{e^{2\gamma(\varrho,x)}(\varrho-r_H)(\varrho-r_-)}.\label{det3}
\ee
From (\ref{det3}), we notice that the surface $r=\varrho$ describes a $2+1$-dimensional space-time, if $\varrho > r_H$ or $\varrho < r_-$. On the other hand for $r_- < \varrho < r_H$, the surface $r=\varrho$ describes a $3$-dimensional space. Of course for $\varrho=r_-$ or $\varrho =r_H$, the surface becomes a null surface.

We note that due to the presence of the NUT charge as well as the magnetic field, the horizon geometry is a distorted sphere. In fact the two-dimensional horizon is given by the line element
\be\label{MAGHORIZON} 
{\rm{d}}s^2_H  = \frac{ f'^{ - 1}(r_H,x) e^{2\gamma(r_H,x) }}{\Delta _x} {\rm{d}}{x}^2 + f' (r_H,x){\rm{d}}\phi  ^2.
\ee
The two grand circles on the horizon, one at $x=0$ and the other passing through $x=\pm 1$, have two different circumferences. The former circumference is given by
\begin{eqnarray}
&&\frac{C_{x=0}^2}{4\pi^2( {3\,{a}^{2}{l}^{2}+2\,{a}^{2}mr_H-{a}^{2}{q}^{2}+{
a}^{2}{r}^{2}_H+{l}^{4}+2\,{l}^{2}{r}^{2}_H+{r}^{4}_H})}=
9\,{a}^{4}{b}^{4}{l}^
{2}+4\,{a}^{4}{b}^{4}{m}^{2}+4\,{a}^{4}{b}^{4}mr_H\nn\\
&+&6\,{a}^{2}{b}^{4}{l}^{4}+12\,{a}^{2}{b}^{4}{l}^{2}mr_H+4\,{a}^{2}{b}^{4
}m{r}^{3}_H+2\,{a}^{2}{b}^{4}{r}^{4}_H+{b}^{4}{l}^{6}+7\,{b}^{4}{l}^{4}{r_H}
^{2}+7\,{b}^{4}{l}^{2}{r}^{4}_H+{b}^{4}{r}^{6}_H+{a}^{4}{b}^{4}{r}^{2}_H\nn\\
&+&8\,{a}^{3}{b}^{3}mq+4\,{a
}^{3}{b}^{3}qr_H+12\,a{b}^{3}{l}^{2}qr_H+4\,a{b}^{3}q{r}^{3}_H+6\,{a}^{2}{b}
^{2}{l}^{2}+4\,{a}^{2}{b}^{2}mr_H+4\,{a}^{2}{b}^{2}{q}^{2}+2\,{a}^{2}{b}
^{2}{r}^{2}_H\nn\\
&+&4\,{b}^{2}{l}^{2}{r}^{2}-H+2\,{b}^{2}{l}^{4}+2\,{b}^{2}{r}^{
4}_H+4\,abqr_H+{l}^{2}+{r}^{2}_H.
\end{eqnarray}
In figure \ref{fig:Cx0}, we plot the typical behaviour of the equatorial circumference versus the NUT charge $l$ and the magnetic field $b$, where we set $a=6,\,m=7$ and $q=1$.

\begin{figure}[H]
\centering
\includegraphics[width=0.325\textwidth]{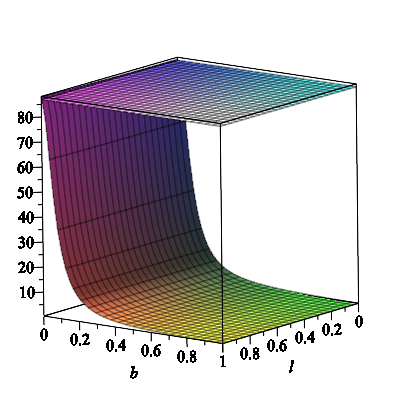}
\caption{The equatorial circumference $C_{x=0}$ versus the NUT charge $l$ and the magnetic field $b$, where we set the other black hole parameters $a=6,\,m=7$ and $q=1$. The horizontal plane shows the equatorial circumference (which is equal to $87.4$ in arbitrary unites), where the NUT charge and the magnetic field are zero.}
\label{fig:Cx0}
\end{figure}
The latter circumference involves an integral, which we can't find it explicitly as an exact form. In figure \ref{fig:Cx1b}, we plot the result of numerical integration for the circumference, as a function of the magnetic field, where we set the other black hole parameters as $a=6,\,l=1,\,m=7$ and $q=1$.
\begin{figure}[H]
\centering
\includegraphics[width=0.325\textwidth]{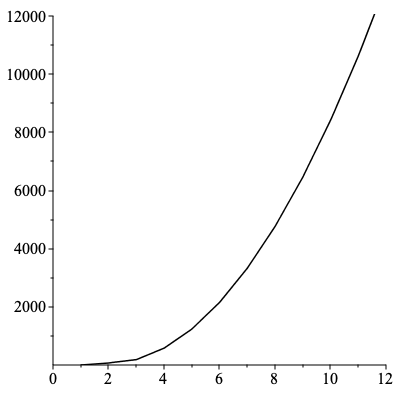}
\caption{The circumference of the great circle, passing through the north and south pole, versus the magnetic field $b$, where we set the other black hole parameters $a=6,\,l=1,\,m=7$ and $q=1$. The  circumference with no magnetic field is equal to $71.9$ in arbitrary units.}
\label{fig:Cx1b}
\end{figure}

The other interesting surfaces for the black hole (\ref{MAG}) are the stationary limit surfaces. The stationary limit surfaces $r=r_s(x)$ are the roots of eqution
\be
\rho (r_s(x),x)\pm f'(r_s(x),x)\omega'(r_s(x),x)=0.\label{stationary}
\ee
For a generic black hole (\ref{MAG}), the equation (\ref{stationary}) turns out to be 
\be
S(r_s,x)\equiv\sum \limits_{n=0}^{10}\alpha_n(a,b,l,m,q,x) r_s^n=0,\label{eqforrs}
\ee
which of course, is not solvable by radicals. Hence, we consider the black hole with the same parameters $a=6,\,b=2,\,l=1,\,m=7$ and $q=1$, which we considered before in this section.  The inner and outer event horizons are at $r_-=3.4$ and $r_H= 10.6$, respectively.

In figure \ref{fig:ergo}, we plot the function $S(r,x)$ versus $r \geq r_H$ and $-1 \leq x \leq 1$. The outer stationary limit surfaces  $r=r_s(x)$ are the intersection of the curve with horizontal plane at $0$, where $S(r_s,x)=0$. The outer ergoregion for the black hole (\ref{MAG}), is the region between $r=r_H$ and $r=r_s(x)$, where $S(r,x) > 0$.  

In figure \ref{fig:ergoinn}, we plot the function $S(r,x)$ versus $r \leq r_-$ and $-1 \leq x \leq 1$. The inner stationary limit surfaces  $r_s(x)$ are the intersection of the curve $S(r,x)$  with the horizontal plane at $0$, where $S(r_s,x)=0$. The inner ergoregion for the black hole (\ref{MAG}), is the region between $r=r_s(x)$ and $r=r_-$, where $S(r,x) > 0$.  We also note that for $ r_- < r < r_H$, the function $S(r,x)$ is positive everywhere, as shown in figure \ref{fig:between}.

\begin{figure}[H]
\centering
\includegraphics[width=0.325\textwidth]{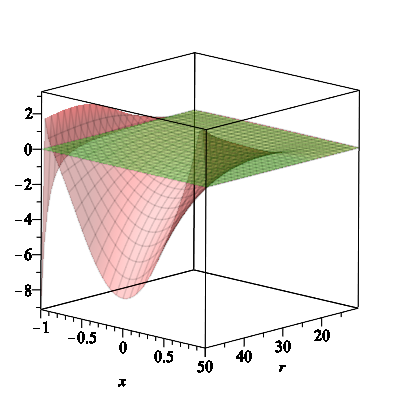}
\caption{The scaled function $S(r,x)$ versus $r \geq r_H$ and $-1 \leq x \leq 1$, where we set the black hole parameters $a=6,\,b=2,\,l=1,\,m=7$ and $q=1$. The stationary limit surfaces $r_s(x)$ are the intersection of the curve $S(r,x)$ with the horizontal plane at zero.}
\label{fig:ergo}
\end{figure}

\begin{figure}[H]
\centering
\includegraphics[width=0.325\textwidth]{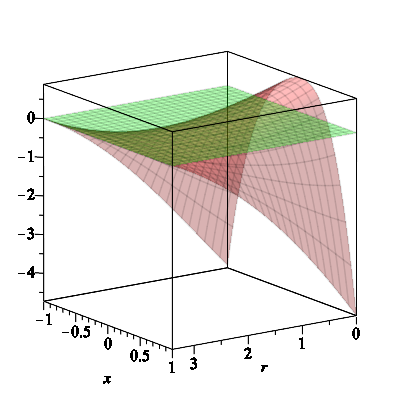}
\caption{The scaled function $S(r,x)$ versus $0 \leq r \geq r_H$ and $-1 \leq x \leq 1$, where we set the black hole parameters $a=6,\,b=2,\,l=1,\,m=7$ and $q=1$. The stationary limit surface $r_s(x)$ is the intersection of the function $S(r,x)$ with the horizontal plane at zero.}
\label{fig:ergoinn}
\end{figure}

\begin{figure}[H]
\centering
\includegraphics[width=0.325\textwidth]{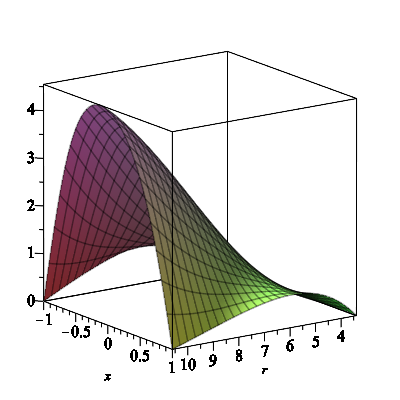}
\caption{The scaled function $S(r,x)$ versus $r_- \leq  r \leq r_H$ and $-1 \leq x \leq 1$, where we set the black hole parameters $a=6,\,b=2,\,l=1,\,m=7$ and $q=1$.}
\label{fig:between}
\end{figure}

Now, we consider the rich structure of the electromagnetic fields, on and outside the event horizon. The electromagnetic field components are given by
\begin{eqnarray}
E^i=F^{i\mu}u_\mu,\\
B^i={\tilde F}^{i\mu}u_\mu
\label{EMcompo}
\end{eqnarray}
where ${\tilde F}$, the Hodge dual of two-form $F$, is given in (\ref{dualF}), and $u^{\mu}$ is the 4-velocity of the observer. 
We find the exact forms for the electromagnetic fields, though their expressions are quite long, and so we do not present them here. In figure \ref{fig:B}, we plot the typical behaviour of the components of the magnetic field, outside the event horizon, for a black hole with parameters $a=6,\, q=1,\,m=7,\,l=1$ and $b=2$. Moreover in figure \ref{fig:E}, we plot the typical behaviour of the components of the electric field, outside the event horizon, for the same black hole parameters. 
\begin{figure}[H]
\centering
\includegraphics[width=0.325\textwidth]{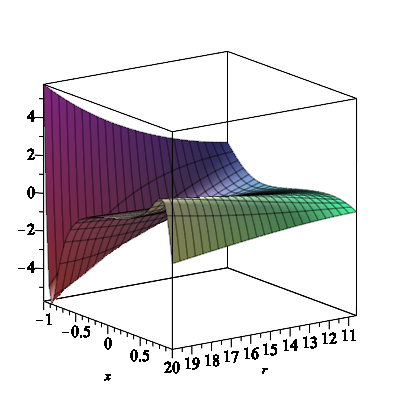}\includegraphics[width=0.325\textwidth]{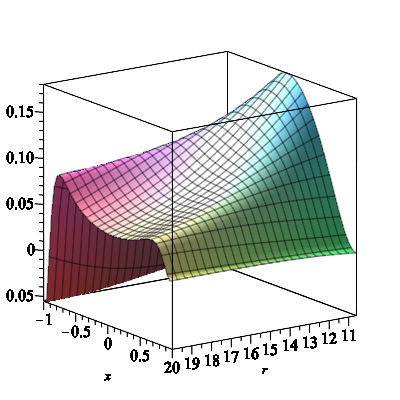}
\caption{The $r$ (left) and $x$ (right) components of the magnetic field, versus $r$ and $x$, for a black hole which the event horizon is located at $r_+=10.6$. We set the black hole parameters as $a=6,\, q=1,\,m=7,\,l=1$ and $b=2$.}
\label{fig:B}
\end{figure}

\begin{figure}[H]
\centering
\includegraphics[width=0.325\textwidth]{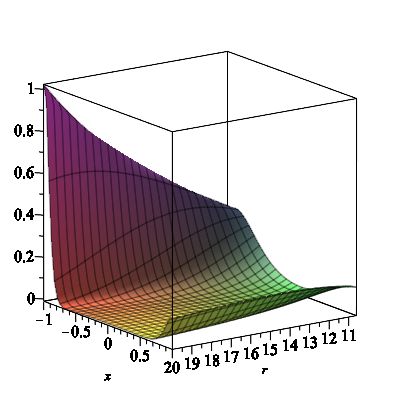}\includegraphics[width=0.325\textwidth]{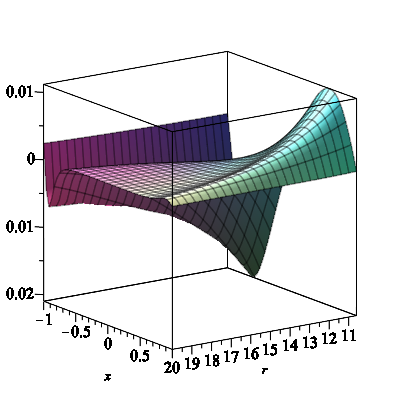}
\caption{The $r$ (left) and $x$ (right) components of the electric field, versus $r$ and $x$, for a black hole which the event horizon is located at $r_+=10.6$. We set the black hole parameters as $a=6,\, q=1,\,m=7,\,l=1$ and $b=2$.}
\label{fig:E}
\end{figure}

In figures \ref{fig:Bh} and \ref{fig:Eh}, we plot the behaviour of the electromagnetic fields on the event horizon.  We notice the minimum and maximum of $B_r$ and $E_x$ appear quite away from the equatorial plane, however the maximum of $B_x$ and the minimum of $E_r$ occurs almost on the equatorial plane. 
\begin{figure}[H]
\centering
\includegraphics[width=0.325\textwidth]{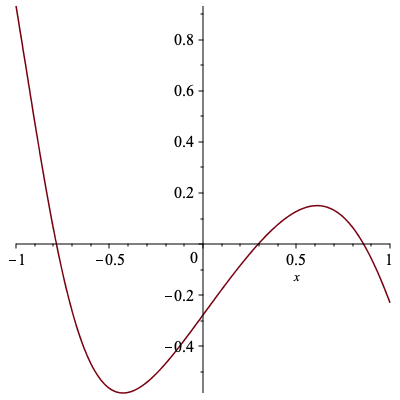}\includegraphics[width=0.325\textwidth]{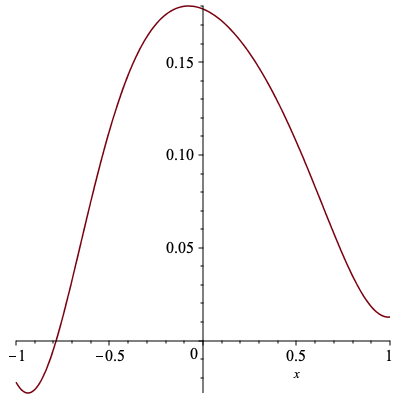}
\caption{The $r$ (left) and $x$ (right) components of the magnetic field, on the event horizon, versus $x$, for a black hole which the event horizon is located at $r_+=10.6$. We set the black hole parameters as $a=6,\, q=1,\,m=7,\,l=1$ and $b=2$.}
\label{fig:Bh}
\end{figure}

\begin{figure}[H]
\centering
\includegraphics[width=0.325\textwidth]{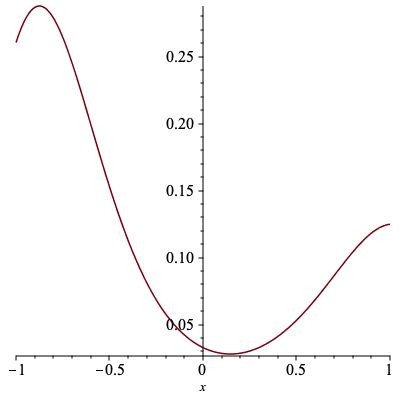}\includegraphics[width=0.325\textwidth]{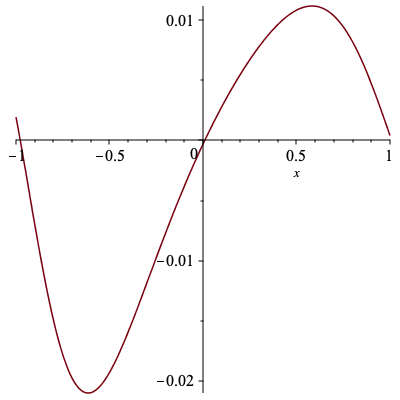}
\caption{The $r$ (left) and $x$ (right) components of the electric field, on the event horizon, versus $x$, for a black hole which the event horizon is located at $r_+=10.6$. We set the black hole parameters as $a=6,\, q=1,\,m=7,\,l=1$ and $b=2$.}
\label{fig:Eh}
\end{figure}

We also plot the polar electromagnetic fields outside the even horizon in figures \ref{fig:Bpol} and \ref{fig:Epol}.

\begin{figure}[H]
\centering
\includegraphics[width=0.325\textwidth]{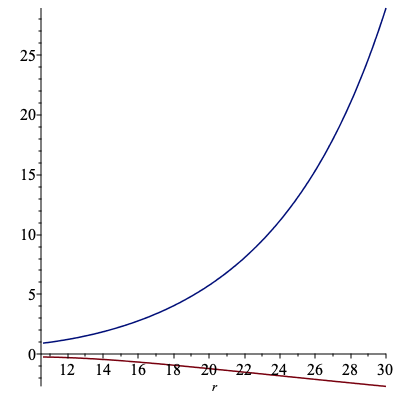}\includegraphics[width=0.325\textwidth]{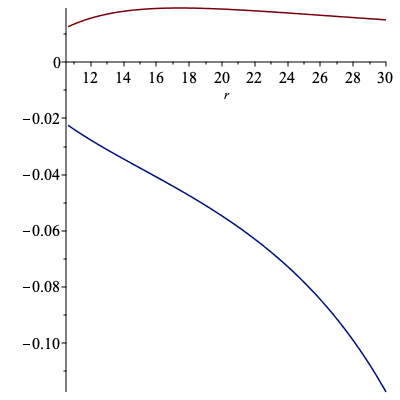}
\caption{The $r$ (left) and $x$ (right) components of the polar magnetic field, outside the event horizon, versus $r$, for a black hole which the event horizon is located at $r_+=10.6$. We set the black hole parameters as $a=6,\, q=1,\,m=7,\,l=1$ and $b=2$. In the left figure, the down curve is for $x=1$ and the up curve is for $x=-1$. In the right figure, the down curve is for $x=-1$ and the up curve is for $x=1$. }
\label{fig:Bpol}
\end{figure}

\begin{figure}[H]
\centering
\includegraphics[width=0.325\textwidth]{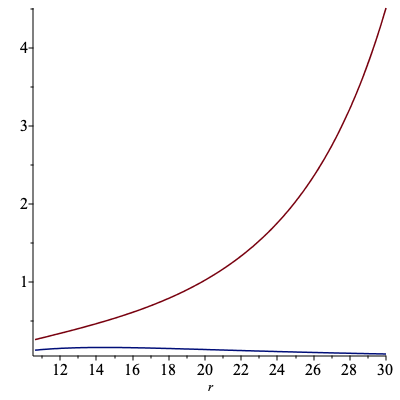}\includegraphics[width=0.325\textwidth]{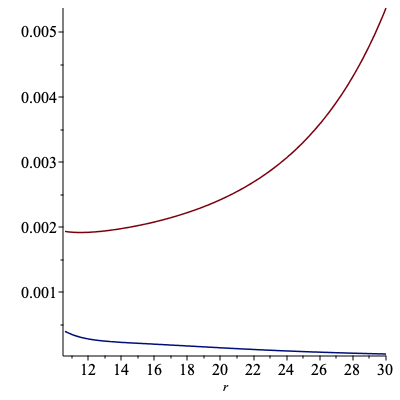}
\caption{The $r$ (left) and $x$ (right) components of he polar electric field, outside the event horizon, versus $r$, for a black hole which the event horizon is located at $r_+=10.6$. We set the black hole parameters as $a=6,\, q=1,\,m=7,\,l=1$ and $b=2$. In both figures, the down curve is for $x=1$ and the up curve is for $x=-1$.}
\label{fig:Epol}
\end{figure}

We should notice that increasing behaviour in the polar electromagnetic fields doesn't extend to large values of the radial coordinate. In fact, we find the asymptotic behaviour of the electromagnetic fields for $r\rightarrow \infty$, as
\be
B_r = {\cal O}({r^{-2}}),\,
B_x ={\cal O}({r^{-3}}),\,
E_r ={\cal O}({r^{-8}}),\,
E_x ={\cal O}({r^{-9}}),
\ee
for an observer with the 4-velocity $u_\mu=(1,0,0,0)$.


\section{Thermodynamics of the Melvin-Kerr-Newman-Taub-NUT spacetimes}
\label{sec:thermo}
In this section, we discuss the thermodynamical quantities for the black hole (\ref{MAG}) and then construct the mass of the black hole according to the Smarr relation.  

{\textcolor{black}{ We should emphasis that using the term ``black hole'' for the metric (\ref{MAG}), with the  NUT charge parameter $l$ must be taken with some cautions, to avoid contradictions with the black hole uniqueness theorems.  The event horizon is a global concept and, therefore, requires asymptotic flatness to be well defined.  On the other hand, the NUT charge parameter $l$ makes the spacetime (\ref{MAG}) asymptotically locally flat, and so violates the global flatness condition. Hence, in general, a global event horizon does not exist. However, since the spacetime (\ref{MAG}) possesses many physical quantities, similar to a black hole (such as event horizons,$\cdots$), we often refer to the spacetime (\ref{MAG}), as a black hole \footnote{ {\textcolor{black}{We would like to thank anonymous referee for the above-mentioned comment.}}}}.

The surface gravity for the black hole  (\ref{MAG}) is given by 
\be 
\kappa=\sqrt{-\frac{1}{2}\nabla _\nu\xi _\mu\nabla ^\nu\xi ^\mu}, \label{kappa}
\ee
where $\xi^\mu$ is the Killing vector $\xi^\mu=\frac{\partial}{\partial t}+\Omega _H\frac{\partial}{\partial \phi }$ and $\Omega_H$ is the angular velocity of the  horizon, which is given by
\be
\Omega_H=\left. \left(\frac{\sum\limits _{i=0}^4o_ib^i}{\Xi}\right) \right|_{r = r_ +  } ,\label{OMH}
\ee
where
\begin{eqnarray}
o_0&=&2\,a{m}^{2}+2\,a{l}^{2}-a{q}^{2}+2\,am \left( r-m \right), \nn\\
o_1&=&\left( -
16\,{l}^{2}qm+12\,{q}^{3}m-16\,q{m}^{3}+8\,{a}^{2}mq-8\,{l}^{2}q
 \left( r-m \right) -16\,{m}^{2}q \left( r-m \right) +4\,{q}^{3}
 \left( r-m \right)  \right), \nn\\
 o_2&=&-12\,a{m}^{2}{q}^{2}-12\, \left( r-m \right) am{q}^{2},\nn\\
 o_3&=&\left( -32\,{a}^{2}{l}^{2}mq+16\,{l}^{4}qm-80\,{q}^{3}{m}^{3}+20\,{
q}^{5}m+64\,q{m}^{5}-64\,{a}^{2}{m}^{3}q+24\,{a}^{2}m{q}^{3}+80\,{l}^{
2}{m}^{3}q \right.  \nn\\
&-&\left.40\,{l}^{2}m{q}^{3}+\left( r-m \right) (64\,q{m}^{4}-8\,{l}
^{2}{q}^{3} -48m^2q^3-32qa^2m^2+48\,{l}^{2}{m}^{2}q  +4\,{q}^{5}) \right),\nn
\end{eqnarray}
\begin{eqnarray}
o_4&=&\left( -88
\,{a}^{3}{l}^{2}{m}^{2}+16\,{a}^{3}{l}^{2}{q}^{2}-32\,{a}^{3}{l}^{4}-8
\,a{l}^{6}-4\,a{l}^{4}{q}^{2}-104\,a{m}^{2}{q}^{2}{l}^{2}-80\,{a}^{3}{
m}^{4}+96\,a{m}^{6}\right.\nn\\
&+&\left.28\,{a}^{3}{m}^{2}{q}^{2}+64\,a{m}^{2}{l}^{4}+168
\,a{m}^{4}{l}^{2}+4\,a{l}^{2}{q}^{4}-120\,a{m}^{4}{q}^{2}+30\,a{m}^{2}
{q}^{4} \right.\nn\\
&+&\left.\, \left( r-m
 \right) (-32a{l}^{2}m{q}^{2}-32{a}^{3}{m}^{3}+96  a{m}^{5} -32{a}^{3}{l}^{2}m+16a{l}^{4}m+120 a{l}^{2}{m}^{3} -72\,a{m}^{3}{q}^{2} +6\,am{q}^{4}  )\right. ,\nn\\ 
 &&
\end{eqnarray}
and
\begin{eqnarray}
\Xi&=&8\left( r-m \right) (8{l}^{2}m+8\,{m}^{3} -4\,m{q}
^{2}  -4\,{a}^{2}{m}^{2}+4\,{l}^{4}+12\,{l}^{2}{m}^{
2}-4\,{l}^{2}{q}^{2}+8\,{m}^{4}-8\,{m}^{2}{q}^{2}+{q}^{4}.\nn\\
&&
\end{eqnarray}

We note that for $b=0$, we recover the angular velocity for the Kerr-Newman-NUT black hole \cite{Booth:2015nwa}. In figures \ref{fig:OM} and \ref{fig:OM2}, we plot the angular velocity of the horizon versus different black hole parameters.
\begin{figure}[H]
\centering
\includegraphics[width=0.325\textwidth]{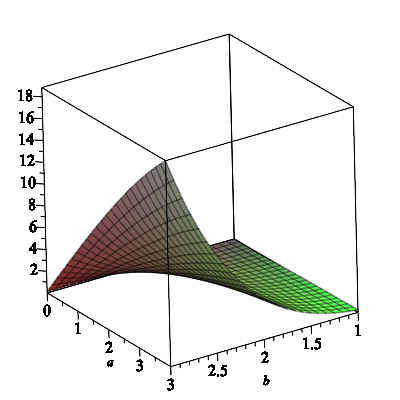}\includegraphics[width=0.325\textwidth]{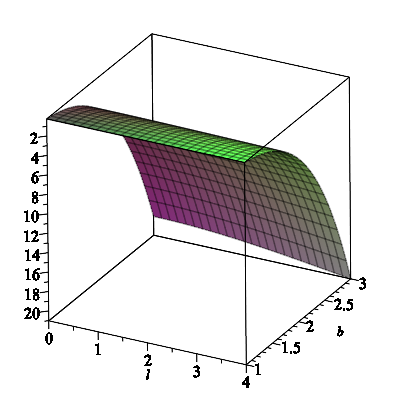}
\caption{The angular velocity of the horizon as function of $a,\,b$ (left) and $l,\,b$ (right), where we set the other black hole parameters to a set of fixed numbers.}
\label{fig:OM}
\end{figure}
We find the surface gravity of the black hole 
\be
\kappa=\frac{1}{2r_+}\frac{r_+^2+l^2-a^2-q^2}{r_+^2+l^2+a^2},\label{KAPPA}
\ee
is the same as the Kerr-Newman-NUT black hole, and so the Hawking temperature  is 
\be
T_H=\frac{\kappa}{2\pi}.
\ee
\begin{figure}[H]
\centering
\includegraphics[width=0.325\textwidth]{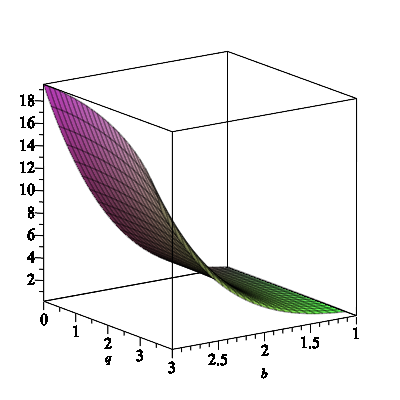}\includegraphics[width=0.325\textwidth]{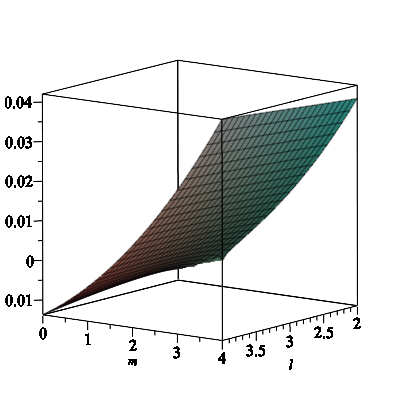}
\caption{The angular velocity of the horizon as function of $q,\,b$ (left) and $m, \,l$ (right), where we set the other black hole parameters to a set of fixed numbers.}
\label{fig:OM2}
\end{figure}
The Coulomb potential on the horizon is given by
\be
\Phi_H=-\left(A'_\mu \xi^\mu\right)\vert _{r=r_+} =\left. \left(- A'_t -\frac{\Omega_H}{\vert\Lambda_0\vert^2} A'_\phi\right) \right|_{r = r_ +  }  -  \Phi_H^{(0)},\label{PHIH}
\ee
where we add a constant term  $\Phi_H^{(0)}$ to the potential to make it regular at $r \rightarrow \infty$. The expression for the $\Phi_H$ is long and so we do not present it explicitly here. 
The area of the horizon is given by
\be
{\cal A}_H={\cal A}_0\vert{\Lambda_0}\vert^2,\label{AH}
\ee
where ${\cal A}_0=4\pi (2l^2+2mr_+-q^2)$ is the horizon area for the Kerr-Newman-NUT black hole.
To find the electric charge of the black hole (\ref{MAG}), we use  the well-known equation \cite{Booth:2015nwa}
\be
Q=\frac{1}{4 \pi}\int _\Sigma E_{\perp}\label{QQ},
\ee
where $\Sigma$ is the two-dimensional hypersurface, parameterized with $(\theta,\phi)$ in (\ref{MAG}) and $E_{\perp}$ is the normal component of the electric field on $\Sigma$.
We find 
\be
Q=\left. \left(\frac{\sum \limits_{i=0}^{6} q_i b^i}{\sum \limits_{i=0}^{4} s_i b^i}\right) \right|_{r = r_ +  }, \label{QMAG}
\ee 
where
\begin{eqnarray}
q_0&=&-{a}^{4}q-2\,{a}^{2}q{r}^{2}+{l}^{4}q-q{r}^{4},\nonumber\\
q_1&=&-4\,{a}^{5}m-8
\,{a}^{3}m{r}^{2}+4\,a{l}^{4}m-16\,a{l}^{4}r-16\,a{l}^{2}m{r}^{2}+8\,a
{l}^{2}{q}^{2}r-4\,am{r}^{4} ,\nonumber
\end{eqnarray}
\begin{eqnarray}
q_2&=&4\,{a}^{4}{l}^{2}q-5\,{
a}^{4}{q}^{3}-40\,{a}^{2}{l}^{4}q-52\,{a}^{2}{l}^{2}mqr+20\,{a}^{2}{l}
^{2}{q}^{3}-16\,{a}^{2}{l}^{2}q{r}^{2}-10\,{a}^{2}{q}^{3}{r}^{2}+4\,{l
}^{6}q\nn\\
&-&4\,{l}^{4}mqr+{l}^{4}{q}^{3}+16\,{l}^{4}q{r}^{2}+28\,{l}^{2}mq{
r}^{3}-20\,{l}^{2}{q}^{3}{r}^{2}-20\,{l}^{2}q{r}^{4}-5\,{q}^{3}{r}^{4},\nonumber\\
q_3&=&32\,{a}^{5}{l}^{2}m-40\,{a}^{5}m{q}^{2}-32\,{
a}^{3}{l}^{4}m+64\,{a}^{3}{l}^{4}r-64\,{a}^{3}{l}^{2}{m}^{2}r-16\,{a}^
{3}{l}^{2}m{q}^{2}+64\,{a}^{3}{l}^{2}m{r}^{2}\nn\\
&-&96\,{a}^{3}{l}^{2}{q}^{2
}r-80\,{a}^{3}m{q}^{2}{r}^{2}-64\,a{l}^{6}r-8\,a{l}^{4}m{q}^{2}-160\,a
{l}^{4}m{r}^{2}+128\,a{l}^{4}{q}^{2}r+64\,a{l}^{4}{r}^{3}-64\,a{l}^{2}
{m}^{2}{r}^{3}\nn\\
&+&144\,a{l}^{2}m{q}^{2}{r}^{2}+32\,a{l}^{2}m{r}^{4}-80\,a
{l}^{2}{q}^{4}r
-96\,a{l}^{2}{q}^{2}{r}^{3}-40\,am{q}^{2}{r}^{4},\nonumber\\
q_4&=&-80\,{a}^{6}{m}^{2}q+16\,{a}^{4}{l}^{4}q-116
\,{a}^{4}{l}^{2}{m}^{2}q-208\,{a}^{4}{l}^{2}mqr-24\,{a}^{4}{l}^{2}{q}^
{3}+48\,{a}^{4}{l}^{2}q{r}^{2}-160\,{a}^{4}{m}^{2}q{r}^{2}\nn\\
&+&5\,{a}^{4}{
q}^{5}-48\,{a}^{2}{l}^{4}{m}^{2}q-128\,{a}^{2}{l}^{4}mqr+80\,{a}^{2}{l
}^{4}{q}^{3}-96\,{a}^{2}{l}^{4}q{r}^{2}-8\,{a}^{2}{l}^{2}{m}^{2}q{r}^{
2}+40\,{a}^{2}{l}^{2}m{q}^{3}r\nn\\
&-&352\,{a}^{2}{l}^{2}mq{r}^{3}-40\,{a}^{2
}{l}^{2}{q}^{5}+64\,{a}^{2}{l}^{2}{q}^{3}{r}^{2}+96\,{a}^{2}{l}^{2}q{r
}^{4}-80\,{a}^{2}{m}^{2}q{r}^{4}+10\,{a}^{2}{q}^{5}{r}^{2}-16\,{l}^{8}
q\nn\\
&-&12\,{l}^{6}{m}^{2}q-48\,{l}^{6}mqr+8\,{l}^{6}{q}^{3}+80\,{l}^{6}q{r}
^{2}-32\,{l}^{4}{m}^{2}q{r}^{2}+8\,{l}^{4}m{q}^{3}r+192\,{l}^{4}mq{r}^
{3}+3\,{l}^{4}{q}^{5}-96\,{l}^{4}{q}^{3}{r}^{2}\nn\\
&-&112\,{l}^{4}q{r}^{4}+
108\,{l}^{2}{m}^{2}q{r}^{4}-120\,{l}^{2}m{q}^{3}{r}^{3}-144\,{l}^{2}mq
{r}^{5}+40\,{l}^{2}{q}^{5}{r}^{2}+88\,{l}^{2}{q}^{3}{r}^{4}+48\,{l}^{2
}q{r}^{6}+5\,{q}^{5}{r}^{4},\nonumber\\
q_5&=&-64\,{a}^{7}{m}^{3
}-64\,{a}^{5}{l}^{4}m-112\,{a}^{5}{l}^{2}{m}^{3}-192\,{a}^{5}{l}^{2}{m
}^{2}r-32\,{a}^{5}{l}^{2}m{q}^{2}+64\,{a}^{5}{l}^{2}m{r}^{2}-128\,{a}^
{5}{m}^{3}{r}^{2}\nn\\
&+&12\,{a}^{5}m{q}^{4}-128\,{a}^{3}{l}^{6}m-256\,{a}^{3
}{l}^{6}r-64\,{a}^{3}{l}^{4}{m}^{3}-384\,{a}^{3}{l}^{4}{m}^{2}r+224\,{
a}^{3}{l}^{4}m{q}^{2}-384\,{a}^{3}{l}^{4}m{r}^{2}\nn\\
&+&64\,{a}^{3}{l}^{4}{q
}^{2}r-224\,{a}^{3}{l}^{2}{m}^{3}{r}^{2}+256\,{a}^{3}{l}^{2}{m}^{2}{q}
^{2}r-384\,{a}^{3}{l}^{2}{m}^{2}{r}^{3}-80\,{a}^{3}{l}^{2}m{q}^{4}+64
\,{a}^{3}{l}^{2}m{q}^{2}{r}^{2}\nn\\
&+&128\,{a}^{3}{l}^{2}m{r}^{4}+32\,{a}^{3
}{l}^{2}{q}^{4}r-64\,{a}^{3}{m}^{3}{r}^{4}+24\,{a}^{3}m{q}^{4}{r}^{2}-
64\,a{l}^{8}m-16\,a{l}^{6}{m}^{3}-128\,a{l}^{6}{m}^{2}r+64\,a{l}^{6}m{
q}^{2}\nn\\
&+&64\,a{l}^{6}m{r}^{2}+64\,a{l}^{6}{q}^{2}r-64\,a{l}^{4}{m}^{3}{r
}^{2}+32\,a{l}^{4}{m}^{2}{q}^{2}r+128\,a{l}^{4}{m}^{2}{r}^{3}+4\,a{l}^
{4}m{q}^{4}+96\,a{l}^{4}m{q}^{2}{r}^{2}\nn\\
&-&64\,a{l}^{4}m{r}^{4}-48\,a{l}^
{4}{q}^{4}r-64\,a{l}^{4}{q}^{2}{r}^{3}+144\,a{l}^{2}{m}^{3}{r}^{4}-128
\,a{l}^{2}{m}^{2}{q}^{2}{r}^{3}-192\,a{l}^{2}{m}^{2}{r}^{5}+32\,a{l}^{
2}m{q}^{4}{r}^{2}\nn\\
&+&96\,a{l}^{2}m{q}^{2}{r}^{4}+64\,a{l}^{2}m{r}^{6}+8\,
a{l}^{2}{q}^{6}r+32\,a{l}^{2}{q}^{4}{r}^{3}+12\,am{q}^{4}{r}^{4},
\end{eqnarray}
and 
\begin{eqnarray}
q_6&=&-64\,{a}^{6}{l}^{2}{m}^{2}q+16\,{a}^{6}{m}^{2
}{q}^{3}-64\,{a}^{4}{l}^{6}q-16\,{a}^{4}{l}^{4}{m}^{2}q-320\,{a}^{4}{l
}^{4}mqr+48\,{a}^{4}{l}^{4}{q}^{3}-64\,{a}^{4}{l}^{4}q{r}^{2}\nn\\
&+&64\,{a}^
{4}{l}^{2}{m}^{3}qr+4\,{a}^{4}{l}^{2}{m}^{2}{q}^{3}-256\,{a}^{4}{l}^{2
}{m}^{2}q{r}^{2}+144\,{a}^{4}{l}^{2}m{q}^{3}r-12\,{a}^{4}{l}^{2}{q}^{5
}+16\,{a}^{4}{l}^{2}{q}^{3}{r}^{2}+32\,{a}^{4}{m}^{2}{q}^{3}{r}^{2}\nn\\
&+&{a
}^{4}{q}^{7}+128\,{a}^{2}{l}^{8}q-32\,{a}^{2}{l}^{6}{m}^{2}q+448\,{a}^
{2}{l}^{6}mqr-192\,{a}^{2}{l}^{6}{q}^{3}-128\,{a}^{2}{l}^{6}q{r}^{2}-
16\,{a}^{2}{l}^{4}{m}^{3}qr\nn\\
&+&32\,{a}^{2}{l}^{4}{m}^{2}{q}^{3}+704\,{a}^
{2}{l}^{4}{m}^{2}q{r}^{2}-512\,{a}^{2}{l}^{4}m{q}^{3}r-320\,{a}^{2}{l}
^{4}mq{r}^{3}+88\,{a}^{2}{l}^{4}{q}^{5}+32\,{a}^{2}{l}^{4}{q}^{3}{r}^{
2}\nn\\
&+&320\,{a}^{2}{l}^{2}{m}^{3}q{r}^{3}-344\,{a}^{2}{l}^{2}{m}^{2}{q}^{3
}{r}^{2}-192\,{a}^{2}{l}^{2}{m}^{2}q{r}^{4}+140\,{a}^{2}{l}^{2}m{q}^{5
}r+96\,{a}^{2}{l}^{2}m{q}^{3}{r}^{3}-12\,{a}^{2}{l}^{2}{q}^{7}\nn\\
&+&16\,{a}
^{2}{l}^{2}{q}^{5}{r}^{2}+32\,{a}^{2}{l}^{2}{q}^{3}{r}^{4}+16\,{a}^{2}
{m}^{2}{q}^{3}{r}^{4}+2\,{a}^{2}{q}^{7}{r}^{2}-64\,{l}^{10}q-16\,{l}^{
8}{m}^{2}q-256\,{l}^{8}mqr+144\,{l}^{8}{q}^{3}\nn\\
&+&192\,{l}^{8}q{r}^{2}-16
\,{l}^{6}{m}^{3}qr+12\,{l}^{6}{m}^{2}{q}^{3}-320\,{l}^{6}{m}^{2}q{r}^{
2}+368\,{l}^{6}m{q}^{3}r+512\,{l}^{6}mq{r}^{3}-108\,{l}^{6}{q}^{5}-272
\,{l}^{6}{q}^{3}{r}^{2}\nn\\
&-&192\,{l}^{6}q{r}^{4}-144\,{l}^{4}{m}^{3}q{r}^{
3}+240\,{l}^{4}{m}^{2}{q}^{3}{r}^{2}+336\,{l}^{4}{m}^{2}q{r}^{4}-132\,
{l}^{4}m{q}^{5}r-320\,{l}^{4}m{q}^{3}{r}^{3}-256\,{l}^{4}mq{r}^{5}\nn\\
&+&27
\,{l}^{4}{q}^{7}+80\,{l}^{4}{q}^{5}{r}^{2}+112\,{l}^{4}{q}^{3}{r}^{4}+
64\,{l}^{4}q{r}^{6}+36\,{l}^{2}{m}^{2}{q}^{3}{r}^{4}-36\,{l}^{2}m{q}^{
5}{r}^{3}-48\,{l}^{2}m{q}^{3}{r}^{5}+12\,{l}^{2}{q}^{7}{r}^{2}\nn\\
&+&28\,{l}
^{2}{q}^{5}{r}^{4}+16\,{l}^{2}{q}^{3}{r}^{6}+{q}^{7}{r}^{4}.
\end{eqnarray}
Moreover, the functions $s_i,\,i=0,\cdots,4$ are given by
\begin{eqnarray}
s_0&=&\left( {a}^{2}+2\,al+{l}^{2}+{r}^{2} \right)  \left( {a}^{2}-2\,al+{
l}^{2}+{r}^{2} \right), \nn\\
s_1&=&8\, \left( {a}^{2}+2\,al+{l}^{2}+{r}^{2}
 \right) lqr,\nn\\
s_2&=&\left( {a}^{2}+2\,al+{l}^{2}+{r}^{2} \right)  \left( -8
\,{a}^{2}{l}^{2}+6\,{a}^{2}{q}^{2}+12\,al{q}^{2}+8\,{l}^{4}+16\,{l}^{2
}mr-2\,{l}^{2}{q}^{2}-8\,{l}^{2}{r}^{2}+6\,{q}^{2}{r}^{2} \right), \nn
\end{eqnarray}
\begin{eqnarray}
s_3&=& \left( {a}^{2}+2\,al+{l}^{2}+{r}^{2} \right)  \left( 16\,{a}^{3}m
q+24\,{a}^{2}lmq-16\,{a}^{2}lqr+16\,a{l}^{2}mq+16\,amq{r}^{2}+8\,{l}^{
3}mq \right .  \nn\\
&+&\left .16\,{l}^{3}qr+24\,lmq{r}^{2}-8\,l{q}^{3}r-16\,lq{r}^{3} \right),\nn\\
s_4&=&\left( {a}^{2}+2\,al+{l}^{2}+{r}^{2} \right)  \left( 16\,{a}^{
4}{m}^{2}+16\,{a}^{3}l{m}^{2}-32\,{a}^{3}lmr+16\,{a}^{2}{l}^{4}+20\,{a
}^{2}{l}^{2}{m}^{2}+16\,{a}^{2}{l}^{2}mr\right .\nn\\
&-&\left .8\,{a}^{2}{l}^{2}{q}^{2}+16\,
{a}^{2}{l}^{2}{r}^{2}+16\,{a}^{2}{m}^{2}{r}^{2}+{a}^{2}{q}^{4}+32\,a{l
}^{5}+8\,a{l}^{3}{m}^{2}+64\,a{l}^{3}mr-32\,a{l}^{3}{q}^{2}-32\,a{l}^{
3}{r}^{2}\right.\nn\\
&+&\left.48\,al{m}^{2}{r}^{2}-32\,alm{q}^{2}r-32\,alm{r}^{3}+6\,al{q}
^{4}+16\,{l}^{6}+4\,{m}^{2}{l}^{4}+48\,{l}^{4}mr-24\,{l}^{4}{q}^{2}-32
\,{l}^{4}{r}^{2}\right.\nn\\
&+&\left.36\,{l}^{2}{m}^{2}{r}^{2}-32\,{l}^{2}m{q}^{2}r-48\,{l
}^{2}m{r}^{3}+9\,{l}^{2}{q}^{4}+24\,{l}^{2}{q}^{2}{r}^{2}+16\,{l}^{2}{
r}^{4}+{q}^{4}{r}^{2} \right).
\end{eqnarray}
In the limit of $l\rightarrow 0$, we recover exactly the results of \cite{Booth:2015nwa} for the electric charge. In figures \ref{fig:Q} and \ref{fig:Q2} we plot the electric charge (\ref{QMAG})  of the black hole, versus different black hole parameters.
\begin{figure}[H]
\centering
\includegraphics[width=0.325\textwidth]{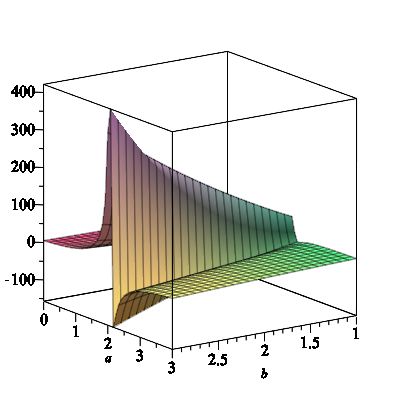}\includegraphics[width=0.325\textwidth]{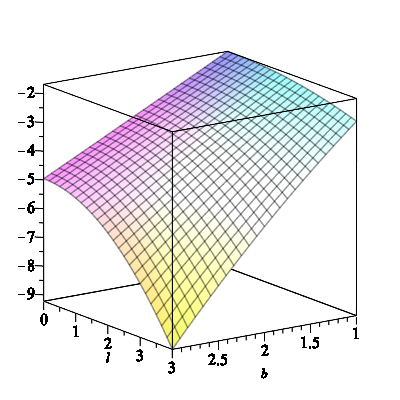}
\caption{The electric charge of the black hole as function of $a,\,b$ (left) and $l,\,b$ (right), where we set the other black hole parameters to a set of fixed numbers.}
\label{fig:Q}
\end{figure}
\begin{figure}[H]
\centering
\includegraphics[width=0.325\textwidth]{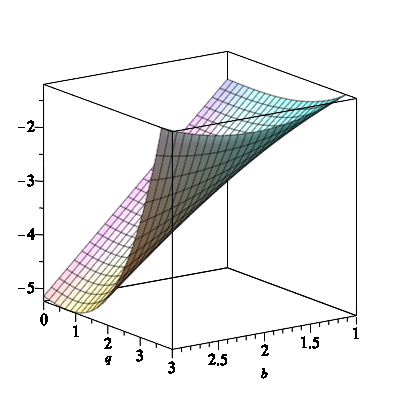}\includegraphics[width=0.325\textwidth]{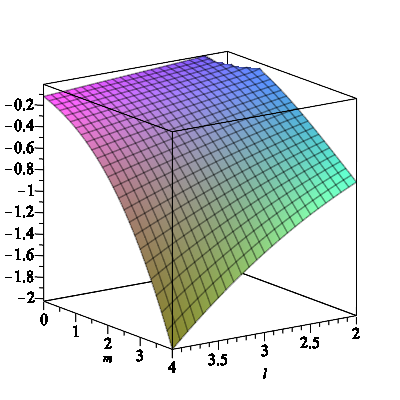}
\caption{The electric charge of the black hole as function of $q,\,b$ (left), and $m, \,l$ (right), where we set the other black hole parameters to a set of fixed numbers.}
\label{fig:Q2}
\end{figure}
We then find the angular momentum $J$ of the black hole (\ref{MAG}), according to \cite{Booth:2015nwa}
\be
J =\frac{1}{16\pi}\int_\Sigma *d{\cal P},\label{JJ}
\ee
where ${\cal P}^\mu=\frac{\partial}{\partial \phi}$ is the Killing vector in $\phi$-direction. We find the integrand in (\ref{JJ}) is given by
\be
(*d{\cal P})_{x\phi}=-{\frac { { f'}^2 ({\frac {\partial }{\partial r}}{w'} ) }{2
{ \rho} }{{\sqrt {{\frac { \Delta_r }{
 { \Delta _x}  }}}}}}.
\ee
A straightforward and lengthy calculation shows that we get
\be
J=\frac{\sum \limits_{i=0}^{12} j_i(a,l,m,q) b^i}{\sum \limits_{i=0}^{8} k_i(a,l,m,q) b^i},\label{JMAG}
\ee
where $j_i$ and $k_i$ depend on black hole parameters $a$, $l$, $m$ and $q$. The expressions for $j_i$ and $k_i$ are very long to present, so we don't present them here. We verify that the expression (\ref{JMAG}) recovers exactly the results of \cite{Booth:2015nwa} for Melvin-Kerr-Newman black holes, where the NUT charge $l$ goes to zero. In figure  \ref{fig:J} and \ref{fig:J2} we plot the angular momentum (\ref{JMAG}) of the black hole, versus different black hole parameters.
\begin{figure}[H]
\centering
\includegraphics[width=0.325\textwidth]{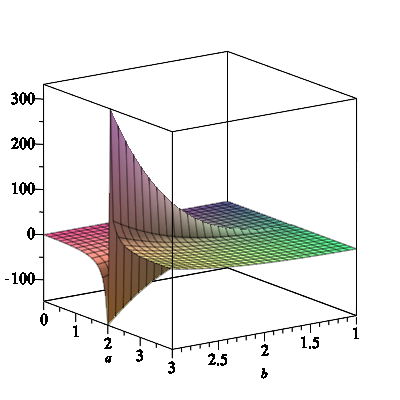}\includegraphics[width=0.325\textwidth]{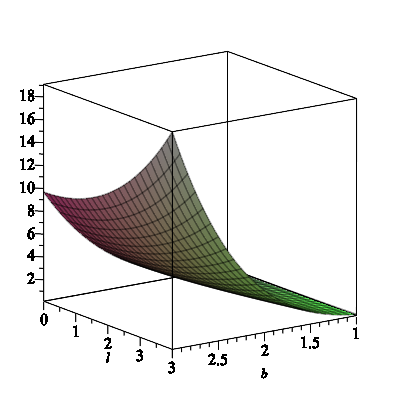}
\caption{The angular momentum of the black hole as function of $a,\,b$ (left) and $l,\,b$ (right), where we set the other black hole parameters to a set of fixed numbers.}
\label{fig:J}
\end{figure}
\begin{figure}[H]
\centering
\includegraphics[width=0.325\textwidth]{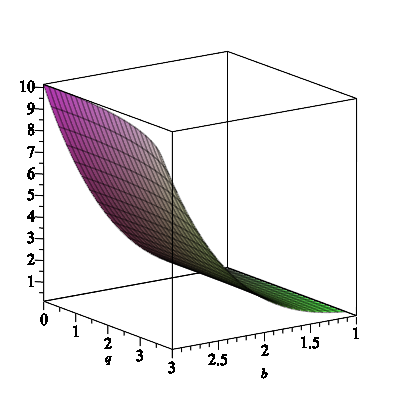}\includegraphics[width=0.325\textwidth]{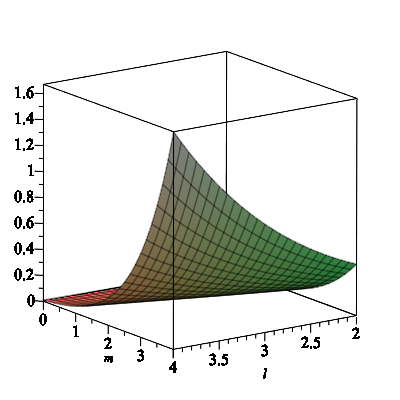}
\caption{The angular momentum of the black hole as function of $q,\,b$ (left) and $m, \,l$ (right), where we set the other black hole parameters to a set of fixed numbers.}
\label{fig:J2}
\end{figure}
Inspired by the dimensions of physical quantities for the black hole, as well as the presence of the NUT charge \cite{Bordo}, we consider the total mass of the black hole (\ref{MAG}) as
\be
{\cal M}=\Phi_HQ+2\Omega_H {\cal J}+\frac{\kappa}{4\pi}{\cal A}_H+2\psi_N N_N+2\psi_S N_S,\label{totM}
\ee
where $N_N$ and $\psi_N$ ($N_S$ and $\psi_S$) are the thermodynamic charge and the potential, due to the presence of the NUT charge on north (south) pole, and 
$\Phi_H,\, Q,\, \Omega_H,\,\kappa$ and ${\cal A}_H$ are given by equations (\ref{PHIH}),(\ref{QMAG}),(\ref{OMH}),(\ref{KAPPA}) and (\ref{AH}), respectively. The total angular momentum ${\cal J}$ in equation (\ref{totM}) is 
\be
{\cal J}=J+J_N+J_S,\label{totJ}
\ee
where $J$ is given by (\ref{JMAG}), and $J_N$ and $J_S$ are contributions to the angular momentum from NUT charges on the north and south poles.  
The total mass for the black hole (\ref{MAG}) is given by
\be
{\cal M}=-\frac{1}{8\pi}\int_{r \rightarrow \infty} *d {\cal T},\label{masss}
\ee
where ${\cal T}^\mu=\frac{\partial}{\partial t}$ is the time-like Killing vector. Evaluating the integral in equation (\ref{masss}), where $b=0$, we find
\be
{\cal M}=m.\label{MM}
\ee 
Moreover, the thermodynamic potentials $\psi_N$ and $\psi_S$  are given by \cite{Bordo}
\begin{eqnarray}
\psi_N&=&\frac{1}{4\pi}\sqrt{\frac{1}{2}\nabla_\mu \xi _{N\nu}\nabla^\mu \xi_N^\nu},\\
\psi_S&=&\frac{1}{4\pi}\sqrt{\frac{1}{2}\nabla_\mu \xi _{S\nu}\nabla^\mu \xi_S^\nu},
\end{eqnarray}
where $\xi_N^\mu$ is the Killing vector 
\be
\xi_N^\mu=\frac{\partial}{\partial t} +\Omega_N\frac{\partial}{\partial \phi},\label{xiN}
\ee
which generates the north pole Killing horizon, due to the presence of the NUT charge.  The other Killing vector $\xi_S^\mu$ is  
\be
\xi_S^\mu=\frac{\partial}{\partial t} +\Omega_S\frac{\partial}{\partial \phi},\label{xiS}
\ee
which generates the south pole Killing horizon, due to the presence of the NUT charge. The functions $\Omega_N$ and $\Omega_S$ are given  by
\begin{eqnarray}
\Omega_N&=&-{\frac {1}{2\,l}}-3\,{\frac {{q}^{2}}{l}}{b}^{2}-{\frac {8mq\,(2a-l)}{2\,l}}{b}^{3}-{\frac {16\,{a}^{2}\,({l}^{2}+m^2)+4l^2\,(al+m^2)\,-8\,al{m}^{2}+6\,al{q}^{2}+{q}^{4}}{2\,l}}{b}^
{4} ,\nn\\
&&\\
\Omega_S&=&{\frac {1}{2\,l}}+3\,{\frac {{q}^{2}}{l}}{b}^{2}+{\frac {8mq\,(2a+l)}{2\,l}}{b}^{3}+{\frac {16\,{a}^{2}\,({l}^{2}+m^2)-4l^2\,(al-m^2)\,+8\,al{m}^{2}-6\,al{q}^{2}+{q}^{4}}{2\,l}}{b}^
{4} .\nn\\
&&
\end{eqnarray}

Using the form of the Killing vectors on the north and south pole horizons, we find
\be
\psi_N=\lim_{x\rightarrow 1,r \rightarrow \infty,\Omega\rightarrow \Omega_N}\Psi(r,x,\Omega),\label{PNN}
\ee
and
\be
\psi_S=\lim_{x\rightarrow -1,r \rightarrow \infty,\Omega\rightarrow \Omega_S}\Psi(r,x,\Omega),\label{PSS}
\ee
where $\Psi(r,x,\Omega)$ is given by
\begin{eqnarray}
&&
64\pi^2{f'^{2}e^{2\gamma}  
 \rho ^2}\Psi^2(r,x,\Omega)=\nn\\
&-&{\Omega}^{2}\Delta_x 
  \left( {\frac {\partial }{\partial x}}\omega' \right) ^{2} f'^{6}-{\Omega}^{2}\Delta _r   
f'^{6} \left( {\frac {\partial }{\partial r}}\omega' \right) ^{2}+2\,\Omega\,\Delta_x 
     \left( {\frac {\partial }{\partial x}}\omega' 
     \right) ^{2} f'^{6}\omega'  +2\,\Omega\,\Delta _r  
 f'^{6}\omega' \left( {\frac {\partial }{\partial r}}\omega' 
   \right) ^{2} \nn\\
&-&\Delta_x   
 \left( {\frac {\partial }{\partial x}}\omega'   \right) ^{2}
 f'^{6} {\omega'} ^{2}-\Delta _r   f'^{6}{ \omega'}^{2} 
 \left( {\frac {\partial }{\partial r}}\omega' \right) ^{2}+{\Omega}^{2}\Delta_x 
     \left( {\frac {\partial }{\partial x}}f'
     \right) ^{2}\rho ^2   
   f'^{2}+{\Omega}^{2}\Delta _r 
    \rho ^2   \left( {\frac {\partial }{\partial r}}f'  \right) ^{2} 
   f'^{2}\nn\\
&-&2\,\Omega\,\Delta_x   \left( {\frac {\partial }{\partial x}}f'\right) ^{2}\rho ^2   f'^{2}\omega'  -2\,
   \Omega\,\Delta_x   \left( {\frac {\partial }{\partial x}}\omega'   \right)  \left( {\frac {\partial }{\partial x}}\rho ^2   \right)  
   f'^{4}-2\,\Omega\,\Delta _r  \rho ^2   
   \left( {\frac {\partial }{\partial r}}f'  \right) ^{2} f'^{2}\omega'  \nn\\
&-&2\,\Omega\,\Delta _r   f'^{4} \left( {\frac {\partial }{\partial r}}\omega' 
     \right) {\frac {\partial }{\partial r}}\rho ^2  +\Delta_x   
   \left( {\frac {\partial }{\partial x}}f'  \right) ^{2}\rho ^2
   f'^{2}
\omega'^{2}+\Delta_x   \left( {\frac {\partial }{\partial x}}\omega' 
   \right) ^{2}\rho ^2   f'^{4} \nn\\
 &+&\Delta _r  \rho ^2
   \left( {\frac {\partial }{\partial r}}f'\right) ^{2} f'^{2}  \omega'^{2}+
   \Delta _r  \rho ^2   f'^{4} \left( {\frac {\partial }{\partial r}}\omega'   \right) ^{2}+2\,\Delta _r 
     f'^{4}\omega'
   \left( {\frac {\partial }{\partial r}}\omega' \right) {\frac {\partial }{\partial r}}\rho ^2
  -\Delta_x   
  \left( {\frac {\partial }{\partial x}}f'  \right) ^{2}
  \rho ^4 \nn\\
 &+&2\,\Delta_x
   \left( {\frac {\partial }{\partial x}}f'\right)  \left( {\frac {\partial }{\partial x}}{\rho^2}   \right) \rho ^2 
  {f'}  -\Delta_x   
  \left( {\frac {\partial }{\partial x}}\rho ^2   \right) ^{2}
 f'^{2}-\Delta _r \rho^{4} 
 \left( {\frac {\partial }{\partial r}}f'  \right) ^{2
}+2\,\Delta _r  \rho ^2   
\left( {\frac {\partial }{\partial r}}f'  \right) 
{f'} \left( {\frac {\partial }{\partial r}}\rho ^2\right)\nn\\
  &-&\Delta _r  
   f'^{2} \left( {\frac {\partial }{\partial r}}\rho ^2   \right) ^{2}+2\,\Delta_x  
 \left( {\frac {\partial }{\partial x}}\omega'  \right)  \left( {\frac {\partial }{\partial x}}\rho ^2   \right)  f'^{4}
 \omega' .\label{PPSI}
\end{eqnarray}
After substituting all the known functions $f'$, $\omega'$, $\rho$, $\Delta_r$ and $\Delta_x$ in (\ref{PPSI}) and taking the limits in equations (\ref{PNN}) and (\ref{PSS}), we find
\be
\psi_N=\frac{1}{8\pi l},\label{PSIN}
\ee
and 
\be
\psi_S=\frac{1}{8\pi l}.\label{PSIS}
\ee

As expected, we notice that neither the magnetic field nor the other parameters of the black hole (except the NUT charge) contribute to the thermodynamic potentials $\psi_N$ and $\psi_S$, where $r\rightarrow \infty$. The independence of $\psi_N$ and $\psi_S$ from other parameters of the black hole and the magnetic field, is consistent with the notion of considering them as the ``surface gravity'' over the NUT tubes \cite{Bordo}, similar to $\kappa$ as the surface gravity (\ref{kappa}) over the horizon $r_H$.
The thermodynamic charges $N_N$ and $N_S$ are given by 
\begin{eqnarray}
N_N\Psi_N&=&\frac{1}{16\pi}\int_{T_N} *d {\cal T}, \label{NpsiN}\\
N_S\Psi_S&=&-\frac{1}{16\pi}\int_{T_S} *d {\cal T},\label{NpsiS}
\end{eqnarray}
where the integrals are over the very narrow NUT tubes $T_N$ and $T_S$, along the positive and negative $z$-axis, where $r_H \leq  z <\infty,\,x\simeq 1$ and $-\infty < z \leq -r_H, x\simeq-1$, respectively.  We find the integrand in equations (\ref{NpsiN}) and (\ref{NpsiS}), is given by
\be
(*d {\cal T})_{\phi r} =\,{\frac {  { w'}  {f'}  ^{3}({\frac {
\partial }{\partial x}}{ w'})  -{ f'}( {\frac {\partial }{\partial x}}{ \rho ^2})+{\rho ^2} ({\frac {\partial }{\partial x}
}{ f'})  }{2 { f'}\rho }{{
 \sqrt {\frac{
    \Delta_x }{\Delta_r} 
}
}}}.
\ee
Evaluating the integrals in (\ref{NpsiN}) and (\ref{NpsiS}), we find very long expressions for the $N_N$ and $N_S$ in terms of the black hole parameters $a,l,m$ and $q$ and the magnetic field parameter $b$ which are unfeasible to present here. We verify that in the special cases, the thermodynamic NUT potentials reduce exactly to the well-known results in \cite{Bordo}.

As we notice from equation (\ref{MAGINF}), the NUT charge contributes to an off-diagonal term in the asymptotic of MKNTN black holes (\ref{MAG}). As a result, we find two contributions to the total angular momentum (\ref{totJ}) from the north and south NUT tubes, which are given by
\begin{eqnarray}
J_N&=&\frac{1}{16\pi}\int _{T_N}*d{\cal P}\label{JN},\\
J_S&=&\frac{-1}{16\pi}\int _{T_S}*d{\cal P}\label{JS},
\end{eqnarray}
respectively, where ${\cal P}^\mu=\frac{\partial}{\partial \phi}$ is the space-like Killing vector.  We find the integrand in equations (\ref{JN}) and (\ref{JS}), is given  by
\be
(*d{\cal P})_{\phi r}=-{\frac { { f'}^2 ({\frac {\partial }{\partial x}}{w'} ) }{2
{ \rho} }{{\sqrt {{\frac { \Delta_x }{
 { \Delta _r}  }}}}}}.
\ee
Evaluating the integrals in (\ref{JN}) and (\ref{JS}), we find very long expressions for the $J_N$ and $J_S$ in terms of the black hole parameters $a,l,m$ and $q$ and the magnetic field parameter $b$ which are unfeasible to present here. We verify that in the special cases, the thermodynamic NUT potentials reduce exactly to the well-known results in \cite{Bordo}.

To complete the calculation, we also mention that the area ${\cal S}_N$ and ${\cal S}_S$ of the north and south NUT tubes are given by
\begin{eqnarray}
{\cal S}_N&=&n\int _{T_N}*d{\xi_N}\label{AreaN},\\
{\cal S}_S&=&-n\int _{T_S}*d{\xi_S}\label{AreaS},
\end{eqnarray}
respectively, where $\xi_N$ and $\xi_S$ are the Killing vectors (\ref{xiN}) and (\ref{xiS}), which generate the north and south pole tubes.
The integrand in equations (\ref{AreaN}) and (\ref{AreaS}), is given by
\be  
(*d{\xi_Z})_{\phi r}={{\sqrt {{\frac { \Delta_x }{ {\Delta_r}  
  }}}}
{\frac {1 }
{2\rho 
 {f'}  } 
 \left( ({w'-\Omega_Z})
  {f'}^{3}({
\frac {\partial }{\partial x}}{w'}) 
-{f'} ( {\frac {\partial }
{\partial x}}{\rho ^2} ) +{\rho ^2}  ({\frac {\partial }{\partial x}}{ f'} ) 
 \right) 
}
 },
\ee 
where $Z$ stands for $N$ and $S$, respectively.
After calculating the integrals in (\ref{AreaN}), (\ref{AreaS}), we find that ${\cal S}_N$ and ${\cal S}_S$, are independent of the magnetic field parameter $b$, and are given by
\be
{\cal S}_N={\cal S}_S=4\pi l (r_\infty -r_H),
\ee
where $r_\infty - r_H$ is the length of the north (south) NUT tube along the positive 
(negative) $z$-axis.

Furnished by the results  (\ref{totM}), (\ref{totJ}), (\ref{PSIN}), (\ref{PSIS}), (\ref{NpsiN}), (\ref{NpsiS}), (\ref{JN}) and (\ref{JS}), we can verify that equation (\ref{totM}) is indeed the Smarr relation for the MKNTN black hole (\ref{MAG}). Moreover, by construction of different thermodynamical quantities for the black hole (\ref{MAG}), and the Smarr equation (\ref{totM}), we find
the first law of thermodynamics, as given by
\be
d{\cal M}=\Phi_HdQ+\Omega_H d{\cal J}+T_HdS+\psi_N dN_N+\psi_S dN_S.\label{firstlaw}
\ee
In fact, a tedious but straightforward calculation shows that equation (\ref{totM}) implies that for ${\cal M}={\cal M}(Q,{\cal J},S,N_N,N_S)$, we have the first law of thermodynamics, as it is given by (\ref{firstlaw}). We also verify that equation (\ref{firstlaw}) reduces exactly to the well-know first law of thermodynamics for the Kerr-NUT and Melvin-Kerr-Newman black holes in \cite{Bordo} and \cite{Booth:2015nwa}.


\section{Concluding Remarks}
\label{sec:con}

In this work, we have constructed a new class of the exact solutions to the Einstein-Maxwell theory in four dimensions which describes the immersion of the Kerr-Newman-Taub-NUT spacetimes in an external magnetic field. The solutions are obtained by applying the Ernst magnetization procedure to the four dimensional Kerr-Newman-Taub-NUT spacetime as the seed. We discuss the properties of the MKNTN spacetimes and show that they are completely regular at $r=0,\,x=0$. Particularly, in addition to the extensive investigation on the space-time structure and the ergoregions for the MKNTN black hole, we also study the behaviour of the electromagnetic fields in the magnetized spacetime solutions. We find that the horizon has a non-trivial topology which leads to the eccentric horizon.  The thermodynamical quasi-local conserved  quantities of the spacetime are obtained, though they are generally quite complicated functions of the five independent parameters of MKNTN black holes, namely $m,a,l,q$ and $b$. In this paper, we also establish the Smarr formula for the MKNTN black holes. Finally, we study the thermodynamics of the MKNTN spacetimes, and show the corresponding first law of thermodynamics.


\bigskip
{\Large Acknowledgments}

This work was supported by the Natural Sciences and Engineering Research Council of Canada. 
{\textcolor{black}{
\appendix 
\section{The differential equations for $\gamma\left(\rho,z\right)$ in the LPW metric}\label{app.gamma}
Let us first consider the LPW metric (\ref{metricLPW}) with $d\chi d\chi^*$ as given in (\ref{metric2rho.z}), namely
\be 
ds^2  = - \frac{\rho ^2}{f} dt^2  + \frac{e^{2\gamma}}{f}\left(d\rho^2 + dz^2\right) + f\left( {\omega dt-d\phi } \right)^2.
\ee 
If the system is electrovacuum, the accompanying vector field $A_\mu$ obeys eq. (\ref{eqA}), which then constructs the Ernst electromagnetic potential $\Phi$, as defined in (\ref{Ernst.potential.EM}). Now let us write the Einstein equations for the electrovacuum system as
\be \label{EMeqtns}
{E_{\mu \nu }} \equiv {R_{\mu \nu }} - \left( {2{F_{\mu \alpha }}F_\nu ^\alpha  - \frac{1}{2}{g_{\mu \nu }}{F_{\alpha \beta }}{F^{\alpha \beta }}} \right) = 0.
\ee 
In fact, the Ernst equations (\ref{eq.Ernst.grav}) and (\ref{eq.Ernst.EM}) are obtained from the last equations. Particularly, from the $E_{\rho z}$ component of (\ref{EMeqtns}), we  have
\be \label{dzgamma}
{\partial _z}\gamma  = \frac{1}{{2{f^2}\rho }}\left\{ {{\rho ^2}{\partial _\rho }f{\partial _z}f - {f^4}{\partial _\rho }\omega {\partial _z}\omega  + 2f{\rho ^2}\left( {{\partial _\rho }\Phi {\partial _z}{\Phi ^*} + {\partial _\rho }{\Phi ^*}{\partial _z}\Phi } \right)} \right\},
\ee 
while from $E_{\rho\rho}$ and $E_{zz}$, we find
\be \label{drhogamma}
{\partial _\rho }\gamma  = \frac{1}{{4{f^2}\rho }}\left\{ {{\rho ^2}\left( {{{\left( {{\partial _\rho }f} \right)}^2} - {{\left( {{\partial _z}f} \right)}^2}} \right) + {f^4}\left( {{{\left( {{\partial _z}\omega } \right)}^2} - {{\left( {{\partial _\rho }\omega } \right)}^2}} \right) + 4f{\rho ^2}\left( {{\partial _\rho }\Phi {\partial _\rho}{\Phi ^*} - {\partial _z }{\Phi ^*}{\partial _z}\Phi } \right)} \right\}.
\ee 
Interestingly, despite the other metric functions $f$ and $\omega$ transform under the Ernst magnetization (as reviewed in section \ref{sec:ErnstMag}), the $\gamma$ function remains invariant after the transformation. In other words, setting $\omega \to \omega'$, $f \to f'$, $\Phi \to \Phi '$ according to (\ref{magnetization}), (\ref{fp}), and (\ref{wp}) in (\ref{dzgamma}) and (\ref{drhogamma}), leaves the two differential equations (\ref{dzgamma}) and (\ref{drhogamma}), for $\gamma$ unchanged.
Indeed, the invariance of equations (\ref{dzgamma}) and (\ref{drhogamma}) under the Ernst magnetization (\ref{magnetization}), (\ref{fp}), and (\ref{wp})  is not too obvious. Therefore, we provide a simple example which illustrate the invariance of the  $\gamma$ function in LPW metric (\ref{metricLPW}) under the Ernst transformation. We consider magnetizing Minkowski spacetime to get the magnetic Melvin spacetime. In the cylindrical type coordinate $\left(t,\rho,z,\phi\right)$, Minkowski spacetime can be expressed as
\be 
d{s^2} =  - d{t^2} + d{\rho ^2} + d{z^2} + {\rho ^2}d{\phi ^2}.
\ee 
Obviously, the LPW metric functions for the Minkowski spacetime are $f=\rho^2$, $\gamma = \ln \rho$, and $\omega =0$.  Accordingly, the Ernst potentials for this seed solution are ${\cal E} =\rho^2$ and $\Phi =0$, which give us $\Lambda = 1 + b^2 \rho^2$. The magnetized Ernst potentials then read 
\be \label{Ernst.Melvin}
{\cal E}' = \Lambda^{-1}\rho^2 ~~{\rm and}~~{\Phi}' = -\Lambda^{-1}\rho^2 b,
\ee 
while the metric functions are 
\be \label{fw.Melvin}
f ' = \Lambda^{-2}\rho^2~~{\rm and}~~\omega'=0.
\ee 
On the other hand, the associated vector solution obtained from $\Phi'$ has the components $A'_\phi = -\Lambda^{-1} b\rho^2 $ and ${\tilde A}'_\phi =0$, which lead to $A'_t =0$, according to (\ref{eqA}). Suppose we consider the function $\gamma'$ in the Melvin spacetime, changes under the Ernst magnetization. To find $\gamma'$, we need to solve the remaining non-zero equations (\ref{EMeqtns}) for the Melvin spacetime with (\ref{Ernst.Melvin}) and (\ref{fw.Melvin}). They are given by
\be 
{E'_{\rho \rho }} =  - {\rho ^{ - 2}}\left\{ {{\rho ^2}\left( {\partial _z^2\gamma'  + \partial _\rho ^2\gamma' } \right) - \rho {\partial _\rho }\gamma'  - 2} \right\},
\ee 
\be 
{E'_{\rho z}} =  - {\rho ^{ - 1}}{\partial _z}\gamma',
\ee 
and
\be 
{E'_{zz}} =  - {\rho ^{ - 1}}\left\{ {\rho \left( {\partial _z^2\gamma' + \partial _\rho ^2\gamma' } \right) + {\partial _\rho }\gamma' } \right\}.
\ee 
We notice that the absence of external magnetic parameter $b$ in the last three equations tells us that $\gamma'$ obeys the same differential equations as $\gamma$. Hence, we can conclude $\gamma'$ for the Melvin spacetime is the same as $\gamma$ for the  Minkowski spacetime. We also note that we get the same result $\gamma'=\gamma$, by plugging equations (\ref{Ernst.Melvin}) and (\ref{fw.Melvin}) in (\ref{dzgamma}) and (\ref{drhogamma}).
In regard to the Melvin-Kerr-Newman-Taub-NUT spacetime (\ref{MAG}) with the Boyer-Lindquist type coordinate $\left(t,r,x=\cos\theta,\phi\right)$, we find the following differential equations for  $\gamma\left(r,x\right)$  from the Einstein equations,
\[{f^2}\left( {2{\Delta _r}x{\partial _r}\gamma  - {\Delta _x}{\partial _r}{\Delta _r}{\partial _x}\gamma } \right) + 2f\rho^2\left( {{\partial _r}\Phi {\partial _x}{\Phi ^*} + {\partial _r}{\Phi ^*}{\partial _x}\Phi } \right)\]
\be \label{eq.dgamma1} - {f^2}\left( {{f^2}{\partial _r}\omega {\partial _x}\omega  + x{\partial _r}{\Delta _r}} \right) + \rho^2{\partial _r}f{\partial _x}f = 0,\ee
and
\[ - 2{\rho ^2}{f^2}\left( {{\partial _r}{\Delta _r}{\partial _r}\gamma  + 2x{\partial _x}\gamma } \right) + {\rho ^2}\left( {{\Delta _r}{{\left( {{\partial _r}f} \right)}^2} - {\Delta _x}{{\left( {{\partial _x}f} \right)}^2}} \right)\]
\be \label{eq.dgamma2}
 - {f^4}\left( {{\Delta _r}{{\left( {{\partial _r}\omega } \right)}^2} - {\Delta _x}{{\left( {{\partial _x}\omega } \right)}^2}} \right) + 4{\rho ^2}f\left( {{\Delta _r}{\partial _r}\Phi {\partial _r}{\Phi ^*} - {\Delta _x}{\partial _x}\Phi {\partial _x}{\Phi ^*}} \right) = 0.
\ee 
We have explicitly checked that the seed solution $\left\{f,\omega,\Phi \right\}$ that belongs to Kerr-Newman-Taub-NUT system, and the magnetized version $\left\{f',\omega',\Phi' \right\}$ (as presented in section \ref{sec:MKNTNsol}), obey (\ref{eq.dgamma1}) and (\ref{eq.dgamma2}), with the same $\gamma$ function, as given by (\ref{e2gamma}).
}

\section{The magnetized metric functions for the Melvin-Kerr-Newman-Taub-NUT spacetimes}
The coefficients that appear in the metric function $f'$ for the Melvin-Kerr-Newman-NUT black hole,  in equation (\ref{FPMAG}), are given by

\[
c_4 = -a^2 \Delta_r,
\]
\[
c_3 = -4la \Delta_r,
\]
\[
c_2 = 3{l}^{4}+ \left( 8rm-8{a}^{2}-4{q}^{2}-6{r}^{2} \right) {l}^{2}+{a}^{4}-4{a}^{2}mr+2{a}^{2}{q}^{2}-{r}^{4},
\]
\[
c_1 = 4la \Delta_r,
\]
\[
c_0 = 3{a}^{2}{l}^{2}+2{a}^{2}mr-{a}^{2}{q}^{2}+{a}^{2}{r}^{2}+{l}^{4}+2{l}^{2}{r}^{2}+{r}^{4},
\]
\[
d_6 = a^2 b^4 \Delta_r^2,
\]
\[
d_5 = 6 a b^4 l\Delta_r^2 ,
\]
\[
d_4 = -b\left\{  2{a}^{6}{b}^{2}-21{a}^{4}{b}^{2}{l}^{2}-4{a}^{4}{b}^{2}{m}^{2}-
12{a}^{4}{b}^{2}mr+6{a}^{4}{b}^{2}{q}^{2}+3{a}^{4}{b}^{2}{r}^{2}
+28{a}^{2}{b}^{2}{l}^{4}\right.
\]
\[
-8{a}^{2}{b}^{2}{l}^{2}{m}^{2}+84{a}^{2}
{b}^{2}{l}^{2}mr-32{b}^{2}{a}^{2}{l}^{2}{q}^{2}-36{a}^{2}{b}^{2}{l}^{2}{r}^{2}+24{a}^{2}{b}^{2}{m}^{2}{r}^{2}
\]
\[
-24{a}^{2}{b}^{2}m{q}^{2}r-12{a}^{2}{b}^{2}m{r}^{3}+4{a}^{2}{b}^{2}{q}^{4}+8{a}^{2}{b}^{2}{q}^{2}{r}^{2}-9{b}^{2}{l}^{6}-4{b}^{2}{l}^{4}{m}^{2}-24{b}^{2}{l}^{4}mr
\]
\[
+18{b}^{2}{l}^{4}{q}^{2}+9{b}^{2}{l}^{4}{r}^{2}-36{b}^{2}{l}^{2}{m}^{2}{r}^{2}+32{b}^{2}{l}^{2}m{q}^{2}r+40{b}^{2}{l}^{2}m{r}^{3}-9{b}^{2}{l}^{2}{q}^{4}
\]
\[
-12{b}^{2}{l}^{2}{q}^{2}{r}^{2}-15{b}^{2}{l}^{2}{r}^{4}-{b}^{2}{q}^{4}{r}^{2}+2{b}^{2}{q}^{2}{r}^{4}-{b}^{2}{r}^{6}-4{a}^{3}bqr+4ab{l}^{2}qr
\]
\[
\left. +8abmq{r}^{2}-4ab{q}^{3}r-4abq{r}^{3}+2{a}^{4}-2{a}^{2}{l}^{2}-4{a}^{2}mr+2{a}^{2}{q}^{2}+2{a}^{2}{r}^{2} \right\},
\]
\[
d_3 = -4b^2 l\left\{ 3{a}^{5}{b}^{2}-10{a}^{3}{b}^{2}{l}^{2}+2{a}^{3}{b}^{2}{m}^{2}-
20{a}^{3}{b}^{2}mr+6{a}^{3}{b}^{2}{q}^{2} +6{a}^{3}{b}^{2}{r}^{2}\right.
\]
\[
+7a{b}^{2}{l}^{4}+2a{b}^{2}{l}^{2}{m}^{2}+16a{b}^{2}{l}^{2}mr-10
a{b}^{2}{l}^{2}{q}^{2}-2a{b}^{2}{l}^{2}{r}^{2}+18a{b}^{2}{m}^{2}{r}^{2}
\]
\[
-14a{b}^{2}m{q}^{2}r-12a{b}^{2}m{r}^{3}+3a{b}^{2}{q}^{4}+
2a{b}^{2}{q}^{2}{r}^{2}+3a{b}^{2}{r}^{4}+2{a}^{2}bmq-4{a}^{2}b
qr
\]
\[
\left. +2b{l}^{2}mq+6bmq{r}^{2}-2b{q}^{3}r-4bq{r}^{3}+2{a}^{3}-2
a{l}^{2}-4amr+2a{q}^{2}+2a{r}^{2} \right\},
\]
\[
d_2 = {a}^{6}{b}^{4}-28{a}^{4}{b}^{4}{l}^{2}+8{a}^{4}{b}^{4}{m}^{2}-12
{a}^{4}{b}^{4}mr+4{a}^{4}{b}^{4}{q}^{2}+37{a}^{2}{b}^{4}{l}^{4}+12
{a}^{2}{b}^{4}{l}^{2}{m}^{2}
\]
\[
+84{a}^{2}{b}^{4}{l}^{2}mr-38{a}^{2}
{b}^{4}{l}^{2}{q}^{2}-18{a}^{2}{b}^{4}{l}^{2}{r}^{2}+36{a}^{2}{b}^
{4}{m}^{2}{r}^{2}-20{a}^{2}{b}^{4}m{q}^{2}r
\]
\[
-12{a}^{2}{b}^{4}m{r}^{3}+4{a}^{2}{b}^{4}{q}^{4}+6{a}^{2}{b}^{4}{q}^{2}{r}^{2}-3{a}^{2}
{b}^{4}{r}^{4}+6{b}^{4}{l}^{6}+24{b}^{4}{l}^{4}mr-6{b}^{4}{l}^{4
}{q}^{2}
\]
\[
-30{b}^{4}{l}^{4}{r}^{2}-8{b}^{4}{l}^{2}m{r}^{3}+12{b}^{4}{l}^{2}{q}^{2}{r}^{2}-6{b}^{4}{l}^{2}{r}^{4}+2{b}^{4}{q}^{2}{r}^
{4}-2{b}^{4}{r}^{6}+8{a}^{3}{b}^{3}mq
\]
\[
-8{a}^{3}{b}^{3}qr+16a{b}
^{3}{l}^{2}mq-8a{b}^{3}{l}^{2}qr+24a{b}^{3}mq{r}^{2}-4a{b}^{3}{q
}^{3}r-8a{b}^{3}q{r}^{3}+2{a}^{4}{b}^{2}
\]
\[
-16{a}^{2}{b}^{2}{l}^{2}
-8{a}^{2}{b}^{2}mr+4{a}^{2}{b}^{2}{q}^{2}+6{b}^{2}{l}^{4}+16{b}^{2}{l}^{2}mr-2{b}^{2}{l}^{2}{q}^{2}-12{b}^{2}{l}^{2}{r}^{2}
\]
\[
+6{b}^{2}{q}^{2}{r}^{2}-2{b}^{2}{r}^{4}-4abqr+{a}^{2},
\]
\[
d_1 = 2l\left\{ 3{a}^{5}{b}^{4}-14{a}^{3}{b}^{4}{l}^{2}-4{a}^{3}{b}^{4}{m}^{2}-
12{a}^{3}{b}^{4}mr+6{a}^{3}{b}^{4}{q}^{2}+6{a}^{3}{b}^{4}{r}^{2}
-5a{b}^{4}{l}^{4} \right.
\]
\[
-12a{b}^{4}{l}^{2}mr+2a{b}^{4}{l}^{2}{q}^{2}+18
a{b}^{4}{l}^{2}{r}^{2}+4a{b}^{4}m{r}^{3}-2a{b}^{4}{q}^{2}{r}^{2}
+3a{b}^{4}{r}^{4}-8{a}^{2}{b}^{3}mq
\]
\[
\left. -8{b}^{3}{l}^{2}qr+4{b}^{2}
{a}^{3}-4a{b}^{2}{l}^{2}-8a{b}^{2}mr-2a{b}^{2}{q}^{2}+4a{b}^{2
}{r}^{2}-4bqr+a \right\},
\]
\[
d_0 = 9{a}^{4}{b}^{4}{l}^{2}+4{a}^{4}{b}^{4}{m}^{2}+4{a}^{4}{b}^{4}mr+
{a}^{4}{b}^{4}{r}^{2}+6{a}^{2}{b}^{4}{l}^{4}+12{a}^{2}{b}^{4}{l}^{2}mr+4{a}^{2}{b}^{4}m{r}^{3}
\]
\[
+2{a}^{2}{b}^{4}{r}^{4}+{b}^{4}{l}^{6}
+7{b}^{4}{l}^{4}{r}^{2}+7{b}^{4}{l}^{2}{r}^{4}+{b}^{4}{r}^{6}+8{a}^{3}{b}^{3}mq+4{a}^{3}{b}^{3}qr+12a{b}^{3}{l}^{2}qr
\]
\[
+4a{b}^{3}q
{r}^{3}+6{a}^{2}{b}^{2}{l}^{2}+4{a}^{2}{b}^{2}mr+4{a}^{2}{b}^{2}
{q}^{2}+2{a}^{2}{b}^{2}{r}^{2}+2{b}^{2}{l}^{4}+4{b}^{2}{l}^{2}{r}^{2}
\]
\be
+2{b}^{2}{r}^{4}+4abqr+{l}^{2}+{r}^{2}.
\ee
\\
The coefficients that appear in the metric function $\omega'$ for the Melvin-Kerr-Newman-NUT black hole,  in equation (\ref{OPMAG}), are given by

\[
{\tilde c}_6 = a{b}^{4} \Delta_r \left( 
2{a}^{2}{l}^{2}-4{m}^{2}{a}^{2}+6{a}^{2}mr-{a}^{2}{q}^{2}-2{l}
^{4}-4{l}^{2}{m}^{2}+6{l}^{2}mr+3{l}^{2}{q}^{2} \right. 
\]
\[
\left. -6{l}^{2}{r}^{2
}-2m{r}^{3}-{q}^{4}+3{q}^{2}{r}^{2} \right),
\]
\[
{\tilde c}_5 = -2{b}^{3}l \Delta_r 
\left( 3{a}^{4}b-6{a}^{2}b{l}^{2}+4{a}^{2}b{q}^{2}-6{a}^{2}b{
	r}^{2}+3b{l}^{4}+4b{l}^{2}{m}^{2} -4b{l}^{2}mr-4b{l}^{2}{q}^{2}
+6b{l}^{2}{r}^{2} \right.
\]
\[
\left. +4bm{r}^{3}+b{q}^{4}-4b{q}^{2}{r}^{2}-b{r}^{4}-
4qma+4qar \right),
\]
\[
{\tilde c}_4 = -{b}^{2} \Delta_r  \left( 
32{a}^{3}{b}^{2}{l}^{2}+8{a}^{3}{b}^{2}{m}^{2}-6{a}^{3}{b}^{2}mr
-30a{b}^{2}{l}^{4}-20a{b}^{2}{l}^{2}{m}^{2}+18a{b}^{2}{l}^{2}mr+
25{q}^{2}{l}^{2}{b}^{2}a \right.
\]
\[
-42a{b}^{2}{l}^{2}{r}^{2}-14a{b}^{2}m{r}
^{3}-a{b}^{2}{q}^{4}+9{r}^{2}a{b}^{2}{q}^{2}+16{a}^{2}bmq-12r{a}
^{2}bq-16b{l}^{2}mq
\]
\[
\left. +12rb{l}^{2}q-4{r}^{3}bq+6a{q}^{2} \right),
\]
\[
{\tilde c}_3 = 4{b}^{2}l \Delta_r 
\left( {a}^{4}{b}^{2}-20{a}^{2}{b}^{2}{l}^{2}-8{a}^{2}{b}^{2}{m}^
{2}+6{a}^{2}{b}^{2}mr+3{a}^{2}{b}^{2}{q}^{2}-6{a}^{2}{b}^{2}{r}^
{2}+{b}^{2}{l}^{4}-2{b}^{2}{l}^{2}mr \right.
\]
\[
\left. -2{b}^{2}{l}^{2}{q}^{2}+6{b}
^{2}{l}^{2}{r}^{2}+2{b}^{2}m{r}^{3}-2{b}^{2}{q}^{2}{r}^{2}+{b}^{2}
{r}^{4}-10abmq+8abqr-3{q}^{2} \right) ,
\]
\[
{\tilde c}_2 = 24{a}^{5}{b}^{4}{l}^{2}+8{a}^{5}{b}^{4}{m}^{2}-14r{a}^{5}{b}^{4}
m-92{a}^{3}{b}^{4}{l}^{4}-16{a}^{3}{b}^{4}{l}^{2}{m}^{2}-36r{a}^
{3}{b}^{4}{l}^{2}m+46{a}^{3}{b}^{4}{l}^{2}{q}^{2}
\]
\[
-12{r}^{2}{a}^{3}
{b}^{4}{l}^{2}-24r{a}^{3}{b}^{4}{m}^{3}+12{a}^{3}{b}^{4}{m}^{2}{q}
^{2}+48{a}^{3}{b}^{4}{m}^{2}{r}^{2}-20{a}^{3}{b}^{4}m{q}^{2}r-20
{r}^{3}{b}^{4}{a}^{3}m
\]
\[
+6{r}^{2}{a}^{3}{b}^{4}{q}^{2}+28a{b}^{4}{l}
^{6}+58ra{b}^{4}{l}^{4}m-46a{b}^{4}{l}^{4}{q}^{2}-24{r}^{2}a{b}^
{4}{l}^{2}{m}^{2}-8ra{b}^{4}{l}^{2}m{q}^{2}
\]
\[
+108a{b}^{4}{l}^{2}m{r}^{3}+10a{b}^{4}{l}^{2}{q}^{4}-36{r}^{2}a{b}^{4}{l}^{2}{q}^{2}-36
a{b}^{4}{l}^{2}{r}^{4}+24{b}^{4}a{m}^{2}{r}^{4}-24{r}^{3}{b}^{4}am
{q}^{2}
\]
\[
-6a{b}^{4}m{r}^{5}+6{r}^{2}a{b}^{4}{q}^{4}+6{b}^{4}a{q}^{2}{r}^{4}+8{a}^{4}{b}^{3}mq-16r{a}^{4}{b}^{3}q-24{a}^{2}{b}^{3}{l}^{2}mq+32r{a}^{2}{b}^{3}{l}^{2}q
\]
\[
-32r{a}^{2}{b}^{3}{m}^{2}q+16{a}^{2}{b}^{3}m{q}^{3}+48{a}^{2}{b}^{3}mq{r}^{2}-20r{a}^{2}{b}^{3}{q}^{3}-16{r}^{3}{b}^{3}{a}^{2}q-16r{b}^{3}{l}^{4}q
\]
\[
-24{r}^{2}{b}^{3}{l}^{2}mq+12r{b}^{3}{l}^{2}{q}^{3}+16{r}^{3}{b}^{3}{l}^{2}q+8
{b}^{3}mq{r}^{4}-4{r}^{3}{b}^{3}{q}^{3}-12{q}^{2}{l}^{2}{b}^{2}a
\]
\[
-12ra{b}^{2}m{q}^{2}+6a{b}^{2}{q}^{4}-4r{a}^{2}bq-4rb{l}^{2}q-4
{r}^{3}bq+2a{l}^{2}+2amr-a{q}^{2},
\]
\[
{\tilde c}_1 = 2l \Delta_r  \left( {a}^
{4}{b}^{4}+18{a}^{2}{b}^{4}{l}^{2}-12{a}^{2}{b}^{4}mr-2{a}^{2}{b
}^{4}{q}^{2}+6{a}^{2}{r}^{2}{b}^{4}+{b}^{4}{l}^{4}-6{b}^{4}{l}^{2}
{r}^{2}-3{b}^{4}{r}^{4}-12a{b}^{3}qr-1 \right),
\]
\[
{\tilde c}_0 = 6{a}^{5}{b}^{4}{l}^{2}+4{a}^{5}{b}^{4}{m}^{2}+2r{a}^{5}{b}^{4}m+
{a}^{5}{b}^{4}{q}^{2}+26{a}^{3}{b}^{4}{l}^{4}+4{a}^{3}{b}^{4}{l}^{2}{m}^{2}-7{a}^{3}{b}^{4}{l}^{2}{q}^{2}+6{r}^{2}{a}^{3}{b}^{4}{l}^{2}
\]
\[
-12{a}^{3}{b}^{4}{m}^{2}{r}^{2}+6{a}^{3}{b}^{4}m{q}^{2}r-4{r}
^{3}{b}^{4}{a}^{3}m+{a}^{3}{b}^{4}{q}^{4}+{r}^{2}{a}^{3}{b}^{4}{q}^{2}
+8a{b}^{4}{l}^{6}+2ra{b}^{4}{l}^{4}m-12a{b}^{4}{l}^{2}m{r}^{3}+4
{r}^{2}a{b}^{4}{l}^{2}{q}^{2}
\]
\[
-6a{b}^{4}m{r}^{5}+8{a}^{4}{b}^{3}m
q+4r{a}^{4}{b}^{3}q+8{a}^{2}{b}^{3}{l}^{2}mq-8r{a}^{2}{b}^{3}{l}^{2}q-8{a}^{2}{b}^{3}mq{r}^{2}+8r{a}^{2}{b}^{3}{q}^{3}+4r{b}^{3}
{l}^{4}q-4{b}^{3}q{r}^{5}
\]
\be
+6{a}^{3}{b}^{2}{q}^{2}+6{q}^{2}{l}^{2}
{b}^{2}a+6{r}^{2}a{b}^{2}{q}^{2}+4r{a}^{2}bq+4rb{l}^{2}q+4{r}^{3}bq-2a{l}^{2}-2amr+a{q}^{2}.
\ee
\\
The coefficients that appear in the Maxwell's field  $A'_\phi$ for the Melvin-Kerr-Newman-NUT black hole,  in equation (\ref{APMAG}), are given by

\begin{eqnarray}
a_0&=& -9\,{a}^{4}{b}^{3}{l}^{2}-4\,{a}^{4}{b}^{3}{m}^{2}-4\,{a}^{4}{b}^{3}m
r-{a}^{4}{b}^{3}{r}^{2}-6\,{a}^{2}{b}^{3}{l}^{4}-12\,{a}^{2}{b}^{3}{l}
^{2}mr-4\,{a}^{2}{b}^{3}m{r}^{3}-2\,{a}^{2}{b}^{3}{r}^{4}-{b}^{3}{l}^{
6}\nonumber\\
&-&7\,{b}^{3}{l}^{4}{r}^{2}-7\,{b}^{3}{l}^{2}{r}^{4}-{b}^{3}{r}^{6}-6
\,{a}^{3}{b}^{2}mq-3\,{a}^{3}{b}^{2}qr-9\,a{b}^{2}{l}^{2}qr-3\,a{b}^{2
}q{r}^{3}-3\,{a}^{2}b{l}^{2}-2\,{a}^{2}bmr-2\,{a}^{2}b{q}^{2}\nonumber\\
&-&{a}^{2}b
{r}^{2}-b{l}^{4}-2\,b{l}^{2}{r}^{2}-b{r}^{4}-aqr,      \nonumber\nonumber\\
a_1&=& -6\,{a}^{5}{b
}^{3}l+28\,{a}^{3}{b}^{3}{l}^{3}+8\,{a}^{3}{b}^{3}l{m}^{2}+24\,{a}^{3}
{b}^{3}lmr-12\,{a}^{3}{b}^{3}l{q}^{2}-12\,{a}^{3}{b}^{3}l{r}^{2}+10\,a
{b}^{3}{l}^{5}+24\,a{b}^{3}{l}^{3}mr\nonumber\\
&-&4\,a{b}^{3}{l}^{3}{q}^{2}-36\,a{b
}^{3}{l}^{3}{r}^{2}-8\,a{b}^{3}lm{r}^{3}+4\,a{b}^{3}l{q}^{2}{r}^{2}-6
\,a{b}^{3}l{r}^{4}+12\,{a}^{2}{b}^{2}lmq
+12\,{b}^{2}{l}^{3}qr-4\,{a}^{
3}bl+4\,ab{l}^{3}\nonumber\\
&+&8\,ablmr+2\,abl{q}^{2}-4\,abl{r}^{2}+2\,lqr,       \nonumber\nonumber\\
a_2&=& -{a}^{6}{b}^{3}+28\,{a}^{4}{b}^{3}{l}^{2}-8\,{a}^{4}{b}^{3}{
m}^{2}+12\,{a}^{4}{b}^{3}mr-4\,{a}^{4}{b}^{3}{q}^{2}-37\,{a}^{2}{b}^{3
}{l}^{4}-12\,{a}^{2}{b}^{3}{l}^{2}{m}^{2}-84\,{a}^{2}{b}^{3}{l}^{2}mr\nonumber\\
&+&
38\,{a}^{2}{b}^{3}{l}^{2}{q}^{2}+18\,{a}^{2}{b}^{3}{l}^{2}{r}^{2}-36\,
{a}^{2}{b}^{3}{m}^{2}{r}^{2}+20\,{a}^{2}{b}^{3}m{q}^{2}r+12\,{a}^{2}{b
}^{3}m{r}^{3}-4\,{a}^{2}{b}^{3}{q}^{4}-6\,{a}^{2}{b}^{3}{q}^{2}{r}^{2}
+3\,{a}^{2}{b}^{3}{r}^{4}\nonumber\\
&-&6\,{b}^{3}{l}^{6}-24\,{b}^{3}{l}^{4}mr+6\,{b
}^{3}{l}^{4}{q}^{2}+30\,{b}^{3}{l}^{4}{r}^{2}+8\,{b}^{3}{l}^{2}m{r}^{3
}-12\,{b}^{3}{l}^{2}{q}^{2}{r}^{2}+6\,{b}^{3}{l}^{2}{r}^{4}-2\,{b}^{3}
{q}^{2}{r}^{4}+2\,{b}^{3}{r}^{6}\nonumber\\
&-&6\,{a}^{3}{b}^{2}mq+6\,{a}^{3}{b}^{2}
qr-12\,a{b}^{2}{l}^{2}mq+6\,a{b}^{2}{l}^{2}qr-18\,a{b}^{2}mq{r}^{2}+3
\,a{b}^{2}{q}^{3}r+6\,a{b}^{2}q{r}^{3}-{a}^{4}b+8\,{a}^{2}b{l}^{2}\nonumber\\
&+&4\,
{a}^{2}bmr-2\,{a}^{2}b{q}^{2}-3\,b{l}^{4}-8\,b{l}^{2}mr+b{l}^{2}{q}^{2
}+6\,b{l}^{2}{r}^{2}-3\,b{q}^{2}{r}^{2}+b{r}^{4}+aqr,     \nonumber\nonumber\\
a_3&=&  12\,{a}^{5}{b}^{3}l-40\,{a}^{3}{b}^{3}{l}^{3}+8\,{a}^{3}{b}^{3
}l{m}^{2}-80\,{a}^{3}{b}^{3}lmr+24\,{a}^{3}{b}^{3}l{q}^{2}+24\,{a}^{3}
{b}^{3}l{r}^{2}+28\,a{b}^{3}{l}^{5}+8\,a{b}^{3}{l}^{3}{m}^{2}\nonumber\\
&+&64\,a{b}
^{3}{l}^{3}mr-40\,a{b}^{3}{l}^{3}{q}^{2}-8\,a{b}^{3}{l}^{3}{r}^{2}+72
\,a{b}^{3}l{m}^{2}{r}^{2}-56\,a{b}^{3}lm{q}^{2}r-48\,a{b}^{3}lm{r}^{3}
+12\,a{b}^{3}l{q}^{4}\nonumber\\
&+&8\,a{b}^{3}l{q}^{2}{r}^{2}+12\,a{b}^{3}l{r}^{4}+
6\,{a}^{2}{b}^{2}lmq-12\,{a}^{2}{b}^{2}lqr+6\,{b}^{2}{l}^{3}mq+18\,{b}
^{2}lmq{r}^{2}-6\,{b}^{2}l{q}^{3}r-12\,{b}^{2}lq{r}^{3}\nonumber\\
&+&4\,{a}^{3}bl-4
\,ab{l}^{3}-8\,ablmr+4\,abl{q}^{2}+4\,abl{r}^{2} ,     \nonumber\nonumber
\end{eqnarray}
\begin{eqnarray}
a_4&=&   2\,{a}^{6}{b}^{3}-21\,{a}^{4}{b}^{3}{l}^{2}-4\,{a}^{4}{b}^{3}{
m}^{2}-12\,{a}^{4}{b}^{3}mr+6\,{a}^{4}{b}^{3}{q}^{2}+3\,{a}^{4}{b}^{3}
{r}^{2}+28\,{a}^{2}{b}^{3}{l}^{4}-8\,{a}^{2}{b}^{3}{l}^{2}{m}^{2}\nonumber\\
&+&84\,
{a}^{2}{b}^{3}{l}^{2}mr-32\,{a}^{2}{b}^{3}{l}^{2}{q}^{2}-36\,{a}^{2}{b
}^{3}{l}^{2}{r}^{2}+24\,{a}^{2}{b}^{3}{m}^{2}{r}^{2}-24\,{a}^{2}{b}^{3
}m{q}^{2}r-12\,{a}^{2}{b}^{3}m{r}^{3}+4\,{a}^{2}{b}^{3}{q}^{4}\nonumber\\
&+&8\,{a}^
{2}{b}^{3}{q}^{2}{r}^{2}-9\,{b}^{3}{l}^{6}-4\,{b}^{3}{l}^{4}{m}^{2}-24
\,{b}^{3}{l}^{4}mr+18\,{b}^{3}{l}^{4}{q}^{2}+9\,{b}^{3}{l}^{4}{r}^{2}-
36\,{b}^{3}{l}^{2}{m}^{2}{r}^{2}+32\,{b}^{3}{l}^{2}m{q}^{2}r\nonumber\\
&+&40\,{b}^{
3}{l}^{2}m{r}^{3}-9\,{b}^{3}{l}^{2}{q}^{4}-12\,{b}^{3}{l}^{2}{q}^{2}{r
}^{2}-15\,{b}^{3}{l}^{2}{r}^{4}-{b}^{3}{q}^{4}{r}^{2}+2\,{b}^{3}{q}^{2
}{r}^{4}-{b}^{3}{r}^{6}-3\,{a}^{3}{b}^{2}qr+3\,a{b}^{2}{l}^{2}qr\nonumber\\
&+&6\,a{
b}^{2}mq{r}^{2}-3\,a{b}^{2}{q}^{3}r-3\,a{b}^{2}q{r}^{3}+{a}^{4}b-{a}^{
2}b{l}^{2}-2\,{a}^{2}bmr+{a}^{2}b{q}^{2}+{a}^{2}b{r}^{2},  \nonumber\\
a_5&=&-6ab^3l\Delta_r^2,\nn\\
a_6&=&-a^2b^3\Delta_r^2,
\end{eqnarray}
and
\begin{eqnarray}
b_0&=& (9\,{a}^{4}{b}^{4}{l}^{2}+4\,{a}^{4}{b}^{4}{m}^{2}+4\,{a}^{4}{b}^{4}mr
+{a}^{4}{b}^{4}{r}^{2}+6\,{a}^{2}{b}^{4}{l}^{4}+12\,{a}^{2}{b}^{4}{l}^
{2}mr+4\,{a}^{2}{b}^{4}m{r}^{3}+2\,{a}^{2}{b}^{4}{r}^{4}+{b}^{4}{l}^{6
}\nonumber\nonumber\\
&+&7\,{b}^{4}{l}^{4}{r}^{2}+7\,{b}^{4}{l}^{2}{r}^{4}+{b}^{4}{r}^{6}+8\,
{a}^{3}{b}^{3}mq+4\,{a}^{3}{b}^{3}qr+12\,a{b}^{3}{l}^{2}qr+4\,a{b}^{3}
q{r}^{3}+6\,{a}^{2}{b}^{2}{l}^{2}+4\,{a}^{2}{b}^{2}mr\nonumber\\
&+&4\,{a}^{2}{b}^{2
}{q}^{2}+2\,{a}^{2}{b}^{2}{r}^{2}+2\,{b}^{2}{l}^{4}+4\,{b}^{2}{l}^{2}{
r}^{2}+2\,{b}^{2}{r}^{4}+4\,abqr+{l}^{2}+{r}^{2} ,     \nonumber\nonumber\\
b_1&=&  6\,{a}^{5}{b}
^{4}l-28\,{a}^{3}{b}^{4}{l}^{3}-8\,{a}^{3}{b}^{4}l{m}^{2}-24\,{a}^{3}{
b}^{4}lmr+12\,{a}^{3}{b}^{4}l{q}^{2}+12\,{a}^{3}{b}^{4}l{r}^{2}-10\,a{
b}^{4}{l}^{5}-24\,a{b}^{4}{l}^{3}mr\nonumber\\
&+&4\,a{b}^{4}{l}^{3}{q}^{2}+36\,a{b}
^{4}{l}^{3}{r}^{2}+8\,a{b}^{4}lm{r}^{3}-4\,a{b}^{4}l{q}^{2}{r}^{2}+6\,
a{b}^{4}l{r}^{4}-16\,{a}^{2}{b}^{3}lmq-16\,{b}^{3}{l}^{3}qr+8\,{a}^{3}
{b}^{2}l\nonumber\\
&-&8\,a{b}^{2}{l}^{3}-16\,a{b}^{2}lmr-4\,a{b}^{2}l{q}^{2}+8\,a{b
}^{2}l{r}^{2}-8\,blqr+2\,al,     \nonumber\nonumber\\
b_2&=&{a}^{6}{b}^{4}-28\,{a}^{
4}{b}^{4}{l}^{2}+8\,{a}^{4}{b}^{4}{m}^{2}-12\,{a}^{4}{b}^{4}mr+4\,{a}^
{4}{b}^{4}{q}^{2}+37\,{a}^{2}{b}^{4}{l}^{4}+12\,{a}^{2}{b}^{4}{l}^{2}{
m}^{2}+84\,{a}^{2}{b}^{4}{l}^{2}mr\nonumber\\
&-&38\,{a}^{2}{b}^{4}{l}^{2}{q}^{2}-18
\,{a}^{2}{b}^{4}{l}^{2}{r}^{2}+36\,{a}^{2}{b}^{4}{m}^{2}{r}^{2}-20\,{a
}^{2}{b}^{4}m{q}^{2}r-12\,{a}^{2}{b}^{4}m{r}^{3}+4\,{a}^{2}{b}^{4}{q}^
{4}+6\,{a}^{2}{b}^{4}{q}^{2}{r}^{2}\nonumber\\
&-&3\,{a}^{2}{b}^{4}{r}^{4}+6\,{b}^{4
}{l}^{6}+24\,{b}^{4}{l}^{4}mr-6\,{b}^{4}{l}^{4}{q}^{2}-30\,{b}^{4}{l}^
{4}{r}^{2}-8\,{b}^{4}{l}^{2}m{r}^{3}+12\,{b}^{4}{l}^{2}{q}^{2}{r}^{2}-
6\,{b}^{4}{l}^{2}{r}^{4}+2\,{b}^{4}{q}^{2}{r}^{4}\nonumber\\
&-&2\,{b}^{4}{r}^{6}+8
\,{a}^{3}{b}^{3}mq-8\,{a}^{3}{b}^{3}qr+16\,a{b}^{3}{l}^{2}mq-8\,a{b}^{
3}{l}^{2}qr+24\,a{b}^{3}mq{r}^{2}-4\,a{b}^{3}{q}^{3}r-8\,a{b}^{3}q{r}^
{3}+2\,{a}^{4}{b}^{2}\nonumber\\
&-&16\,{a}^{2}{b}^{2}{l}^{2}-8\,{a}^{2}{b}^{2}mr+4
\,{a}^{2}{b}^{2}{q}^{2}+6\,{b}^{2}{l}^{4}+16\,{b}^{2}{l}^{2}mr-2\,{b}^
{2}{l}^{2}{q}^{2}-12\,{b}^{2}{l}^{2}{r}^{2}+6\,{b}^{2}{q}^{2}{r}^{2}-2
\,{b}^{2}{r}^{4}\nonumber\\
&-&4\,abqr+{a}^{2},         \nonumber\nonumber\\
b_3&=&  -12\,{a}^{5}{
b}^{4}l+40\,{a}^{3}{b}^{4}{l}^{3}-8\,{a}^{3}{b}^{4}l{m}^{2}+80\,{a}^{3
}{b}^{4}lmr-24\,{a}^{3}{b}^{4}l{q}^{2}-24\,{a}^{3}{b}^{4}l{r}^{2}-28\,
a{b}^{4}{l}^{5}-8\,a{b}^{4}{l}^{3}{m}^{2}\nonumber\\
&-&64\,a{b}^{4}{l}^{3}mr+40\,a{
b}^{4}{l}^{3}{q}^{2}+8\,a{b}^{4}{l}^{3}{r}^{2}-72\,a{b}^{4}l{m}^{2}{r}
^{2}+56\,a{b}^{4}lm{q}^{2}r+48\,a{b}^{4}lm{r}^{3}-12\,a{b}^{4}l{q}^{4}
\nonumber\\&-&8\,a{b}^{4}l{q}^{2}{r}^{2}-12\,a{b}^{4}l{r}^{4}-8\,{a}^{2}{b}^{3}lmq+
16\,{a}^{2}{b}^{3}lqr-8\,{b}^{3}{l}^{3}mq-24\,{b}^{3}lmq{r}^{2}+8\,{b}
^{3}l{q}^{3}r+16\,{b}^{3}lq{r}^{3}\nonumber\\
&-&8\,{a}^{3}{b}^{2}l+8\,a{b}^{2}{l}^{
3}+16\,a{b}^{2}lmr-8\,a{b}^{2}l{q}^{2}-8\,a{b}^{2}l{r}^{2},      \nonumber\nonumber\\
b_4&=&  -2\,{a}^{6}{b}^{4}+21\,{a}^{4}{b}^{4}{l}^{2}+4\,{a}^{4}{b
}^{4}{m}^{2}+12\,{a}^{4}{b}^{4}mr-6\,{a}^{4}{b}^{4}{q}^{2}-3\,{a}^{4}{
b}^{4}{r}^{2}-28\,{a}^{2}{b}^{4}{l}^{4}+8\,{a}^{2}{b}^{4}{l}^{2}{m}^{2
}\nonumber\\
&-&84\,{a}^{2}{b}^{4}{l}^{2}mr+32\,{a}^{2}{b}^{4}{l}^{2}{q}^{2}+36\,{a}
^{2}{b}^{4}{l}^{2}{r}^{2}-24\,{a}^{2}{b}^{4}{m}^{2}{r}^{2}+24\,{a}^{2}
{b}^{4}m{q}^{2}r+12\,{a}^{2}{b}^{4}m{r}^{3}-4\,{a}^{2}{b}^{4}{q}^{4}\nonumber\\
&-&8
\,{a}^{2}{b}^{4}{q}^{2}{r}^{2}+9\,{b}^{4}{l}^{6}+4\,{b}^{4}{l}^{4}{m}^
{2}+24\,{b}^{4}{l}^{4}mr-18\,{b}^{4}{l}^{4}{q}^{2}-9\,{b}^{4}{l}^{4}{r
}^{2}+36\,{b}^{4}{l}^{2}{m}^{2}{r}^{2}-32\,{b}^{4}{l}^{2}m{q}^{2}r\nonumber\\
&-&40
\,{b}^{4}{l}^{2}m{r}^{3}+9\,{b}^{4}{l}^{2}{q}^{4}+12\,{b}^{4}{l}^{2}{q
}^{2}{r}^{2}+15\,{b}^{4}{l}^{2}{r}^{4}+{b}^{4}{q}^{4}{r}^{2}-2\,{b}^{4
}{q}^{2}{r}^{4}+{b}^{4}{r}^{6}+4\,{a}^{3}{b}^{3}qr-4\,a{b}^{3}{l}^{2}q
r\nonumber\\
&-&8\,a{b}^{3}mq{r}^{2}+4\,a{b}^{3}{q}^{3}r+4\,a{b}^{3}q{r}^{3}-2\,{a}^
{4}{b}^{2}+2\,{a}^{2}{b}^{2}{l}^{2}+4\,{a}^{2}{b}^{2}mr-2\,{a}^{2}{b}^
{2}{q}^{2}-2\,{a}^{2}{b}^{2}{r}^{2}, \nonumber\nonumber\\
b_5&=&6ab^4l\Delta_r^2,\nn\\
b_6&=&a^2b^4\Delta_r^2.
\end{eqnarray}
Finally, the coefficients that appear in the Maxwell's field  $A'_t$ for the Melvin-Kerr-Newman-NUT black hole,  in equation (\ref{ATMAG}), are given by

\begin{eqnarray}
e_0&=& -2\,{a}^{6}{b}^{6}mq-{a}^{6}{b}^{6}qr+8\,{a}^{4}{b}^{6}{l}^{2}mq+19\,
{a}^{4}{b}^{6}{l}^{2}qr-8\,{a}^{4}{b}^{6}{m}^{2}qr-2\,{a}^{4}{b}^{6}m{
q}^{3}-4\,{a}^{4}{b}^{6}mq{r}^{2}-{a}^{4}{b}^{6}{q}^{3}r\nonumber\\
&-&{a}^{4}{b}^{6
}q{r}^{3}-2\,{a}^{2}{b}^{6}{l}^{4}mq+23\,{a}^{2}{b}^{6}{l}^{4}qr-12\,{
a}^{2}{b}^{6}{l}^{2}mq{r}^{2}-3\,{a}^{2}{b}^{6}{l}^{2}{q}^{3}r-8\,{a}^
{2}{b}^{6}{l}^{2}q{r}^{3}-2\,{a}^{2}{b}^{6}mq{r}^{4}\nonumber\\
&-&{a}^{2}{b}^{6}{q}
^{3}{r}^{3}+{a}^{2}{b}^{6}q{r}^{5}-{b}^{6}{l}^{6}qr-5\,{b}^{6}{l}^{4}q
{r}^{3}+5\,{b}^{6}{l}^{2}q{r}^{5}+{b}^{6}q{r}^{7}+6\,{a}^{5}{b}^{5}{l}
^{2}+4\,{a}^{5}{b}^{5}{m}^{2}+2\,{a}^{5}{b}^{5}mr\nonumber\\
&-&2\,{a}^{5}{b}^{5}{q}
^{2}+26\,{a}^{3}{b}^{5}{l}^{4}+4\,{a}^{3}{b}^{5}{l}^{2}{m}^{2}+5\,{a}^
{3}{b}^{5}{l}^{2}{q}^{2}+6\,{a}^{3}{b}^{5}{l}^{2}{r}^{2}-12\,{a}^{3}{b
}^{5}{m}^{2}{r}^{2}-18\,{a}^{3}{b}^{5}m{q}^{2}r-4\,{a}^{3}{b}^{5}m{r}^
{3}\nonumber\\
&-&2\,{a}^{3}{b}^{5}{q}^{4}-5\,{a}^{3}{b}^{5}{q}^{2}{r}^{2}+8\,a{b}^{
5}{l}^{6}+2\,a{b}^{5}{l}^{4}mr-3\,a{b}^{5}{l}^{4}{q}^{2}-12\,a{b}^{5}{
l}^{2}m{r}^{3}-14\,a{b}^{5}{l}^{2}{q}^{2}{r}^{2}-6\,a{b}^{5}m{r}^{5}\nonumber\\
&-&3
\,a{b}^{5}{q}^{2}{r}^{4}+4\,{a}^{4}{b}^{4}mq-{a}^{4}{b}^{4}qr+4\,{a}^{
2}{b}^{4}{l}^{2}mq+8\,{a}^{2}{b}^{4}{l}^{2}qr-28\,{a}^{2}{b}^{4}mq{r}^
{2}-11\,{a}^{2}{b}^{4}{q}^{3}r-6\,{a}^{2}{b}^{4}q{r}^{3}\nonumber\\
&-&{b}^{4}{l}^{4
}qr-18\,{b}^{4}{l}^{2}q{r}^{3}-5\,{b}^{4}q{r}^{5}+8\,{a}^{3}{b}^{3}{l}
^{2}+4\,{a}^{3}{b}^{3}{m}^{2}+4\,{a}^{3}{b}^{3}mr+10\,a{b}^{3}{l}^{4}+
4\,a{b}^{3}{l}^{2}mr-a{b}^{3}{l}^{2}{q}^{2}\nonumber\\
&+&2\,a{b}^{3}{l}^{2}{r}^{2}-
4\,a{b}^{3}m{r}^{3}-17\,a{b}^{3}{q}^{2}{r}^{2}+6\,{a}^{2}{b}^{2}mq+{a}
^{2}{b}^{2}qr+{b}^{2}{l}^{2}qr-5\,{b}^{2}q{r}^{3}\nonumber\\
&+&2\,ab{l}^{2}+2\,abmr
+2\,ab{q}^{2}+qr,      \nonumber\nonumber\\
e_1&=& 16\,{a}^{5}{b}^{6}lmq+18\,{a}^{5}{b}^{6}lqr-
24\,{a}^{3}{b}^{6}{l}^{3}mq-20\,{a}^{3}{b}^{6}{l}^{3}qr+8\,{a}^{3}{b}^
{6}l{m}^{2}qr+4\,{a}^{3}{b}^{6}lm{q}^{3}-64\,{a}^{3}{b}^{6}lmq{r}^{2}\nonumber\\
&+&
24\,{a}^{3}{b}^{6}l{q}^{3}r+20\,{a}^{3}{b}^{6}lq{r}^{3}-38\,a{b}^{6}{l
}^{5}qr-16\,a{b}^{6}{l}^{3}mq{r}^{2}+12\,a{b}^{6}{l}^{3}{q}^{3}r+28\,a
{b}^{6}{l}^{3}q{r}^{3}+16\,a{b}^{6}lmq{r}^{4}\nonumber\\
&-&8\,a{b}^{6}l{q}^{3}{r}^{
3}+2\,a{b}^{6}lq{r}^{5}+2\,{a}^{6}{b}^{5}l+34\,{a}^{4}{b}^{5}{l}^{3}-
28\,{a}^{4}{b}^{5}lmr+22\,{a}^{4}{b}^{5}l{q}^{2}+14\,{a}^{4}{b}^{5}l{r
}^{2}-34\,{a}^{2}{b}^{5}{l}^{5}\nonumber\\
&-&48\,{a}^{2}{b}^{5}{l}^{3}mr+4\,{a}^{2}
{b}^{5}{l}^{3}{q}^{2}+12\,{a}^{2}{b}^{5}{l}^{3}{r}^{2}+48\,{a}^{2}{b}^
{5}l{m}^{2}{r}^{2}+8\,{a}^{2}{b}^{5}lm{q}^{2}r-48\,{a}^{2}{b}^{5}lm{r}
^{3}+2\,{a}^{2}{b}^{5}l{q}^{4}\nonumber\\
&-&16\,{a}^{2}{b}^{5}l{q}^{2}{r}^{2}+6\,{a
}^{2}{b}^{5}l{r}^{4}-2\,{b}^{5}{l}^{7}-4\,{b}^{5}{l}^{5}mr+2\,{b}^{5}{
l}^{5}{q}^{2}+14\,{b}^{5}{l}^{5}{r}^{2}+24\,{b}^{5}{l}^{3}m{r}^{3}-12
\,{b}^{5}{l}^{3}{q}^{2}{r}^{2}\nonumber\\
&-&6\,{b}^{5}{l}^{3}{r}^{4}+12\,{b}^{5}lm{
r}^{5}-6\,{b}^{5}l{q}^{2}{r}^{4}-6\,{b}^{5}l{r}^{6}+12\,{a}^{3}{b}^{4}
lqr-24\,a{b}^{4}{l}^{3}qr+48\,a{b}^{4}lmq{r}^{2}+6\,a{b}^{4}l{q}^{3}r\nonumber\\
&-&
36\,a{b}^{4}lq{r}^{3}+4\,{a}^{4}{b}^{3}l+4\,{a}^{2}{b}^{3}{l}^{3}-32\,
{a}^{2}{b}^{3}lmr+10\,{a}^{2}{b}^{3}l{q}^{2}+16\,{a}^{2}{b}^{3}l{r}^{2
}-4\,{b}^{3}{l}^{5}-8\,{b}^{3}{l}^{3}mr+4\,{b}^{3}{l}^{3}{q}^{2}\nonumber\\
&+&8\,{b
}^{3}{l}^{3}{r}^{2}+8\,{b}^{3}lm{r}^{3}+12\,{b}^{3}l{q}^{2}{r}^{2}-4\,
{b}^{3}l{r}^{4}-6\,a{b}^{2}lqr+2\,{a}^{2}bl-2\,b{l}^{3}-4\,blmr+2\,bl{
q}^{2}+2\,bl{r}^{2} ,      \nonumber\nonumber\\
e_2&=& 4\,{a}^{6}{b}^{6}mq+7\,{a}^{6}{b
}^{6}qr-12\,{a}^{4}{b}^{6}{l}^{2}mq-91\,{a}^{4}{b}^{6}{l}^{2}qr-8\,{a}
^{4}{b}^{6}{m}^{3}q+4\,{a}^{4}{b}^{6}{m}^{2}qr+6\,{a}^{4}{b}^{6}m{q}^{
3}\nonumber\\
&-&10\,{a}^{4}{b}^{6}mq{r}^{2}
+15\,{a}^{4}{b}^{6}{q}^{3}r+11\,{a}^{4}{
b}^{6}q{r}^{3}+44\,{a}^{2}{b}^{6}{l}^{4}mq-47\,{a}^{2}{b}^{6}{l}^{4}qr
-12\,{a}^{2}{b}^{6}{l}^{2}{m}^{2}qr-8\,{a}^{2}{b}^{6}{l}^{2}m{q}^{3}\nonumber\\
&+&
144\,{a}^{2}{b}^{6}{l}^{2}mq{r}^{2}-44\,{a}^{2}{b}^{6}{l}^{2}{q}^{3}r-
2\,{a}^{2}{b}^{6}{l}^{2}q{r}^{3}+28\,{a}^{2}{b}^{6}{m}^{2}q{r}^{3}-42
\,{a}^{2}{b}^{6}m{q}^{3}{r}^{2}-16\,{a}^{2}{b}^{6}mq{r}^{4}\nonumber\\
&+&9\,{a}^{2}
{b}^{6}{q}^{5}r+18\,{a}^{2}{b}^{6}{q}^{3}{r}^{3}+{a}^{2}{b}^{6}q{r}^{5
}+11\,{b}^{6}{l}^{6}qr+6\,{b}^{6}{l}^{4}mq{r}^{2}-9\,{b}^{6}{l}^{4}{q}
^{3}r-{b}^{6}{l}^{4}q{r}^{3}-20\,{b}^{6}{l}^{2}mq{r}^{4}\nonumber\\
&+&14\,{b}^{6}{l
}^{2}{q}^{3}{r}^{3}-7\,{b}^{6}{l}^{2}q{r}^{5}-2\,{b}^{6}mq{r}^{6}+3\,{
b}^{6}{q}^{3}{r}^{5}-3\,{b}^{6}q{r}^{7}+12\,{a}^{5}{b}^{5}{l}^{2}-18\,
{a}^{5}{b}^{5}mr+7\,{a}^{5}{b}^{5}{q}^{2}\nonumber\\
&-&144\,{a}^{3}{b}^{5}{l}^{4}-
24\,{a}^{3}{b}^{5}{l}^{2}{m}^{2}-36\,{a}^{3}{b}^{5}{l}^{2}mr+6\,{a}^{3
}{b}^{5}{l}^{2}{q}^{2}-24\,{a}^{3}{b}^{5}{l}^{2}{r}^{2}-24\,{a}^{3}{b}
^{5}{m}^{3}r-12\,{a}^{3}{b}^{5}{m}^{2}{q}^{2}\nonumber\\
&+&72\,{a}^{3}{b}^{5}{m}^{2
}{r}^{2}+4\,{a}^{3}{b}^{5}m{q}^{2}r-12\,{a}^{3}{b}^{5}m{r}^{3}+10\,{a}
^{3}{b}^{5}{q}^{4}+22\,{a}^{3}{b}^{5}{q}^{2}{r}^{2}+12\,a{b}^{5}{l}^{6
}+54\,a{b}^{5}{l}^{4}mr\nonumber\\
&-&7\,a{b}^{5}{l}^{4}{q}^{2}-24\,a{b}^{5}{l}^{2}{
m}^{2}{r}^{2}-20\,a{b}^{5}{l}^{2}m{q}^{2}r+132\,a{b}^{5}{l}^{2}m{r}^{3
}+a{b}^{5}{l}^{2}{q}^{4}+16\,a{b}^{5}{l}^{2}{q}^{2}{r}^{2}-36\,a{b}^{5
}{l}^{2}{r}^{4}\nonumber\\
&+&24\,a{b}^{5}{m}^{2}{r}^{4}-36\,a{b}^{5}m{q}^{2}{r}^{3}
+6\,a{b}^{5}m{r}^{5}-3\,a{b}^{5}{q}^{4}{r}^{2}+15\,a{b}^{5}{q}^{2}{r}^
{4}-4\,{a}^{4}{b}^{4}mq-20\,{a}^{2}{b}^{4}{l}^{2}mq\nonumber\\
&-&66\,{a}^{2}{b}^{4}
{l}^{2}qr-52\,{a}^{2}{b}^{4}{m}^{2}qr-4\,{a}^{2}{b}^{4}m{q}^{3}+80\,{a
}^{2}{b}^{4}mq{r}^{2}+27\,{a}^{2}{b}^{4}{q}^{3}r+16\,{a}^{2}{b}^{4}q{r
}^{3}\nonumber\\
&-&36\,{b}^{4}{l}^{2}mq{r}^{2}+6\,{b}^{4}{l}^{2}{q}^{3}r+80\,{b}^{4
}{l}^{2}q{r}^{3}+4\,{b}^{4}mq{r}^{4}-14\,{b}^{4}{q}^{3}{r}^{3}+16\,{b}
^{4}q{r}^{5}-6\,{a}^{3}{b}^{3}{l}^{2}+4\,{a}^{3}{b}^{3}{m}^{2}\nonumber\\
&-&24\,{a}
^{3}{b}^{3}mr+3\,{a}^{3}{b}^{3}{q}^{2}-12\,a{b}^{3}{l}^{4}+32\,a{b}^{3
}{l}^{2}mr-10\,a{b}^{3}{l}^{2}{q}^{2}-28\,a{b}^{3}{l}^{2}{r}^{2}+24\,a
{b}^{3}{m}^{2}{r}^{2}-52\,a{b}^{3}m{q}^{2}r\nonumber\\
&-&8\,a{b}^{3}m{r}^{3}+4\,a{b
}^{3}{q}^{4}+58\,a{b}^{3}{q}^{2}{r}^{2}-9\,{a}^{2}{b}^{2}qr+3\,{b}^{2}
{l}^{2}qr+6\,{b}^{2}mq{r}^{2}\nonumber\\
&-&9\,{b}^{2}{q}^{3}r+9\,{b}^{2}q{r}^{3}-6
\,ab{l}^{2}-6\,abmr-2\,qr,       \nonumber\nonumber\\
e_3&=&  2\,{a}^{5}{b}^{6}lmq
-74\,{a}^{5}{b}^{6}lqr+8\,{a}^{3}{b}^{6}{l}^{3}mq+124\,{a}^{3}{b}^{6}{
l}^{3}qr+8\,{a}^{3}{b}^{6}l{m}^{3}q-120\,{a}^{3}{b}^{6}l{m}^{2}qr+26\,
{a}^{3}{b}^{6}lm{q}^{3}\nonumber\\
&+&300\,{a}^{3}{b}^{6}lmq{r}^{2}-104\,{a}^{3}{b}^
{6}l{q}^{3}r-84\,{a}^{3}{b}^{6}lq{r}^{3}-34\,a{b}^{6}{l}^{5}mq+70\,a{b
}^{6}{l}^{5}qr-32\,a{b}^{6}{l}^{3}{m}^{2}qr+18\,a{b}^{6}{l}^{3}m{q}^{3
}\nonumber\\
&-&24\,a{b}^{6}{l}^{3}mq{r}^{2}+16\,a{b}^{6}{l}^{3}{q}^{3}r-36\,a{b}^{6
}{l}^{3}q{r}^{3}-32\,a{b}^{6}l{m}^{2}q{r}^{3}+54\,a{b}^{6}lm{q}^{3}{r}
^{2}-22\,a{b}^{6}lmq{r}^{4}\nonumber
\end{eqnarray}
\begin{eqnarray}
&-&18\,a{b}^{6}l{q}^{5}r+8\,a{b}^{6}l{q}^{3}{
r}^{3}-10\,a{b}^{6}lq{r}^{5}-152\,{a}^{4}{b}^{5}{l}^{3}-32\,{a}^{4}{b}
^{5}l{m}^{2}+72\,{a}^{4}{b}^{5}lmr-16\,{a}^{4}{b}^{5}l{q}^{2}\nonumber\\
&-&48\,{a}^
{4}{b}^{5}l{r}^{2}+152\,{a}^{2}{b}^{5}{l}^{5}+32\,{a}^{2}{b}^{5}{l}^{3
}{m}^{2}+224\,{a}^{2}{b}^{5}{l}^{3}mr-108\,{a}^{2}{b}^{5}{l}^{3}{q}^{2
}-56\,{a}^{2}{b}^{5}{l}^{3}{r}^{2}+64\,{a}^{2}{b}^{5}l{m}^{3}r\nonumber\\
&-&8\,{a}^
{2}{b}^{5}l{m}^{2}{q}^{2}-176\,{a}^{2}{b}^{5}l{m}^{2}{r}^{2}-112\,{a}^
{2}{b}^{5}lm{q}^{2}r+176\,{a}^{2}{b}^{5}lm{r}^{3}+20\,{a}^{2}{b}^{5}l{
q}^{4}+60\,{a}^{2}{b}^{5}l{q}^{2}{r}^{2}\nonumber\\
&-&32\,{a}^{2}{b}^{5}l{r}^{4}+8
\,{b}^{5}{l}^{5}mr-4\,{b}^{5}{l}^{5}{q}^{2}-48\,{b}^{5}{l}^{5}{r}^{2}+
16\,{b}^{5}{l}^{3}{m}^{2}{r}^{2}-16\,{b}^{5}{l}^{3}m{q}^{2}r-112\,{b}^
{5}{l}^{3}m{r}^{3}+4\,{b}^{5}{l}^{3}{q}^{4}\nonumber\\
&+&72\,{b}^{5}{l}^{3}{q}^{2}{
r}^{2}+32\,{b}^{5}{l}^{3}{r}^{4}-16\,{b}^{5}l{m}^{2}{r}^{4}+48\,{b}^{5
}lm{q}^{2}{r}^{3}-24\,{b}^{5}lm{r}^{5}-20\,{b}^{5}l{q}^{4}{r}^{2}-4\,{
b}^{5}l{q}^{2}{r}^{4}+16\,{b}^{5}l{r}^{6}\nonumber\\
&-&20\,{a}^{3}{b}^{4}lmq-32\,{a
}^{3}{b}^{4}lqr+20\,a{b}^{4}{l}^{3}mq+56\,a{b}^{4}{l}^{3}qr+64\,a{b}^{
4}l{m}^{2}qr-8\,a{b}^{4}lm{q}^{3}-172\,a{b}^{4}lmq{r}^{2}\nn\\
&-&
32\,a{b}^{4}
l{q}^{3}r+112\,a{b}^{4}lq{r}^{3}-4\,{a}^{4}{b}^{3}l-8\,{a}^{2}{b}^{3}{
l}^{3}-8\,{a}^{2}{b}^{3}l{m}^{2}+88\,{a}^{2}{b}^{3}lmr-20\,{a}^{2}{b}^
{3}l{q}^{2}-48\,{a}^{2}{b}^{3}l{r}^{2}\nonumber\\
&+&4\,{b}^{3}{l}^{5}-8\,{b}^{3}{l}
^{3}{r}^{2}-16\,{b}^{3}l{m}^{2}{r}^{2}+32\,{b}^{3}lm{q}^{2}r
-4\,{b}^{3
}l{q}^{4}-48\,{b}^{3}l{q}^{2}{r}^{2}+4\,{b}^{3}l{r}^{4}
-6\,a{b}^{2}lmq
+18\,a{b}^{2}lqr\nonumber\\
&-&4\,{a}^{2}bl+4\,b{l}^{3}+8\,blmr-4\,bl{q}^{2}-4\,bl{r
}^{2} ,   \nonumber\nonumber\\
e_4&=& 4\,{a}^{6}{b}^{6}mq-18\,{a}^{6}{b}^{6}qr
-68\,{a}^{4}{b}^{6}{l}^{2}mq+206\,{a}^{4}{b}^{6}{l}^{2}qr+8\,{a}^{4}{b
}^{6}{m}^{3}q-16\,{a}^{4}{b}^{6}{m}^{2}qr+14\,{a}^{4}{b}^{6}m{q}^{3}\nonumber\\
&+&
80\,{a}^{4}{b}^{6}mq{r}^{2}-46\,{a}^{4}{b}^{6}{q}^{3}r-34\,{a}^{4}{b}^
{6}q{r}^{3}-28\,{a}^{2}{b}^{6}{l}^{4}mq-50\,{a}^{2}{b}^{6}{l}^{4}qr-8
\,{a}^{2}{b}^{6}{l}^{2}{m}^{3}q\nonumber\\
&+&204\,{a}^{2}{b}^{6}{l}^{2}{m}^{2}qr-60
\,{a}^{2}{b}^{6}{l}^{2}m{q}^{3}-516\,{a}^{2}{b}^{6}{l}^{2}mq{r}^{2}+
192\,{a}^{2}{b}^{6}{l}^{2}{q}^{3}r+104\,{a}^{2}{b}^{6}{l}^{2}q{r}^{3}+
72\,{a}^{2}{b}^{6}{m}^{3}q{r}^{2}\nonumber\\
&-&68\,{a}^{2}{b}^{6}{m}^{2}{q}^{3}r-
160\,{a}^{2}{b}^{6}{m}^{2}q{r}^{3}+14\,{a}^{2}{b}^{6}m{q}^{5}+166\,{a}
^{2}{b}^{6}m{q}^{3}{r}^{2}+84\,{a}^{2}{b}^{6}mq{r}^{4}-30\,{a}^{2}{b}^
{6}{q}^{5}r\nonumber\\
&-&54\,{a}^{2}{b}^{6}{q}^{3}{r}^{3}-14\,{a}^{2}{b}^{6}q{r}^{5
}+8\,{b}^{6}{l}^{6}mq-18\,{b}^{6}{l}^{6}qr+12\,{b}^{6}{l}^{4}{m}^{2}qr
-8\,{b}^{6}{l}^{4}m{q}^{3}-8\,{b}^{6}{l}^{4}mq{r}^{2}+8\,{b}^{6}{l}^{4
}{q}^{3}r\nonumber\\
&+&22\,{b}^{6}{l}^{4}q{r}^{3}+4\,{b}^{6}{l}^{2}{m}^{2}q{r}^{3}-
20\,{b}^{6}{l}^{2}m{q}^{3}{r}^{2}+56\,{b}^{6}{l}^{2}mq{r}^{4}+9\,{b}^{
6}{l}^{2}{q}^{5}r-32\,{b}^{6}{l}^{2}{q}^{3}{r}^{3}-6\,{b}^{6}{l}^{2}q{
r}^{5}\nonumber\\
&-&4\,{b}^{6}m{q}^{3}{r}^{4}+8\,{b}^{6}mq{r}^{6}+3\,{b}^{6}{q}^{5}{r}^{3}-8\,{b}^{6}{q}^{3}{r}^{5}+2\,{b}^{6}q{r}^{7}-74\,{a}^{5}{b}^{5}
{l}^{2}-20\,{a}^{5}{b}^{5}{m}^{2}+36\,{a}^{5}{b}^{5}mr\nonumber\\
&-&5\,{a}^{5}{b}^{
5}{q}^{2}+272\,{a}^{3}{b}^{5}{l}^{4}+64\,{a}^{3}{b}^{5}{l}^{2}{m}^{2}+
112\,{a}^{3}{b}^{5}{l}^{2}mr-114\,{a}^{3}{b}^{5}{l}^{2}{q}^{2}+40\,{a}
^{3}{b}^{5}{l}^{2}{r}^{2}+64\,{a}^{3}{b}^{5}{m}^{3}r\nonumber\\
&+&4\,{a}^{3}{b}^{5}
{m}^{2}{q}^{2}-128\,{a}^{3}{b}^{5}{m}^{2}{r}^{2}+28\,{a}^{3}{b}^{5}m{q
}^{2}r+56\,{a}^{3}{b}^{5}m{r}^{3}-{a}^{3}{b}^{5}{q}^{4}-32\,{a}^{3}{b}
^{5}{q}^{2}{r}^{2}-78\,a{b}^{5}{l}^{6}\nonumber\\
&-&20\,a{b}^{5}{l}^{4}{m}^{2}-156
\,a{b}^{5}{l}^{4}mr+105\,a{b}^{5}{l}^{4}{q}^{2}-12\,a{b}^{5}{l}^{4}{r}
^{2}-40\,a{b}^{5}{l}^{2}{m}^{3}r+8\,a{b}^{5}{l}^{2}{m}^{2}{q}^{2}\nonumber\\
&+&104
\,a{b}^{5}{l}^{2}{m}^{2}{r}^{2}+192\,a{b}^{5}{l}^{2}m{q}^{2}r-344\,a{b
}^{5}{l}^{2}m{r}^{3}-64\,a{b}^{5}{l}^{2}{q}^{4}-18\,a{b}^{5}{l}^{2}{q}
^{2}{r}^{2}+114\,a{b}^{5}{l}^{2}{r}^{4}\nonumber\\
&+&72\,a{b}^{5}{m}^{2}{q}^{2}{r}^
{2}-76\,a{b}^{5}{m}^{2}{r}^{4}-62\,a{b}^{5}m{q}^{4}r+56\,a{b}^{5}m{q}^
{2}{r}^{3}+20\,a{b}^{5}m{r}^{5}+10\,a{b}^{5}{q}^{6}+22\,a{b}^{5}{q}^{4
}{r}^{2}\nonumber\\
&-&27\,a{b}^{5}{q}^{2}{r}^{4}-12\,{a}^{4}{b}^{4}mq+6\,{a}^{4}{b}
^{4}qr+44\,{a}^{2}{b}^{4}{l}^{2}mq+102\,{a}^{2}{b}^{4}{l}^{2}qr+108\,{
a}^{2}{b}^{4}{m}^{2}qr-72\,{a}^{2}{b}^{4}mq{r}^{2}\nonumber\\
&-&27\,{a}^{2}{b}^{4}{
q}^{3}r-12\,{a}^{2}{b}^{4}q{r}^{3}-8\,{b}^{4}{l}^{4}mq+6\,{b}^{4}{l}^{
4}qr-28\,{b}^{4}{l}^{2}{m}^{2}qr+8\,{b}^{4}{l}^{2}m{q}^{3}+116\,{b}^{4
}{l}^{2}mq{r}^{2}\nonumber\\
&-&6\,{b}^{4}{l}^{2}{q}^{3}r-132\,{b}^{4}{l}^{2}q{r}^{3
}+12\,{b}^{4}m{q}^{3}{r}^{2}-12\,{b}^{4}mq{r}^{4}-9\,{b}^{4}{q}^{5}r+
30\,{b}^{4}{q}^{3}{r}^{3}-18\,{b}^{4}q{r}^{5}-12\,{a}^{3}{b}^{3}{l}^{2
}\nonumber\\
&-&16\,{a}^{3}{b}^{3}{m}^{2}+32\,{a}^{3}{b}^{3}mr-6\,{a}^{3}{b}^{3}{q}^
{2}-6\,a{b}^{3}{l}^{4}+4\,a{b}^{3}{l}^{2}{m}^{2}-88\,a{b}^{3}{l}^{2}mr
+23\,a{b}^{3}{l}^{2}{q}^{2}+58\,a{b}^{3}{l}^{2}{r}^{2}\nonumber\\
&-&52\,a{b}^{3}{m}
^{2}{r}^{2}+108\,a{b}^{3}m{q}^{2}r+32\,a{b}^{3}m{r}^{3}-8\,a{b}^{3}{q}
^{4}-69\,a{b}^{3}{q}^{2}{r}^{2}-18\,{a}^{2}{b}^{2}mq+15\,{a}^{2}{b}^{2
}qr\nonumber\\
&-&9\,{b}^{2}{l}^{2}qr-12\,{b}^{2}mq{r}^{2}+18\,{b}^{2}{q}^{3}r-3\,{b
}^{2}q{r}^{3}+6\,ab{l}^{2}+6\,abmr-6\,ab{q}^{2}+qr,      \nonumber\nonumber\\
e_5&=& -72\,{a}^{5}{b}^{6}lmq+116\,{a}^{5}{b}^{6}lqr+112\,{a}^{3}{b}^
{6}{l}^{3}mq-248\,{a}^{3}{b}^{6}{l}^{3}qr-16\,{a}^{3}{b}^{6}l{m}^{3}q+
304\,{a}^{3}{b}^{6}l{m}^{2}qr\nonumber\\
&-&108\,{a}^{3}{b}^{6}lm{q}^{3}-528\,{a}^{3
}{b}^{6}lmq{r}^{2}+184\,{a}^{3}{b}^{6}l{q}^{3}r+136\,{a}^{3}{b}^{6}lq{
r}^{3}+32\,a{b}^{6}{l}^{5}mq+12\,a{b}^{6}{l}^{5}qr\nonumber\\
&-&56\,a{b}^{6}{l}^{3}
{m}^{2}qr+24\,a{b}^{6}{l}^{3}m{q}^{3}+264\,a{b}^{6}{l}^{3}mq{r}^{2}-
132\,a{b}^{6}{l}^{3}{q}^{3}r-56\,a{b}^{6}{l}^{3}q{r}^{3}-96\,a{b}^{6}l
{m}^{3}q{r}^{2}\nonumber\\
&+&96\,a{b}^{6}l{m}^{2}{q}^{3}r+224\,a{b}^{6}l{m}^{2}q{r}
^{3}-24\,a{b}^{6}lm{q}^{5}-240\,a{b}^{6}lm{q}^{3}{r}^{2}-72\,a{b}^{6}l
mq{r}^{4}+62\,a{b}^{6}l{q}^{5}r\nonumber\\
&+&40\,a{b}^{6}l{q}^{3}{r}^{3}+20\,a{b}^{
6}lq{r}^{5}-12\,{a}^{6}{b}^{5}l+220\,{a}^{4}{b}^{5}{l}^{3}+64\,{a}^{4}
{b}^{5}l{m}^{2}-48\,{a}^{4}{b}^{5}lmr-60\,{a}^{4}{b}^{5}l{q}^{2}\nonumber\\
&+&60\,{
a}^{4}{b}^{5}l{r}^{2}-220\,{a}^{2}{b}^{5}{l}^{5}-72\,{a}^{2}{b}^{5}{l}
^{3}{m}^{2}-320\,{a}^{2}{b}^{5}{l}^{3}mr+256\,{a}^{2}{b}^{5}{l}^{3}{q}
^{2}+64\,{a}^{2}{b}^{5}{l}^{3}{r}^{2}-128\,{a}^{2}{b}^{5}l{m}^{3}r\nonumber\\
&+&16
\,{a}^{2}{b}^{5}l{m}^{2}{q}^{2}+208\,{a}^{2}{b}^{5}l{m}^{2}{r}^{2}+264
\,{a}^{2}{b}^{5}lm{q}^{2}r-240\,{a}^{2}{b}^{5}lm{r}^{3}-80\,{a}^{2}{b}
^{5}l{q}^{4}-84\,{a}^{2}{b}^{5}l{q}^{2}{r}^{2}\nonumber\\
&+&60\,{a}^{2}{b}^{5}l{r}^
{4}+12\,{b}^{5}{l}^{7}+8\,{b}^{5}{l}^{5}{m}^{2}-24\,{b}^{5}{l}^{5}{q}^ {2}+60\,{b}^{5}{l}^{5}{r}^{2}+16\,{b}^{5}{l}^{3}{m}^{3}r-8\,{b}^{5}{l}
^{3}{m}^{2}{q}^{2}-56\,{b}^{5}{l}^{3}{m}^{2}{r}^{2}\nonumber
\end{eqnarray}
\begin{eqnarray}
&-&24\,{b}^{5}{l}^{3}
m{q}^{2}r+192\,{b}^{5}{l}^{3}m{r}^{3}+26\,{b}^{5}{l}^{3}{q}^{4}-96\,{b
}^{5}{l}^{3}{q}^{2}{r}^{2}-60\,{b}^{5}{l}^{3}{r}^{4}-48\,{b}^{5}l{m}^{
2}{q}^{2}{r}^{2}+48\,{b}^{5}l{m}^{2}{r}^{4}\nonumber\\
&+&52\,{b}^{5}lm{q}^{4}r-72\,
{b}^{5}lm{q}^{2}{r}^{3}-14\,{b}^{5}l{q}^{6}+22\,{b}^{5}l{q}^{4}{r}^{2}
+24\,{b}^{5}l{q}^{2}{r}^{4}-12\,{b}^{5}l{r}^{6}+44\,{a}^{3}{b}^{4}lmq+
24\,{a}^{3}{b}^{4}lqr\nonumber\\
&-&44\,a{b}^{4}{l}^{3}mq-36\,a{b}^{4}{l}^{3}qr-136
\,a{b}^{4}l{m}^{2}qr+20\,a{b}^{4}lm{q}^{3}+212\,a{b}^{4}lmq{r}^{2}+42
\,a{b}^{4}l{q}^{3}r-120\,a{b}^{4}lq{r}^{3}\nonumber\\
&-&4\,{a}^{4}{b}^{3}l+4\,{a}^{
2}{b}^{3}{l}^{3}+16\,{a}^{2}{b}^{3}l{m}^{2}-80\,{a}^{2}{b}^{3}lmr+10\,
{a}^{2}{b}^{3}l{q}^{2}+48\,{a}^{2}{b}^{3}l{r}^{2}+4\,{b}^{3}{l}^{5}+24
\,{b}^{3}{l}^{3}mr\nonumber\\
&-&12\,{b}^{3}{l}^{3}{q}^{2}-8\,{b}^{3}{l}^{3}{r}^{2}+
32\,{b}^{3}l{m}^{2}{r}^{2}-64\,{b}^{3}lm{q}^{2}r-24\,{b}^{3}lm{r}^{3}+
8\,{b}^{3}l{q}^{4}+60\,{b}^{3}l{q}^{2}{r}^{2}+4\,{b}^{3}l{r}^{4}\nonumber\\
&+&12\,a
{b}^{2}lmq-18\,a{b}^{2}lqr+2\,{a}^{2}bl-2\,b{l}^{3}-4\,blmr+2\,bl{q}^{
2}+2\,bl{r}^{2},      \nonumber\nonumber\\
e_6&=& -16\,{a}^{6}{b}^{6}mq+22\,{a}^
{6}{b}^{6}qr+156\,{a}^{4}{b}^{6}{l}^{2}mq-230\,{a}^{4}{b}^{6}{l}^{2}qr
+8\,{a}^{4}{b}^{6}{m}^{3}q+72\,{a}^{4}{b}^{6}{m}^{2}qr-46\,{a}^{4}{b}^
{6}m{q}^{3}\nonumber\\
&-&140\,{a}^{4}{b}^{6}mq{r}^{2}+58\,{a}^{4}{b}^{6}{q}^{3}r+46
\,{a}^{4}{b}^{6}q{r}^{3}-88\,{a}^{2}{b}^{6}{l}^{4}mq+170\,{a}^{2}{b}^{
6}{l}^{4}qr+16\,{a}^{2}{b}^{6}{l}^{2}{m}^{3}q\nonumber\\
&-&420\,{a}^{2}{b}^{6}{l}^{
2}{m}^{2}qr+168\,{a}^{2}{b}^{6}{l}^{2}m{q}^{3}+716\,{a}^{2}{b}^{6}{l}^
{2}mq{r}^{2}-266\,{a}^{2}{b}^{6}{l}^{2}{q}^{3}r-188\,{a}^{2}{b}^{6}{l}
^{2}q{r}^{3}\nonumber\\
&-&168\,{a}^{2}{b}^{6}{m}^{3}q{r}^{2}+156\,{a}^{2}{b}^{6}{m}
^{2}{q}^{3}r+280\,{a}^{2}{b}^{6}{m}^{2}q{r}^{3}-32\,{a}^{2}{b}^{6}m{q}
^{5}-234\,{a}^{2}{b}^{6}m{q}^{3}{r}^{2}-136\,{a}^{2}{b}^{6}mq{r}^{4}\nonumber\\
&+&
38\,{a}^{2}{b}^{6}{q}^{5}r+64\,{a}^{2}{b}^{6}{q}^{3}{r}^{3}+26\,{a}^{2
}{b}^{6}q{r}^{5}-8\,{b}^{6}{l}^{6}mq-2\,{b}^{6}{l}^{6}qr+4\,{b}^{6}{l}
^{4}{m}^{2}qr-44\,{b}^{6}{l}^{4}mq{r}^{2}\nonumber\\
&+&30\,{b}^{6}{l}^{4}{q}^{3}r-
10\,{b}^{6}{l}^{4}q{r}^{3}+24\,{b}^{6}{l}^{2}{m}^{3}q{r}^{2}-28\,{b}^{
6}{l}^{2}{m}^{2}{q}^{3}r-52\,{b}^{6}{l}^{2}{m}^{2}q{r}^{3}+8\,{b}^{6}{
l}^{2}m{q}^{5}+84\,{b}^{6}{l}^{2}m{q}^{3}{r}^{2}\nonumber\\
&-&32\,{b}^{6}{l}^{2}mq{
r}^{4}-29\,{b}^{6}{l}^{2}{q}^{5}r+12\,{b}^{6}{l}^{2}{q}^{3}{r}^{3}+10
\,{b}^{6}{l}^{2}q{r}^{5}-2\,{b}^{6}m{q}^{5}{r}^{2}+12\,{b}^{6}m{q}^{3}
{r}^{4}-12\,{b}^{6}mq{r}^{6}+{b}^{6}{q}^{7}r\nonumber\\
&-&7\,{b}^{6}{q}^{5}{r}^{3}+
6\,{b}^{6}{q}^{3}{r}^{5}+2\,{b}^{6}q{r}^{7}+90\,{a}^{5}{b}^{5}{l}^{2}+
20\,{a}^{5}{b}^{5}{m}^{2}-20\,{a}^{5}{b}^{5}mr-7\,{a}^{5}{b}^{5}{q}^{2
}-220\,{a}^{3}{b}^{5}{l}^{4}\nonumber\\
&-&72\,{a}^{3}{b}^{5}{l}^{2}{m}^{2}-120\,{a}
^{3}{b}^{5}{l}^{2}mr+202\,{a}^{3}{b}^{5}{l}^{2}{q}^{2}-36\,{a}^{3}{b}^
{5}{l}^{2}{r}^{2}-48\,{a}^{3}{b}^{5}{m}^{3}r+24\,{a}^{3}{b}^{5}{m}^{2}
{q}^{2}\nonumber\\
&+&72\,{a}^{3}{b}^{5}{m}^{2}{r}^{2}+24\,{a}^{3}{b}^{5}m{q}^{2}r-
56\,{a}^{3}{b}^{5}m{r}^{3}-28\,{a}^{3}{b}^{5}{q}^{4}+14\,{a}^{3}{b}^{5
}{q}^{2}{r}^{2}+90\,a{b}^{5}{l}^{6}+44\,a{b}^{5}{l}^{4}{m}^{2}\nonumber\\
&+&140\,a{
b}^{5}{l}^{4}mr-185\,a{b}^{5}{l}^{4}{q}^{2}+28\,a{b}^{5}{l}^{4}{r}^{2}
+88\,a{b}^{5}{l}^{2}{m}^{3}r-20\,a{b}^{5}{l}^{2}{m}^{2}{q}^{2}-152\,a{
b}^{5}{l}^{2}{m}^{2}{r}^{2}\nonumber\\
&-&336\,a{b}^{5}{l}^{2}m{q}^{2}r+360\,a{b}^{5
}{l}^{2}m{r}^{3}+135\,a{b}^{5}{l}^{2}{q}^{4}+44\,a{b}^{5}{l}^{2}{q}^{2 }{r}^{2}-126\,a{b}^{5}{l}^{2}{r}^{4}-156\,a{b}^{5}{m}^{2}{q}^{2}{r}^{2
}\nonumber\\
&+&84\,a{b}^{5}{m}^{2}{r}^{4}+138\,a{b}^{5}m{q}^{4}r-36\,a{b}^{5}m{r}^{
5}-24\,a{b}^{5}{q}^{6}-39\,a{b}^{5}{q}^{4}{r}^{2}+21\,a{b}^{5}{q}^{2}{
r}^{4}+20\,{a}^{4}{b}^{4}mq\nonumber\\
&-&8\,{a}^{4}{b}^{4}qr-44\,{a}^{2}{b}^{4}{l}^
{2}mq-38\,{a}^{2}{b}^{4}{l}^{2}qr-60\,{a}^{2}{b}^{4}{m}^{2}qr+12\,{a}^
{2}{b}^{4}m{q}^{3}+16\,{a}^{2}{b}^{4}mq{r}^{2}+17\,{a}^{2}{b}^{4}{q}^{
3}r\nonumber\\
&+&16\,{b}^{4}{l}^{4}mq-8\,{b}^{4}{l}^{4}qr+56\,{b}^{4}{l}^{2}{m}^{2}
qr-16\,{b}^{4}{l}^{2}m{q}^{3}-124\,{b}^{4}{l}^{2}mq{r}^{2}-6\,{b}^{4}{
l}^{2}{q}^{3}r+96\,{b}^{4}{l}^{2}q{r}^{3}\nonumber\\
&-&24\,{b}^{4}m{q}^{3}{r}^{2}+
12\,{b}^{4}mq{r}^{4}+18\,{b}^{4}{q}^{5}r-18\,{b}^{4}{q}^{3}{r}^{3}+8\,
{b}^{4}q{r}^{5}+10\,{a}^{3}{b}^{3}{l}^{2}+4\,{a}^{3}{b}^{3}{m}^{2}-8\,
{a}^{3}{b}^{3}mr\nonumber\\
&+&3\,{a}^{3}{b}^{3}{q}^{2}+8\,a{b}^{3}{l}^{4}-8\,a{b}^{
3}{l}^{2}{m}^{2}+64\,a{b}^{3}{l}^{2}mr-12\,a{b}^{3}{l}^{2}{q}^{2}-40\,
a{b}^{3}{l}^{2}{r}^{2}+32\,a{b}^{3}{m}^{2}{r}^{2}-60\,a{b}^{3}m{q}^{2}
r\nonumber\\
&-&24\,a{b}^{3}m{r}^{3}+4\,a{b}^{3}{q}^{4}+32\,a{b}^{3}{q}^{2}{r}^{2}+
12\,{a}^{2}{b}^{2}mq-7\,{a}^{2}{b}^{2}qr+5\,{b}^{2}{l}^{2}qr+6\,{b}^{2
}mq{r}^{2}-9\,{b}^{2}{q}^{3}r-{b}^{2}q{r}^{3}\nonumber\\
&-&2\,ab{l}^{2}-2\,abmr+4\,
ab{q}^{2} ,     \nonumber\nonumber\\
e_7&=& 76\,{a}^{5}{b}^{6}lmq-84\,{a}^{5}{b}
^{6}lqr-156\,{a}^{3}{b}^{6}{l}^{3}mq+208\,{a}^{3}{b}^{6}{l}^{3}qr+8\,{
a}^{3}{b}^{6}l{m}^{3}q-288\,{a}^{3}{b}^{6}l{m}^{2}qr\nonumber\\
&+&126\,{a}^{3}{b}^{
6}lm{q}^{3}+424\,{a}^{3}{b}^{6}lmq{r}^{2}-156\,{a}^{3}{b}^{6}l{q}^{3}r
-104\,{a}^{3}{b}^{6}lq{r}^{3}+40\,a{b}^{6}{l}^{5}mq-84\,a{b}^{6}{l}^{5
}qr\nonumber\\
&+&216\,a{b}^{6}{l}^{3}{m}^{2}qr-106\,a{b}^{6}{l}^{3}m{q}^{3}-404\,a{
b}^{6}{l}^{3}mq{r}^{2}+172\,a{b}^{6}{l}^{3}{q}^{3}r+112\,a{b}^{6}{l}^{
3}q{r}^{3}+200\,a{b}^{6}l{m}^{3}q{r}^{2}\nonumber\\
&-&200\,a{b}^{6}l{m}^{2}{q}^{3}r
-368\,a{b}^{6}l{m}^{2}q{r}^{3}+50\,a{b}^{6}lm{q}^{5}+330\,a{b}^{6}lm{q
}^{3}{r}^{2}+156\,a{b}^{6}lmq{r}^{4}-72\,a{b}^{6}l{q}^{5}r\nonumber\\
&-&76\,a{b}^{6
}l{q}^{3}{r}^{3}-20\,a{b}^{6}lq{r}^{5}+16\,{a}^{6}{b}^{5}l-120\,{a}^{4
}{b}^{5}{l}^{3}-32\,{a}^{4}{b}^{5}l{m}^{2}-8\,{a}^{4}{b}^{5}lmr+80\,{a
}^{4}{b}^{5}l{q}^{2}-32\,{a}^{4}{b}^{5}l{r}^{2}\nonumber\\
&+&120\,{a}^{2}{b}^{5}{l}
^{5}+48\,{a}^{2}{b}^{5}{l}^{3}{m}^{2}+160\,{a}^{2}{b}^{5}{l}^{3}mr-204
\,{a}^{2}{b}^{5}{l}^{3}{q}^{2}-8\,{a}^{2}{b}^{5}{l}^{3}{r}^{2}+64\,{a}
^{2}{b}^{5}l{m}^{3}r-8\,{a}^{2}{b}^{5}l{m}^{2}{q}^{2}\nonumber\\
&-&80\,{a}^{2}{b}^{
5}l{m}^{2}{r}^{2}-224\,{a}^{2}{b}^{5}lm{q}^{2}r+144\,{a}^{2}{b}^{5}lm{
r}^{3}+92\,{a}^{2}{b}^{5}l{q}^{4}+52\,{a}^{2}{b}^{5}l{q}^{2}{r}^{2}-48
\,{a}^{2}{b}^{5}l{r}^{4}-16\,{b}^{5}{l}^{7}\nonumber\\
&-&16\,{b}^{5}{l}^{5}{m}^{2}-
8\,{b}^{5}{l}^{5}mr+52\,{b}^{5}{l}^{5}{q}^{2}-32\,{b}^{5}{l}^{5}{r}^{2
}-32\,{b}^{5}{l}^{3}{m}^{3}r+16\,{b}^{5}{l}^{3}{m}^{2}{q}^{2}+64\,{b}^
{5}{l}^{3}{m}^{2}{r}^{2}+96\,{b}^{5}{l}^{3}m{q}^{2}r\nonumber\\
&-&144\,{b}^{5}{l}^{
3}m{r}^{3}-64\,{b}^{5}{l}^{3}{q}^{4}+24\,{b}^{5}{l}^{3}{q}^{2}{r}^{2}+
48\,{b}^{5}{l}^{3}{r}^{4}+96\,{b}^{5}l{m}^{2}{q}^{2}{r}^{2}-48\,{b}^{5
}l{m}^{2}{r}^{4}-104\,{b}^{5}lm{q}^{4}r\nonumber\\
&+&24\,{b}^{5}lm{r}^{5}+28\,{b}^{
5}l{q}^{6}+16\,{b}^{5}l{q}^{4}{r}^{2}
-12\,{b}^{5}l{q}^{2}{r}^{4}-28\,{
a}^{3}{b}^{4}lmq+28\,a{b}^{4}{l}^{3}mq+80\,a{b}^{4}l{m}^{2}qr\nonumber
\end{eqnarray}
\begin{eqnarray}
&-&16\,a{b}
^{4}lm{q}^{3}-100\,a{b}^{4}lmq{r}^{2}
-12\,a{b}^{4}l{q}^{3}r+48\,a{b}^{
4}lq{r}^{3}+4\,{a}^{4}{b}^{3}l-8\,{a}^{2}{b}^{3}l{m}^{2}+24\,{a}^{2}{b
}^{3}lmr\nonumber\\
&-&16\,{a}^{2}{b}^{3}l{r}^{2}-4\,{b}^{3}{l}^{5}-16\,{b}^{3}{l}^{
3}mr+8\,{b}^{3}{l}^{3}{q}^{2}+8\,{b}^{3}{l}^{3}{r}^{2}-16\,{b}^{3}l{m}
^{2}{r}^{2}+32\,{b}^{3}lm{q}^{2}r\nonumber\\
&+&16\,{b}^{3}lm{r}^{3}-4\,{b}^{3}l{q}^
{4}-24\,{b}^{3}l{q}^{2}{r}^{2}-4\,{b}^{3}l{r}^{4}-6\,a{b}^{2}lmq+6\,a{
b}^{2}lqr,\nonumber\nonumber\\
e_8&=&14\,{a}^{6}{b}^{6}mq-13\,{a}^{6}{b}^
{6}qr-100\,{a}^{4}{b}^{6}{l}^{2}mq+111\,{a}^{4}{b}^{6}{l}^{2}qr-8\,{a}
^{4}{b}^{6}{m}^{3}q-72\,{a}^{4}{b}^{6}{m}^{2}qr+36\,{a}^{4}{b}^{6}m{q}
^{3}\nonumber\\
&+&100\,{a}^{4}{b}^{6}mq{r}^{2}-33\,{a}^{4}{b}^{6}{q}^{3}r-29\,{a}^{
4}{b}^{6}q{r}^{3}+94\,{a}^{2}{b}^{6}{l}^{4}mq-117\,{a}^{2}{b}^{6}{l}^{
4}qr-8\,{a}^{2}{b}^{6}{l}^{2}{m}^{3}q\nonumber\\
&+&276\,{a}^{2}{b}^{6}{l}^{2}{m}^{2
}qr-124\,{a}^{2}{b}^{6}{l}^{2}m{q}^{3}-400\,{a}^{2}{b}^{6}{l}^{2}mq{r}
^{2}+147\,{a}^{2}{b}^{6}{l}^{2}{q}^{3}r+112\,{a}^{2}{b}^{6}{l}^{2}q{r}
^{3}+120\,{a}^{2}{b}^{6}{m}^{3}q{r}^{2}\nonumber\\
&-&108\,{a}^{2}{b}^{6}{m}^{2}{q}^
{3}r-192\,{a}^{2}{b}^{6}{m}^{2}q{r}^{3}+22\,{a}^{2}{b}^{6}m{q}^{5}+138
\,{a}^{2}{b}^{6}m{q}^{3}{r}^{2}+94\,{a}^{2}{b}^{6}mq{r}^{4}-22\,{a}^{2
}{b}^{6}{q}^{5}r\nonumber\\
&-&33\,{a}^{2}{b}^{6}{q}^{3}{r}^{3}-19\,{a}^{2}{b}^{6}q{
r}^{5}-8\,{b}^{6}{l}^{6}mq+19\,{b}^{6}{l}^{6}qr-44\,{b}^{6}{l}^{4}{m}^
{2}qr+24\,{b}^{6}{l}^{4}m{q}^{3}+88\,{b}^{6}{l}^{4}mq{r}^{2}\nonumber\\
&-&48\,{b}^{
6}{l}^{4}{q}^{3}r-17\,{b}^{6}{l}^{4}q{r}^{3}-48\,{b}^{6}{l}^{2}{m}^{3}
q{r}^{2}+56\,{b}^{6}{l}^{2}{m}^{2}{q}^{3}r+92\,{b}^{6}{l}^{2}{m}^{2}q{
r}^{3}-16\,{b}^{6}{l}^{2}m{q}^{5}-108\,{b}^{6}{l}^{2}m{q}^{3}{r}^{2}\nonumber\\
&-&
24\,{b}^{6}{l}^{2}mq{r}^{4}+31\,{b}^{6}{l}^{2}{q}^{5}r+16\,{b}^{6}{l}^
{2}{q}^{3}{r}^{3}+{b}^{6}{l}^{2}q{r}^{5}+4\,{b}^{6}m{q}^{5}{r}^{2}-12
\,{b}^{6}m{q}^{3}{r}^{4}+8\,{b}^{6}mq{r}^{6}-2\,{b}^{6}{q}^{7}r\nonumber\\
&+&5\,{b}
^{6}{q}^{5}{r}^{3}-3\,{b}^{6}q{r}^{7}-36\,{a}^{5}{b}^{5}{l}^{2}-6\,{a}
^{5}{b}^{5}mr+11\,{a}^{5}{b}^{5}{q}^{2}+70\,{a}^{3}{b}^{5}{l}^{4}+28\,
{a}^{3}{b}^{5}{l}^{2}{m}^{2}+48\,{a}^{3}{b}^{5}{l}^{2}mr\nonumber\\
&-&111\,{a}^{3}{
b}^{5}{l}^{2}{q}^{2}+18\,{a}^{3}{b}^{5}{l}^{2}{r}^{2}-12\,{a}^{3}{b}^{
5}{m}^{2}{q}^{2}+12\,{a}^{3}{b}^{5}{m}^{2}{r}^{2}-58\,{a}^{3}{b}^{5}m{
q}^{2}r+12\,{a}^{3}{b}^{5}m{r}^{3}+29\,{a}^{3}{b}^{5}{q}^{4}\nonumber\\
&+&5\,{a}^{3
}{b}^{5}{q}^{2}{r}^{2}-34\,a{b}^{5}{l}^{6}-28\,a{b}^{5}{l}^{4}{m}^{2}-
38\,a{b}^{5}{l}^{4}mr+98\,a{b}^{5}{l}^{4}{q}^{2}-20\,a{b}^{5}{l}^{4}{r
}^{2}-56\,a{b}^{5}{l}^{2}{m}^{3}r\nonumber\\
&+&16\,a{b}^{5}{l}^{2}{m}^{2}{q}^{2}+88
\,a{b}^{5}{l}^{2}{m}^{2}{r}^{2}+176\,a{b}^{5}{l}^{2}m{q}^{2}r-156\,a{b
}^{5}{l}^{2}m{r}^{3}-82\,a{b}^{5}{l}^{2}{q}^{4}-28\,a{b}^{5}{l}^{2}{q}
^{2}{r}^{2}\nonumber\\
&+&54\,a{b}^{5}{l}^{2}{r}^{4}+96\,a{b}^{5}{m}^{2}{q}^{2}{r}^{
2}-36\,a{b}^{5}{m}^{2}{r}^{4}-90\,a{b}^{5}m{q}^{4}r-24\,a{b}^{5}m{q}^{
2}{r}^{3}+18\,a{b}^{5}m{r}^{5}+18\,a{b}^{5}{q}^{6}\nonumber\\
&+&24\,a{b}^{5}{q}^{4}
{r}^{2}-6\,a{b}^{5}{q}^{2}{r}^{4}-8\,{a}^{4}{b}^{4}mq+3\,{a}^{4}{b}^{4
}qr+16\,{a}^{2}{b}^{4}{l}^{2}mq-6\,{a}^{2}{b}^{4}{l}^{2}qr+4\,{a}^{2}{
b}^{4}{m}^{2}qr-8\,{a}^{2}{b}^{4}m{q}^{3}\nonumber\\
&+&4\,{a}^{2}{b}^{4}mq{r}^{2}-6
\,{a}^{2}{b}^{4}{q}^{3}r+2\,{a}^{2}{b}^{4}q{r}^{3}-8\,{b}^{4}{l}^{4}mq
+3\,{b}^{4}{l}^{4}qr-28\,{b}^{4}{l}^{2}{m}^{2}qr+8\,{b}^{4}{l}^{2}m{q}
^{3}+44\,{b}^{4}{l}^{2}mq{r}^{2}\nonumber\\
&+&6\,{b}^{4}{l}^{2}{q}^{3}r-26\,{b}^{4}
{l}^{2}q{r}^{3}+12\,{b}^{4}m{q}^{3}{r}^{2}-4\,{b}^{4}mq{r}^{4}-9\,{b}^
{4}{q}^{5}r+2\,{b}^{4}{q}^{3}{r}^{3}-{b}^{4}q{r}^{5}+4\,{a}^{3}{b}^{3}
{m}^{2}-4\,{a}^{3}{b}^{3}mr\nonumber\\
&+&4\,a{b}^{3}{l}^{2}{m}^{2}-12\,a{b}^{3}{l}^
{2}mr+8\,a{b}^{3}{l}^{2}{r}^{2}-4\,a{b}^{3}{m}^{2}{r}^{2}+4\,a{b}^{3}m
{q}^{2}r+4\,a{b}^{3}m{r}^{3}-4\,a{b}^{3}{q}^{2}{r}^{2},\nonumber\nonumber\\
e_9&=&-2\,{b}^{4}l \Delta_r 
 \left( 12\,{a}^{3}{b}^{2}mq-13\,{a}^{3}{b}^{2}qr-20\,a{b}^{2}{l}^{2}m
q+21\,a{b}^{2}{l}^{2}qr-28\,a{b}^{2}{m}^{2}qr+14\,a{b}^{2}m{q}^{3}\right.\nn\\
&+&\left.34
\,a{b}^{2}mq{r}^{2}-15\,a{b}^{2}{q}^{3}r-5\,a{b}^{2}q{r}^{3}+3\,{a}^{4
}b-6\,{a}^{2}b{l}^{2}+10\,{a}^{2}b{q}^{2}-6\,{a}^{2}b{r}^{2}+3\,b{l}^{
4}+4\,b{l}^{2}{m}^{2}\right.\nn\\
&-&\left.4\,b{l}^{2}mr-10\,b{l}^{2}{q}^{2}+6\,b{l}^{2}{r}
^{2}-12\,bm{q}^{2}r+4\,bm{r}^{3}+7\,b{q}^{4}+2\,b{q}^{2}{r}^{2}-b{r}^{
4}-2\,amq+2\,aqr \right),
\nn\\
e_{10}&=&-{b}^{5} \Delta_r \left( 4\,{a}^{4}bmq-3\,{a}^{4}bqr-12\,{a}^{2
}b{l}^{2}mq+12\,{a}^{2}b{l}^{2}qr-12\,{a}^{2}b{m}^{2}qr+4\,{a}^{2}bm{q
}^{3}\right.\nn\\
&+&\left.16\,{a}^{2}bmq{r}^{2}-4\,{a}^{2}b{q}^{3}r-4\,{a}^{2}bq{r}^{3}+8
\,b{l}^{4}mq-9\,b{l}^{4}qr+12\,b{l}^{2}{m}^{2}qr-8\,b{l}^{2}m{q}^{3}-
16\,b{l}^{2}mq{r}^{2}\right.\nn\\
&+&\left.10\,b{l}^{2}{q}^{3}r+2\,b{l}^{2}q{r}^{3}-b{q}^{5
}r+2\,b{q}^{3}{r}^{3}-bq{r}^{5}-2\,{a}^{3}{l}^{2}+4\,{a}^{3}{m}^{2}-6
\,{a}^{3}mr+4\,{a}^{3}{q}^{2}+2\,a{l}^{4}\right.\nn\\
&+&\left.4\,a{l}^{2}{m}^{2}-6\,a{l}^{
2}mr-6\,a{l}^{2}{q}^{2}+6\,a{l}^{2}{r}^{2}-6\,am{q}^{2}r+2\,am{r}^{3}+
4\,a{q}^{4} \right),\nn\\
e_{11}&=&+2\,a{b}^{6}lq \left( m-r \right)  \Delta_r ^{2},
\end{eqnarray}
and
\begin{eqnarray}
f_0&=& 9\,{a}^{4}{b}^{4}{l}^{2}+4\,{a}^{4}{b}^{4}{m}^{2}+4\,{a}^{4}{b}^{4}mr
+{a}^{4}{b}^{4}{r}^{2}+6\,{a}^{2}{b}^{4}{l}^{4}+12\,{a}^{2}{b}^{4}{l}^
{2}mr+4\,{a}^{2}{b}^{4}m{r}^{3}+2\,{a}^{2}{b}^{4}{r}^{4}+{b}^{4}{l}^{6
}\nonumber\\
&+&7\,{b}^{4}{l}^{4}{r}^{2}+7\,{b}^{4}{l}^{2}{r}^{4}+{b}^{4}{r}^{6}+8\,
{a}^{3}{b}^{3}mq+4\,{a}^{3}{b}^{3}qr+12\,a{b}^{3}{l}^{2}qr+4\,a{b}^{3}
q{r}^{3}+6\,{a}^{2}{b}^{2}{l}^{2}+4\,{a}^{2}{b}^{2}mr\nonumber\\
&+&4\,{a}^{2}{b}^{2
}{q}^{2}+2\,{a}^{2}{b}^{2}{r}^{2}+2\,{b}^{2}{l}^{4}+4\,{b}^{2}{l}^{2}{
r}^{2}+2\,{b}^{2}{r}^{4}+4\,abqr+{l}^{2}+{r}^{2},      \nonumber\nonumber\\
f_1&=&   6\,{a}^{5}{b}
^{4}l-28\,{a}^{3}{b}^{4}{l}^{3}-8\,{a}^{3}{b}^{4}l{m}^{2}-24\,{a}^{3}{
b}^{4}lmr+12\,{a}^{3}{b}^{4}l{q}^{2}+12\,{a}^{3}{b}^{4}l{r}^{2}-10\,a{
b}^{4}{l}^{5}-24\,a{b}^{4}{l}^{3}mr\nonumber\\
&+&4\,a{b}^{4}{l}^{3}{q}^{2}+36\,a{b}
^{4}{l}^{3}{r}^{2}+8\,a{b}^{4}lm{r}^{3}-4\,a{b}^{4}l{q}^{2}{r}^{2}+6\,
a{b}^{4}l{r}^{4}-16\,{a}^{2}{b}^{3}lmq-16\,{b}^{3}{l}^{3}qr+8\,{a}^{3}
{b}^{2}l\nonumber\\
&-&8\,a{b}^{2}{l}^{3}-16\,a{b}^{2}lmr-4\,a{b}^{2}l{q}^{2}+8\,a{b
}^{2}l{r}^{2}-8\,blqr+2\,al ,    \nonumber\nonumber
\end{eqnarray}
\begin{eqnarray}
f_2&=&    {a}^{6}{b}^{4}-46\,{a}^{
4}{b}^{4}{l}^{2}-20\,{a}^{4}{b}^{4}mr+4\,{a}^{4}{b}^{4}{q}^{2}-2\,{a}^
{4}{b}^{4}{r}^{2}+25\,{a}^{2}{b}^{4}{l}^{4}+12\,{a}^{2}{b}^{4}{l}^{2}{
m}^{2}+60\,{a}^{2}{b}^{4}{l}^{2}mr\nonumber\\
&-&38\,{a}^{2}{b}^{4}{l}^{2}{q}^{2}-18
\,{a}^{2}{b}^{4}{l}^{2}{r}^{2}+36\,{a}^{2}{b}^{4}{m}^{2}{r}^{2}-20\,{a
}^{2}{b}^{4}m{q}^{2}r-20\,{a}^{2}{b}^{4}m{r}^{3}+4\,{a}^{2}{b}^{4}{q}^
{4}+6\,{a}^{2}{b}^{4}{q}^{2}{r}^{2}\nonumber\\
&-&7\,{a}^{2}{b}^{4}{r}^{4}+4\,{b}^{4
}{l}^{6}+24\,{b}^{4}{l}^{4}mr-6\,{b}^{4}{l}^{4}{q}^{2}-44\,{b}^{4}{l}^
{4}{r}^{2}-8\,{b}^{4}{l}^{2}m{r}^{3}+12\,{b}^{4}{l}^{2}{q}^{2}{r}^{2}-
20\,{b}^{4}{l}^{2}{r}^{4}+2\,{b}^{4}{q}^{2}{r}^{4}\nonumber\\
&-&4\,{b}^{4}{r}^{6}-8
\,{a}^{3}{b}^{3}mq-16\,{a}^{3}{b}^{3}qr+16\,a{b}^{3}{l}^{2}mq-32\,a{b}
^{3}{l}^{2}qr+24\,a{b}^{3}mq{r}^{2}-4\,a{b}^{3}{q}^{3}r-16\,a{b}^{3}q{
r}^{3}\nonumber\\
&+&2\,{a}^{4}{b}^{2}-28\,{a}^{2}{b}^{2}{l}^{2}-16\,{a}^{2}{b}^{2}m
r-4\,{a}^{2}{b}^{2}{q}^{2}-4\,{a}^{2}{b}^{2}{r}^{2}+2\,{b}^{2}{l}^{4}+
16\,{b}^{2}{l}^{2}mr-2\,{b}^{2}{l}^{2}{q}^{2}-20\,{b}^{2}{l}^{2}{r}^{2
}\nonumber\\
&+&6\,{b}^{2}{q}^{2}{r}^{2}-6\,{b}^{2}{r}^{4}-12\,abqr+{a}^{2}-2\,{l}^{
2}-2\,{r}^{2} ,   \nonumber\nonumber\\
f_3&=&   -24\,{a}^{5}{b}^{4}l+96\,{a}^{3}
{b}^{4}{l}^{3}+8\,{a}^{3}{b}^{4}l{m}^{2}+128\,{a}^{3}{b}^{4}lmr-48\,{a
}^{3}{b}^{4}l{q}^{2}-48\,{a}^{3}{b}^{4}l{r}^{2}-8\,a{b}^{4}{l}^{5}-8\,
a{b}^{4}{l}^{3}{m}^{2}\nonumber\\
&-&16\,a{b}^{4}{l}^{3}mr+32\,a{b}^{4}{l}^{3}{q}^{2
}-64\,a{b}^{4}{l}^{3}{r}^{2}-72\,a{b}^{4}l{m}^{2}{r}^{2}+56\,a{b}^{4}l
m{q}^{2}r+32\,a{b}^{4}lm{r}^{3}-12\,a{b}^{4}l{q}^{4}\nonumber\\
&-&24\,a{b}^{4}l{r}^
{4}+24\,{a}^{2}{b}^{3}lmq+16\,{a}^{2}{b}^{3}lqr-8\,{b}^{3}{l}^{3}mq+32
\,{b}^{3}{l}^{3}qr
-24\,{b}^{3}lmq{r}^{2}+8\,{b}^{3}l{q}^{3}r+16\,{b}^{
3}lq{r}^{3}\nonumber\\
&-&24\,{a}^{3}{b}^{2}l+24\,a{b}^{2}{l}^{3}+48\,a{b}^{2}lmr-24
\,a{b}^{2}l{r}^{2}+16\,blqr-4\,al,     \nonumber\nonumber\\
f_4&=&  -4\,{a}^{6}{
b}^{4}+86\,{a}^{4}{b}^{4}{l}^{2}-8\,{a}^{4}{b}^{4}{m}^{2}+40\,{a}^{4}{
b}^{4}mr-14\,{a}^{4}{b}^{4}{q}^{2}-2\,{a}^{4}{b}^{4}{r}^{2}-96\,{a}^{2
}{b}^{4}{l}^{4}
-16\,{a}^{2}{b}^{4}{l}^{2}{m}^{2}\nonumber\\
&-&240\,{a}^{2}{b}^{4}{l
}^{2}mr+108\,{a}^{2}{b}^{4}{l}^{2}{q}^{2}+72\,{a}^{2}{b}^{4}{l}^{2}{r}
^{2}-96\,{a}^{2}{b}^{4}{m}^{2}{r}^{2}+64\,{a}^{2}{b}^{4}m{q}^{2}r+40\,
{a}^{2}{b}^{4}m{r}^{3}-12\,{a}^{2}{b}^{4}{q}^{4}\nonumber\\
&-&20\,{a}^{2}{b}^{4}{q}
^{2}{r}^{2}+8\,{a}^{2}{b}^{4}{r}^{4}-2\,{b}^{4}{l}^{6}+4\,{b}^{4}{l}^{
4}{m}^{2}-24\,{b}^{4}{l}^{4}mr-6\,{b}^{4}{l}^{4}{q}^{2}+58\,{b}^{4}{l}
^{4}{r}^{2}+36\,{b}^{4}{l}^{2}{m}^{2}{r}^{2}\nonumber\\
&-&32\,{b}^{4}{l}^{2}m{q}^{2
}r-24\,{b}^{4}{l}^{2}m{r}^{3}+9\,{b}^{4}{l}^{2}{q}^{4}-12\,{b}^{4}{l}^
{2}{q}^{2}{r}^{2}+34\,{b}^{4}{l}^{2}{r}^{4}+{b}^{4}{q}^{4}{r}^{2}-6\,{
b}^{4}{q}^{2}{r}^{4}+6\,{b}^{4}{r}^{6}-8\,{a}^{3}{b}^{3}mq\nonumber\\
&+&24\,{a}^{3}
{b}^{3}qr-32\,a{b}^{3}{l}^{2}mq+24\,a{b}^{3}{l}^{2}qr-56\,a{b}^{3}mq{r
}^{2}+12\,a{b}^{3}{q}^{3}r+24\,a{b}^{3}q{r}^{3}-6\,{a}^{4}{b}^{2}+40\,
{a}^{2}{b}^{2}{l}^{2}\nonumber\\
&+&24\,{a}^{2}{b}^{2}mr-6\,{a}^{2}{b}^{2}{q}^{2}-10
\,{b}^{2}{l}^{4}-32\,{b}^{2}{l}^{2}mr+4\,{b}^{2}{l}^{2}{q}^{2}+28\,{b}
^{2}{l}^{2}{r}^{2}-12\,{b}^{2}{q}^{2}{r}^{2}+6\,{b}^{2}{r}^{4}+12\,abq
r\nonumber\\
&-&2\,{a}^{2}+{l}^{2}+{r}^{2} ,     \nonumber\nonumber\\
f_5&=& 36\,{a}^{5}{b}^{4
}l-120\,{a}^{3}{b}^{4}{l}^{3}+8\,{a}^{3}{b}^{4}l{m}^{2}-208\,{a}^{3}{b
}^{4}lmr+72\,{a}^{3}{b}^{4}l{q}^{2}+72\,{a}^{3}{b}^{4}l{r}^{2}+52\,a{b
}^{4}{l}^{5}+16\,a{b}^{4}{l}^{3}{m}^{2}\nonumber\\
&+&128\,a{b}^{4}{l}^{3}mr-88\,a{b
}^{4}{l}^{3}{q}^{2}+8\,a{b}^{4}{l}^{3}{r}^{2}+168\,a{b}^{4}l{m}^{2}{r}
^{2}-136\,a{b}^{4}lm{q}^{2}r-112\,a{b}^{4}lm{r}^{3}+30\,a{b}^{4}l{q}^{
4}\nonumber\\
&+&24\,a{b}^{4}l{q}^{2}{r}^{2}+36\,a{b}^{4}l{r}^{4}-32\,{a}^{2}{b}^{3}
lqr+16\,{b}^{3}{l}^{3}mq-16\,{b}^{3}{l}^{3}qr+48\,{b}^{3}lmq{r}^{2}-16
\,{b}^{3}l{q}^{3}r-32\,{b}^{3}lq{r}^{3}\nonumber\\
&+&24\,{a}^{3}{b}^{2}l-24\,a{b}^{
2}{l}^{3}-48\,a{b}^{2}lmr+12\,a{b}^{2}l{q}^{2}+24\,a{b}^{2}l{r}^{2}-8
\,blqr+2\,al,     \nonumber\nonumber\\
f_6&=&  6\,{a}^{6}{b}^{4}-72\,{a}^{4}{b}^
{4}{l}^{2}-40\,{a}^{4}{b}^{4}mr+18\,{a}^{4}{b}^{4}{q}^{2}+8\,{a}^{4}{b
}^{4}{r}^{2}+94\,{a}^{2}{b}^{4}{l}^{4}-4\,{a}^{2}{b}^{4}{l}^{2}{m}^{2}
+256\,{a}^{2}{b}^{4}{l}^{2}mr\nonumber\\
&-&104\,{a}^{2}{b}^{4}{l}^{2}{q}^{2}-92\,{a
}^{2}{b}^{4}{l}^{2}{r}^{2}+88\,{a}^{2}{b}^{4}{m}^{2}{r}^{2}-72\,{a}^{2
}{b}^{4}m{q}^{2}r-40\,{a}^{2}{b}^{4}m{r}^{3}+13\,{a}^{2}{b}^{4}{q}^{4}
+24\,{a}^{2}{b}^{4}{q}^{2}{r}^{2}\nonumber\\
&-&2\,{a}^{2}{b}^{4}{r}^{4}-12\,{b}^{4}
{l}^{6}-8\,{b}^{4}{l}^{4}{m}^{2}-24\,{b}^{4}{l}^{4}mr+30\,{b}^{4}{l}^{
4}{q}^{2}-12\,{b}^{4}{l}^{4}{r}^{2}-72\,{b}^{4}{l}^{2}{m}^{2}{r}^{2}+
64\,{b}^{4}{l}^{2}m{q}^{2}r\nonumber\\
&+&72\,{b}^{4}{l}^{2}m{r}^{3}-18\,{b}^{4}{l}^
{2}{q}^{4}-12\,{b}^{4}{l}^{2}{q}^{2}{r}^{2}-36\,{b}^{4}{l}^{2}{r}^{4}-
2\,{b}^{4}{q}^{4}{r}^{2}+6\,{b}^{4}{q}^{2}{r}^{4}-4\,{b}^{4}{r}^{6}+8
\,{a}^{3}{b}^{3}mq\nonumber\\
&-&16\,{a}^{3}{b}^{3}qr+16\,a{b}^{3}{l}^{2}mq+40\,a{b}
^{3}mq{r}^{2}-12\,a{b}^{3}{q}^{3}r-16\,a{b}^{3}q{r}^{3}+6\,{a}^{4}{b}^
{2}-20\,{a}^{2}{b}^{2}{l}^{2}-16\,{a}^{2}{b}^{2}mr\nonumber\\
&+&8\,{a}^{2}{b}^{2}{q
}^{2}+4\,{a}^{2}{b}^{2}{r}^{2}+6\,{b}^{2}{l}^{4}+16\,{b}^{2}{l}^{2}mr-
2\,{b}^{2}{l}^{2}{q}^{2}-12\,{b}^{2}{l}^{2}{r}^{2}+6\,{b}^{2}{q}^{2}{r
}^{2}-2\,{b}^{2}{r}^{4}-4\,abqr+{a}^{2} ,     \nonumber\nonumber\\
f_7&=&-24\,{
a}^{5}{b}^{4}l+64\,{a}^{3}{b}^{4}{l}^{3}-8\,{a}^{3}{b}^{4}l{m}^{2}+128
\,{a}^{3}{b}^{4}lmr-48\,{a}^{3}{b}^{4}l{q}^{2}-48\,{a}^{3}{b}^{4}l{r}^
{2}-40\,a{b}^{4}{l}^{5}-8\,a{b}^{4}{l}^{3}{m}^{2}\nonumber\\
&-&112\,a{b}^{4}{l}^{3}
mr+64\,a{b}^{4}{l}^{3}{q}^{2}+32\,a{b}^{4}{l}^{3}{r}^{2}-120\,a{b}^{4}
l{m}^{2}{r}^{2}+104\,a{b}^{4}lm{q}^{2}r+96\,a{b}^{4}lm{r}^{3}-24\,a{b}
^{4}l{q}^{4}\nonumber\\
&-&32\,a{b}^{4}l{q}^{2}{r}^{2}-24\,a{b}^{4}l{r}^{4}-8\,{a}^{
2}{b}^{3}lmq+16\,{a}^{2}{b}^{3}lqr
-8\,{b}^{3}{l}^{3}mq-24\,{b}^{3}lmq{
r}^{2}+8\,{b}^{3}l{q}^{3}r+16\,{b}^{3}lq{r}^{3}\nonumber\\
&-&8\,{a}^{3}{b}^{2}l+8\,
a{b}^{2}{l}^{3}+16\,a{b}^{2}lmr
-8\,a{b}^{2}l{q}^{2}-8\,a{b}^{2}l{r}^{2
} ,\nonumber\nonumber\\
f_8&=&-4\,{a}^{6}{b}^{4}+25\,{a}^{4}{b}^{4}{l}^{2}
+4\,{a}^{4}{b}^{4}{m}^{2}+20\,{a}^{4}{b}^{4}mr-10\,{a}^{4}{b}^{4}{q}^{
2}-7\,{a}^{4}{b}^{4}{r}^{2}-30\,{a}^{2}{b}^{4}{l}^{4}+8\,{a}^{2}{b}^{4
}{l}^{2}{m}^{2}\nonumber\\
&-&92\,{a}^{2}{b}^{4}{l}^{2}mr+36\,{a}^{2}{b}^{4}{l}^{2}{
q}^{2}+40\,{a}^{2}{b}^{4}{l}^{2}{r}^{2}-32\,{a}^{2}{b}^{4}{m}^{2}{r}^{
2}+32\,{a}^{2}{b}^{4}m{q}^{2}r+20\,{a}^{2}{b}^{4}m{r}^{3}-6\,{a}^{2}{b
}^{4}{q}^{4}\nonumber\\
&-&12\,{a}^{2}{b}^{4}{q}^{2}{r}^{2}-2\,{a}^{2}{b}^{4}{r}^{4}
+9\,{b}^{4}{l}^{6}+4\,{b}^{4}{l}^{4}{m}^{2}+24\,{b}^{4}{l}^{4}mr-18\,{
b}^{4}{l}^{4}{q}^{2}-9\,{b}^{4}{l}^{4}{r}^{2}+36\,{b}^{4}{l}^{2}{m}^{2
}{r}^{2}\nonumber\\
&-&32\,{b}^{4}{l}^{2}m{q}^{2}r-40\,{b}^{4}{l}^{2}m{r}^{3}+9\,{b}
^{4}{l}^{2}{q}^{4}+12\,{b}^{4}{l}^{2}{q}^{2}{r}^{2}+15\,{b}^{4}{l}^{2}
{r}^{4}+{b}^{4}{q}^{4}{r}^{2}-2\,{b}^{4}{q}^{2}{r}^{4}+{b}^{4}{r}^{6}+
4\,{a}^{3}{b}^{3}qr\nonumber
\end{eqnarray}
\begin{eqnarray}
&-&4\,a{b}^{3}{l}^{2}qr-8\,a{b}^{3}mq{r}^{2}+4\,a{b}^
{3}{q}^{3}r+4\,a{b}^{3}q{r}^{3}-2\,{a}^{4}{b}^{2}+2\,{a}^{2}{b}^{2}{l}
^{2}+4\,{a}^{2}{b}^{2}mr-2\,{a}^{2}{b}^{2}{q}^{2}-2\,{a}^{2}{b}^{2}{r}
^{2}, \nonumber\\
f_{9}&=&6ab^4l\Delta_r^2,\nn\\
f_{10}&=&a^2b^4\Delta_r^2.
\end{eqnarray}

\end{document}